\documentclass[fleqn,usenatbib]{mnras}

\usepackage{newtxtext,newtxmath}

\usepackage[T1]{fontenc}
\usepackage{ae,aecompl}

\usepackage[normalem]{ulem} 
\usepackage{graphicx}	
\usepackage{amsmath}	
\usepackage{amssymb}	
\usepackage{color}

\definecolor{asparagus}{rgb}{0.53, 0.66, 0.42}

\def\del#1{{}}

\newcommand{\mat}{\ensuremath{\mathbfss}}

\newcommand{\dif}[1]{\mathrm{d}#1~}
\newcommand{\dd}{\mathrm{d}}
\newcommand{\dps}{\displaystyle}

\newcommand{\bs}[1]{\boldsymbol{#1}}
\newcommand{\ecr}[0]{\varepsilon_{\text{cr}}}
\newcommand{\fcr}[0]{f_{\text{cr}}}
\newcommand{\kcr}[0]{K_{\text{cr}}}
\newcommand{\pcr}[0]{P_{\text{cr}}}
\newcommand{\ewp}[0]{\varepsilon_{{\rm a}, +}}
\newcommand{\pwp}[0]{P_{{\rm a},+}}
\newcommand{\ewm}[0]{\varepsilon_{{\rm a},-}}
\newcommand{\pwm}[0]{P_{{\rm a},-}}
\newcommand{\ewpm}[0]{\varepsilon_{{\rm a},\pm}}
\newcommand{\pwpm}[0]{P_{{\rm a},\pm}}

\title[Cosmic-ray hydrodynamics]{Cosmic-ray hydrodynamics: Alfv\'en-wave regulated transport of cosmic rays}

\author[Thomas \& Pfrommer]{
T. Thomas,$^{1,2}$\thanks{E-mail: tthomas@aip.de (TT), cpfrommer@aip.de (CP)} and
C. Pfrommer,$^{1}$
\\
$^{1}$Leibniz-Institut f\"ur Astrophysik Potsdam (AIP), An der Sternwarte 16, D-14482 Potsdam, Germany\\
$^{2}$Institut f\"ur Physik und Astronomie, Universit\"at Potsdam, Karl-Liebknecht-Strasse 24/25, D-14476 Potsdam, Germany
}

\date{Accepted XXX. Received YYY; in original form ZZZ}

\pubyear{2018}

\begin{document}
\label{firstpage}
\pagerange{\pageref{firstpage}--\pageref{lastpage}}
\maketitle

\begin{abstract}
Star formation in galaxies appears to be self-regulated by energetic feedback processes. Among the most promising agents of feedback are cosmic rays (CRs), the relativistic ion population of interstellar and intergalactic plasmas. In these environments, energetic CRs are virtually collisionless and interact via collective phenomena mediated by kinetic-scale plasma waves and large-scale magnetic fields. The enormous separation of kinetic and global astrophysical scales requires a hydrodynamic description. Here, we develop a new macroscopic theory for CR transport in the self-confinement picture, which includes CR diffusion and streaming. The interaction between CRs and electromagnetic fields of Alfv\'enic turbulence provides the main source of CR scattering, and causes CRs to stream along the magnetic field with the Alfv\'en velocity if resonant waves are sufficiently energetic. However, numerical simulations struggle to capture this effect with current transport formalisms and adopt regularization schemes to ensure numerical stability. We extent the theory by deriving an equation for the CR momentum density along the mean magnetic field and include a transport equation for the Alfv\'en-wave energy. We account for energy exchange of CRs and Alfv\'en waves via the gyroresonant instability and include other wave damping mechanisms. Using numerical simulations we demonstrate that our new theory enables stable, self-regulated CR transport. The theory is coupled to magneto-hydrodynamics, conserves the total energy and momentum, and correctly recovers previous macroscopic CR transport formalisms in the steady-state flux limit. Because it is free of tunable parameters, it holds the promise to provide predictable simulations of CR feedback in galaxy formation.
\end{abstract}

\begin{keywords}
cosmic rays -- hydrodynamics -- radiative transfer -- methods: analytical -- methods: numerical
\end{keywords}



\section{Introduction}
\label{sec:intro}

CRs are pervasive in galaxies and galaxy clusters and likely play an active role during the formation and evolution of these systems. CRs, magnetic fields, and turbulence are observed to be in pressure equilibrium in the midplane of the Milky Way \citep{1990ApJ...365..544B}, suggesting that CRs have an important dynamical role in maintaining the energy balance of the interstellar medium (ISM).  

This equipartition could be the result of a self-regulated feedback process: provided that CR and magnetic midplane pressures are supercritical, their buoyancy force overcomes the magnetic tension of the dominant toroidal magnetic field, causing it to bend and open up \citep{1966ApJ...145..811P,2016ApJ...816....2R}. CRs stream and diffuse ahead of the gas into the halo along these open field lines and build up a pressure gradient. Once this gradient overcomes the gravitational attraction of the disc, it accelerates the gas, thereby driving a strong galactic outflow as shown in one-dimensional magnetic flux-tube models \citep{1991Breitschwerdt,1996Zirakashvili,1997Ptuskin,Everett2008,2018Samui} and three-dimensional simulations \citep{2012Uhlig,2013Booth,2014Salem,2016PakmorI,Simpson2016,2016Girichidis, 2017PfrommerII,2017Ruszkowski,2018Jacob}. If the CR pressure is subcritical, the thermal gas can quickly radiate away the excess energy, thus approaching equipartition as a dynamical attractor solution. 

Seemingly unrelated, at the centres of dense galaxy clusters the observed gas cooling and star formation rates are reduced to levels substantially below those expected from unimpeded cooling flows \citep{2006Peterson}. Most likely, a heating process associated with radio lobes that are inflated by jets from active galactic nuclei offsets radiative cooling. Apparently, the cooling gas and nuclear activity are tightly coupled to a self-regulated feedback loop \citep{2007McNamara}. A promising heating mechanism can be provided by fast-streaming CRs, which resonantly excite Alfv\'en waves through the ``streaming instability'' \citep{1969KulsrudPearce}. Scattering off of this wave field (partially) isotropizes these CRs in the reference frame of Alfv\'en waves, which causes CRs to stream down their gradient \citep{2013Zweibel}. Damping of these waves transfers CR energy and momentum to the thermal gas at a rate that scales with the CR pressure gradient and provides an efficient means of suppressing the cooling catastrophe in cooling core clusters  \citep{1991Loewenstein,2008Guo,2011Ensslin,2012Fujita,2013Pfrommer,2017JacobPfrommerI,2017JacobPfrommerII,2018Ehlert}. Hence, in sharing energy and momentum with the thermal gas, CRs may play a critical role in galaxy formation and the evolution of galaxy clusters.

CRs interact with the thermal gas through particle collisions as well as through collisionless processes. Low-energy (MeV-to-GeV) CRs are important for collisional ionization and heating of the interstellar medium. In particular the ability of CRs to deeply penetrate into molecular clouds (where ultra-violet and X-ray photons are absorbed) makes them prime drivers of cloud chemistry \citep{2006Dalgarno,2018Ivlev,2018Phan} and responsible for the evolution of these star-forming regions. Hadronic particle interactions generate secondary decay products that emit characteristic signatures from radio to gamma-ray energies, thereby enabling studies of the spatial and spectral CR distribution.

Energetic protons with energies of a few GeV, which dominate the total CR energy density, are mostly collisionless and interact via collective phenomena mediated by the ambient magnetic field. Being charged particles, CRs are bound to follow individual magnetic fields lines, which become modified as a result of the dynamical evolution of the CR distribution. Hence, in combination with the toroidal stretching of magnetic fields due to differential rotation of galactic discs, CR-induced gas motions can twist and fold magnetic structures, thereby amplifying and shaping galactic magnetic fields via a CR-driven dynamo \citep{2004Hanasz}.

Generally, these collective, collisionless interactions can be subdivided into CR transport processes at the microscale, the mesoscale and the macroscale. While CR interactions at the microscale are modelled with kinetic theory, CR transport at the macroscale is treated in the hydrodynamic picture in which the full phase space information of CRs is condensed to a few variables that describe the system such as energy density, pressure, and number density. Interactions at the mesoscale combines elements of both descriptions and enables studies of, e.g., the structure of collisionless shocks \citep{2013Caprioli}. Different scientific questions select the approach that is best suited for a problem at hand. While we always seek for clarity and apply Occam's razor as a basic principle of model building, the richness of physics may force us to move elements from kinetic theory into the hydrodynamic picture to more faithfully capture the physics of CR transport on larger scales.

The kinetic picture of the underlying plasma assumes a sufficiently dense plasma that is well described by a distribution function. This is equivalent to requiring that many particles within a characteristic energy range be present on the plasma scale. Typically, problems such as the growth of kinetic instabilities and damping processes are addressed within kinetic theory. In particular, the non-resonant hybrid instability that excites right-handed circularly polarized Alfv{\'e}n waves by the current of energetic protons, can potentially explain magnetic amplification and CR acceleration to (almost) PeV energies at supernova remnants \citep{2004Bell}. Kinetic instabilities at shocks are important for energy exchange between electrons and protons and in building up the momentum spectrum of energetic particles \citep{2008Spitkovsky,2014Caprioli}. Thus, this approach provides a crucial input to modelling multi-frequency observations across the entire electromagnetic spectrum of supernova remnants \citep[e.g.,][]{2012Morlino,2012Blasi}, galaxies \citep{2002Breitschwerdt,2016Recchia}, and galaxy clusters \citep{2011Brunetti, 2017Pinzke}. However, to obtain a complete (non-linear) picture of a system, the dynamics on the CR gyroscale or at least the growth time-scale of a particular instability needs to be resolved. This requirement prohibits us from directly treating kinetic effects in global simulations of astrophysical objects such as galaxies or jets of active galactic nuclei.

Hence, to model CR transport in the ISM, the circumgalactic medium (CGM) or the intra-cluster medium (ICM), we have to resort to a hydrodynamic prescription. Traditionally, this was done by taking the energy-weighted moment of the Fokker-Planck equation for CR transport, yielding the CR energy equation \citep{1981Drury,1982McKenzie,1984Voelk}. This equation shows that CRs are transported through a combination of advection with the thermal gas as well as streaming and diffusion. In the ideal magneto-hydrodynamic (MHD) approximation, magnetic fields are flux-frozen into the thermal gas and thus advected with the flow. The collisionless CRs are bound to gyrate along magnetic field lines and are also advected alongside the moving gas. As CRs propagate along the mean field, they 
scatter at self-generated Alfv\'en waves, which causes them to stream down their gradient with a macroscopic velocity that is substantially reduced from their intrinsic relativistic speed. MHD turbulence that was driven at larger scales by energetic events and successively cascaded down in scale can also scatter CRs, redistributing their pitch angles, but conserving their energy \citep{2017Zweibel}. This can be described as anisotropic diffusion where the main transport is along the local direction of the magnetic field \citep{2009Shalchi}.

As a closure of these approaches, CR diffusion is modelled with a prescribed coefficient that is usually taken to be constant and not coupled to the physics of turbulence, and CR streaming is always assumed to be in steady state. However, neither of these two approaches is providing the correct prescription of CR transport \citep{2017Wiener}. Moreover, due to the non-linear property of the streaming equation, an ad-hoc regularization is applied that adds numerical diffusion to the solution \citep{2010Sharma}, questioning the results in regime of shallow gradients. Hence, these considerations reinforce the need for a novel description of CR transport that cures these weaknesses.

Recently, \cite{2018Jiang} used an ansatz to reinterpret CR transport as a modification of radiation hydrodynamics. They showed numerically, that their resulting set of equations captures the streaming limit of CR transport while conserving the total energy and momentum. However, in their picture, the conversion between mechanical and thermal energy mediated by CRs is in general not fully accounted for and, as we will show here, they adopt an incomplete treatment of CR scattering. In this work, we provide a first-principle derivation of such an improved CR transport scheme while emphasizing the deep connection between radiation and CR hydrodynamics throughout this work.

This paper is organized as follows. In Section~\ref{sec:crhd}, we show the complete set of MHD and CR transport equations as a reference and derive those in the remainder of this work. In Section~\ref{sec:phase-space}, we use the Eddington approximation for the two-moment approximation of CR transport. In Section~\ref{sec:scattering}, we derive equations accounting for the energy and pitch-angle scattering of CRs by Alfv\'enic turbulence. In Section~\ref{sec:aflven-waves}, we derive transport equations for Alfv\'en waves, which are coupled (i) to the gas via damping mechanisms and (ii) to the CR population by the streaming instability. In Section~\ref{sec:coupling}, we couple the forces and work done by the CR-Alfv\'enic subsystem to the MHD equations and address energy and momentum conservation in the Newtonian limit. In Section~\ref{sec:discussion} we show that the presented theory contains the classical streaming-diffusion equation of CR transport in the steady-state flux limit and discuss spectral extensions of the new theory. We show a numerical demonstration of our coupled transport equations for the energy densities contained in CRs and Alfv\'en waves in Section~\ref{sec:numerics} and compare our theory to other approaches in the literature. We conclude in Section~\ref{sec:conclusion}. In Appendix~\ref{app:diffusion}, we show how pure CR diffusion emerges mathematically by neglecting the electric fields of Alfv\'en waves, thereby emphasizing the need of CR streaming for a full description of CR transport. We present an alternative derivation of the scattering terms in Appendix~\ref{sec:alternative_derivation} that clarifies the approximation used to derive our CR transport equations. In Appendices~\ref{app:Vlasov} and \ref{app:derivation}, we present semi-relativistic derivations of the Vlasov and CR hydrodynamical equations using a covariant formalism. In Appendix~\ref{app:lab}, we derive the lab-frame equations for CR hydrodynamics expressed in comoving quantities and discuss energy and momentum conservation. We denote the frame that is comoving with the gas as the comoving frame and use the Heaviside system of units throughout this paper.

\section{Equations of CR Hydrodynamics}
\label{sec:crhd}

The equations for ideal MHD coupled to non-thermal CR and Alfv\'en wave populations are given by:

\begin{align}
\frac{\upartial \rho}{\upartial t} + \bs{\nabla} \bs{\cdot} (\rho \bs{u}) &= 0, \\
\frac{\upartial \rho \bs{u}}{\upartial t} + \bs{\nabla} \bs{\cdot} \left(\rho \bs{u} \bs{u} + P\mat{1} - \bs{B}\bs{B} \right) &= \bs{g}, \label{eq:final_gas_euler}\\
\frac{\upartial \bs{B}}{\upartial t} + \bs{\nabla} \bs{\cdot} (\bs{B} \bs{u} - \bs{u} \bs{B}) &= 0, \label{eq:induction_law} \\
\frac{\upartial \varepsilon}{\upartial t} + \bs{\nabla} \bs{\cdot} [\bs{u} (\varepsilon + P) - (\bs{u} \bs{\cdot} \bs{B}) \bs{B}] &= \bs{u}\bs{\cdot}\bs{g} + Q_+ + Q_-, \label{eq:final_gas_energy}
\end{align}
where $\mat{1}$ is the unit matrix and $\bs{a}\bs{b}$ is the dyadic product of vectors $\bs{a}$ and $\bs{b}$. Gas density, mean velocity, and the local mean magnetic field are denoted by $\rho$, $\bs{u}$ and $\bs{B}$. The total force exerted by CRs, Alfv\'en waves and the thermal gas is denoted by $\bs{g}$ and will be defined below. The MHD pressure and energy density are given by
\begin{align}
P &= P_{\rm th} + \frac{\bs{B}^2}{2},\\
\varepsilon &= \frac{\rho \bs{u}^2}{2} + \varepsilon_{\rm th} + \varepsilon_{B},
\end{align}
where $P_{\rm th}$ is the thermal pressure, $\varepsilon_{\rm th}$ and  $\varepsilon_{B}= \bs{B}^2/2$ are the thermal and magnetic energy densities, respectively. $Q_\pm$ are the source terms of thermal energy due to Alfv\'en wave energy losses as detailed in Section~\ref{sec:aflven-waves}. All pressures and the respective energy densities are related by equations of states:
\begin{alignat}{4}
P_{\rm th} &= (\gamma_{\rm th} &&- 1) \varepsilon_{\rm th}, \hspace{40pt} & \gamma_{\rm th} &= \frac{5}{3},\\
\pcr &= (\gamma_{\rm cr} &&- 1) \ecr, & \gamma_{\rm cr} &= \frac{4}{3}, \\
\pwpm &= (\gamma_{\rm a} &&- 1) \ewpm, & \gamma_{\rm a} &= \frac{3}{2},
\end{alignat}
where $\pcr$ is the CR pressure and $\pwpm$ are the ponderomotive pressures due to presence of Alfv\'en waves on scales that are resonant with the gyroradii of (pressure-carrying GeV-to-TeV) CRs. This enables a well-defined separation of scales in comparison to the large-scale magnetic field.

We augment these evolution equations of MHD quantities by a CR-Alfv\'enic subsystem, which encompasses the hydrodynamics of CR transport that is mediated by Alfv\'en waves. As we will show in this work, this subsystem describes the transport of CR energy density ($\ecr$), CR momentum density along the mean magnetic field ($\fcr/c^2$), where $\fcr$ denotes the CR energy flux density, and Alfv\'en-wave energy density ($\ewpm$), where the $\pm$ signs denote co- and counter-propagating waves with respect to the large-scale magnetic field. Note that $\ecr$ and $\fcr$ are measured with respect to the comoving frame while $\ewpm$ is measured in the lab frame:
\begin{align}
\label{eq:ecr}
\frac{\upartial \ecr}{\upartial t} + \bs{\nabla} \bs{\cdot} [\bs{u} (\ecr + \pcr) + \bs{b} \fcr] &= \bs{u} \bs{\cdot} \bs{\nabla} \pcr  \nonumber\\ 
&\hspace{-1.5em}- \bs{\varv}_{\rm a} \bs{\cdot} \bs{g}_{\rm gri, +} + \bs{\varv}_{\rm a} \bs{\cdot} \bs{g}_{\rm gri, -} ,\\[.5em]
\label{eq:fcr}
\frac{\upartial (\fcr / c^2)}{\upartial t} + \bs{\nabla} \bs{\cdot} \left( \bs{u} \fcr / c^2 \right) + \bs{b} \bs{\cdot} \bs{\nabla} \pcr &=  - (  \bs{b} \bs{\cdot} \bs{\nabla} \bs{u}) \bs{\cdot} (\bs{b}  \fcr / c^2) \nonumber\\
&\hspace{-1.5em}- \bs{b} \bs{\cdot} \left(\bs{g}_{\rm gri, +} + \bs{g}_{\rm gri, -}\right),\\[.5em]
\label{eq:eaw}
\frac{\upartial \ewpm}{\upartial t} + \bs{\nabla} \bs{\cdot} \left[ \bs{u} (\ewpm + \pwpm) \pm \varv_{\rm a} \bs{b} \ewpm \right] &= \bs{u} \bs{\cdot} \bs{\nabla} \pwpm  \nonumber\\ 
&\hspace{-1.5em}\pm \bs{\varv}_{\rm a} \bs{\cdot} \bs{g}_{\rm gri, \pm} - Q_{\rm \pm}. 
\end{align}
Here, $c$ is the light speed (corresponding to {\em intrinsic} CR velocity in the ultra-relativistic approximation), $\bs{\varv}_{\rm a} = \bs{B}/\sqrt{\rho}$ is the Alfv\'en velocity, $B=\sqrt{\bs{B}^2}$ is the magnetic field strength, and $\bs{b}=\bs{B}/B$ the direction of the mean magnetic field. The exerted forces between CRs, Alfv\'en waves and the thermal gas are given by:
\begin{align}
	\bs{g} &= \bs{g}_{\rm Lorentz} + \bs{g}_{\rmn{ponder}} + \bs{g}_{\rm gri, +} + \bs{g}_{\rm gri, -},\\
    \bs{g}_{\rm Lorentz} &= -\bs{\nabla}_\perp \pcr, \\
    \bs{g}_{\rm ponder} &= -\bs{\nabla} (\pwp + \pwm), \\
    \bs{g}_{\rm gri, \pm} &= \frac{\bs{b}}{3 \kappa_\pm} [\fcr \mp \varv_{\rm a}(\ecr + \pcr)],
\end{align}
where $\bs{g}_{\rm Lorentz}$ is the Lorentz force due the large-scale magnetic field, $\bs{g}_{\rm ponder}$ is the ponderomotive force, $\bs{g}_{\rm gri, \pm}$ are the Lorentz forces due to small-scale magnetic field fluctuations of Alfv\'en waves that affect CRs, and the perpendicular gradient is given by $\bs{\nabla}_{\perp} = (\mat{1} - \bs{b} \bs{b})\bs{\cdot} \bs{\nabla}$.

The CR energy equation~\eqref{eq:ecr} contains source terms on the right-hand side that arise as a result of adiabatic changes and resonant scattering off of Alfv\'en waves via the gyroresonant instability (gri). We refrain from including additional CR source and sink terms, as we focus solely on transport processes of CRs. Equations \eqref{eq:ecr} and \eqref{eq:fcr} fully describe CR diffusion and CR streaming in the self-confinement picture. The right-hand side of the Alfv\'en-wave equation~\eqref{eq:eaw} shows loss terms $Q_\pm$ due to damping processes. 

The CR-Alfv\'enic subsystem is closed by the grey approximation for the CR diffusion coefficient:
\begin{equation}
\frac{1}{\kappa_\pm} = \frac{9\pi}{8} \frac{\Omega}{c^2} \frac{\ewpm/2}{\varepsilon_B}  \left( 1 + \frac{2\varv_{\rm a}^2}{c^2} \right).
\end{equation}
Here, $\Omega=ZeB/(\gamma mc)$ is the relativistic gyrofrequency of a CR population with charge $Ze$ and characteristic Lorentz factor $\gamma$, $e$ is the elementary charge, and $m$ is the particle rest mass. This equation links the transported CR energy density directly to the Alfv\'enic turbulence, described by its energy density $\ewpm$. 

The total pressure and energy density of thermal gas, magnetic fields, CRs, and Alfv\'en waves are given by
\begin{align}
P_{\rm tot} &= P_{\rm th} + \frac{\bs{B}^2}{2} + \pcr + \pwp + \pwm, \label{eq:ptot}\\
\varepsilon_{\rm tot} &= \frac{\rho \bs{u}^2}{2} + \varepsilon_{\rm th} + \varepsilon_B + \ecr + \ewp + \ewm.
\label{eq:etot}
\end{align}
Even in the absence of explicit gain and loss terms, it is not possible to conserve the total energy and momentum in terms of the preceding quantities in every frame. Only the total energy and momentum as measured in an inertial frame (i.e., the `lab' frame) can be manifestly conserved (see Appendix~\ref{app:lab}). The CR energy and momentum densities defined above are measured in the comoving frame and their evolution equations are expressed in the semi-relativistic limit. This semi-relativistic limit prohibits a meaningful Lorentz transformation between both frames so that contributions from pseudo forces do not vanish after a transformation from the comoving frame into the lab frame. Consequently, total momentum and energy are altered by these pseudo forces even in the lab frame. However, if CRs move with non-relativistic bulk velocities, their inertia is negligible and no formal degeneracy between the two frames occurs. In addition, the erroneously transformed pseudo forces vanish. In this case, the total energy $E_{\rm tot}=\int \dif{^3x}\varepsilon_{\rm tot}$ (where $\bs{x}$ denotes the spatial coordinate) is a conserved quantity so that
\begin{align}
\frac{\upartial \varepsilon_{\rm tot}}{\upartial t} + \bs{\nabla} \bs{\cdot} [\bs{u} (\varepsilon_{\rm tot} + P_{\rm tot}) + f_{\rm tot} \bs{b}] = 0 ,
\end{align}
where the total energy flux density along the magnetic field lines is given by
\begin{align}
f_{\rm tot} = \fcr + \varv_{\rm a} \varepsilon_{\rm a, +} - \varv_{\rm a} \varepsilon_{\rm a, -} - B\, (\bs{u} \bs{\cdot}\bs{B}).
\label{eq:ftot}
\end{align}

Likewise, the total momentum is solely given by the mean gas momentum
\begin{align}
\bs{m}_{\rm tot} = \rho \bs{u},
\end{align}
and is a conserved quantity, which follows from the conservation law
\begin{align}
\frac{\upartial \bs{m}_{\rm tot}}{\upartial t} + \bs{\nabla} \bs{\cdot} (\bs{u} \bs{m}_{\rm tot}  + P_{\rm tot}\mat{1} - \bs{B} \bs{B}) = 0.
\end{align}
There is no contribution by either large-scale or small-scale electromagnetic fields because their momenta are assumed to be vanishingly small in the non-relativistic MHD approximation.

\section{CR Phase Space Dynamics}
\label{sec:phase-space}

After summarizing the full set of equations for CR hydrodynamics, we will now derive them. Starting with the Vlasov equation, we discuss the Eddington approximation to the transport of the CR distribution function. In the next step, we will derive the CR fluid equations.

\subsection{Focused CR transport equation}
\label{sec:focused_CRs}

The CR distribution lives in phase space that is spanned by the momentum and spatial coordinates $\bs{p}$ and $\bs{x}$, respectively, and is defined as
\begin{align}
\label{eq:fp}
f\equiv f(\bs{x},\bs{p},t)=\frac{d^6N}{\dif{x^3}\dif{p^3}}.
\end{align}
It evolves according to the comoving Vlasov equation in the semi-relativistic limit,
\begin{align}
\frac{\upartial f}{\upartial t} + (\bs{u} + \bs{\varv}) \bs{\cdot} \bs{\nabla}_{\bs{x}} f + \bs{F} \bs{\cdot} \bs{\nabla}_{\bs{p}} f = 0,
\label{eq:vlasov}
\end{align}
where the mean gas velocity $\bs{u}$ and time $t$ are measured in the lab frame, the CR velocity $\bs{\varv}$ and momentum $\bs{p}$ are measured in the comoving frame, and $\bs{F}$ denotes the total force. The description in the comoving frame introduces pseudo forces (denoted by $\bs{F}_\rmn{pseudo}$) since the momentum measured by an observer in the comoving frame changes for each change of the reference velocity $\bs{u}$. Furthermore, CRs as charged particles are  subject to the Lorentz force, which we split into contributions by large-scale and small-scale electromagnetic fields, $\bs{F}_{\rm macro}$ and $\bs{F}_{\rm micro}$, respectively:
\begin{alignat}{4}
	\bs{F} &=\bs{F}_{\rm pseudo} &+& \bs{F}_{\rm marco} &+& \bs{F}_{\rm micro}\\
    	   &=-m\frac{\rmn{d}\bs{u}}{\rmn{d}t}-\bs{(\bs{p} \bs{\cdot} \bs{\nabla})} \bs{u}  &+& Ze\frac{\bs{\varv} \bs{\times} \bs{B}}{c} &+& Ze \left( \delta\bs{E} + \frac{\bs{\varv} \bs{\times} \delta\bs{B}}{c}\right),
           \label{eq:pseudo}
\end{alignat}
see equation (5.18) in \citet{BookZank} or Appendix~\ref{app:Vlasov} for a covariant derivation. Here, the Lagrangian time derivative is denoted by ${\rm d}/{\rm d}t = \upartial / \upartial t + \bs{u} \bs{\cdot} \bs{\nabla}$ and $\delta\bs{E}$ and $\delta\bs{B}$ are electric and magnetic fluctuations, respectively.

The pseudo forces appear in the Vlasov equation because $\bs{u}$ acts as a reference velocity linking lab and comoving velocities and is itself a dynamical quantity. Both pseudo forces in equation~\eqref{eq:pseudo} have slightly different interpretations: the first pseudo force is the result of an acceleration of the comoving frame itself. A CR at rest in the lab frame is perceived to be accelerated from the point of view of a comoving observer. The second pseudo force is due to spatial inhomogeneities of the flow field. If the CR moves in the lab frame, then a change of its position also causes the reference velocity to change because the comoving frame is now linked by a different velocity to the lab frame. From the perspective of a comoving observer this change in comoving CR velocity is perceived as an acceleration. Dimensional analysis suggests that the first pseudo force corresponds to an acceleration that is smaller by a factor of $\mathcal{O}(u / \varv)$ in comparison to the second pseudo force (i.e., $\mathcal{O}(u / c)$ for relativistic CRs). In the following, we thus neglect the contribution from the first pseudo force.

The small-scale field fluctuations are provided by MHD waves, in particularly by Alfv\'en waves, which are generated by the CR-driven gyroresonant instability. Since these waves are the source of CR scattering, we denote their contribution to the Vlasov equation as:
\begin{align}
\left.\frac{\upartial f}{\upartial t} \right\rvert_{\rm scatt} = \bs{F}_{\rm mirco} \bs{\cdot} \bs{\nabla}_{\bs{p}} f.
\end{align} 
We leave this term unspecified for now and return to it in Section~\ref{sec:scattering}.

CRs gyrate around large-scale magnetic fields on spatial and temporal scales that are small in comparison to any MHD scale. We can thus project out the full phase dynamics of CRs by taking the gyroaverage. Calculating this average of equation~\eqref{eq:vlasov} results in the so called {\em focused} transport equation, which describes the gyroaveraged evolution of CRs. While \citet{1971Skilling} performs this calculation in the Alfv\'en-wave frame, $\bs{u} + \varv_{\rmn{a}} \bs{b}$, the identical result is obtained in the frame comoving with the mean gas velocity $\bs{u}$ \citep{BookZank}. Using the latter result of the focused transport equation, we arrive at:
\begin{align}
\frac{\upartial f}{\upartial t} &+ (\bs{u} + \mu \varv \bs{b}) \bs{\cdot} \bs{\nabla} f \nonumber\\ 
&+ \left[ \frac{1-3\mu^2}{2} (\bs{b} \bs{\cdot} \bs{\nabla} \bs{u} \bs{\cdot} \bs{b}) - \frac{1-\mu^2}{2} \bs{\nabla} \bs{\cdot} \bs{u} \right] p \frac{\upartial f}{\upartial p} \label{eq:fpe_skilling}  \\
&+ \left[ \varv \bs{\nabla} \bs{\cdot} \bs{b} + \mu \bs{\nabla} \bs{\cdot} \bs{u} - 3 \mu (\bs{b} \bs{\cdot} \bs{\nabla} \bs{u} \bs{\cdot} \bs{b}) \right] \frac{1-\mu^2}{2} \frac{\upartial f}{\upartial \mu}  = \left. \frac{\upartial f}{\upartial t}  \right\rvert_{\rm scatt}. \nonumber
\end{align}
Here, we use the conventional mixed coordinate system for phase space. While the ambient gas velocity $\bs{u}$ and the direction of the large scale magnetic field $\bs{b} = \bs{B} / B$ are measured in the lab frame, the particle velocity $\varv$, momentum $p$ and the cosine of the pitch angle $\mu = \bs{\varv} \bs{\cdot} \bs{b} / \varv$ are given with respect to the comoving frame. A general discussion of the adiabatic terms and other pseudo forces of this equation is given in \citet{2012leroux}. 

The complexity of transport terms in equation~\eqref{eq:fpe_skilling} alone precludes a general solution and we have to resort to approximations. In the following, we use a procedure which preserves the large-scale dynamics of the entire distribution in terms of thermodynamical quantities. To this end, we take moments of the momentum space variables $\mu$ and $p$ and describe the energy content in CRs and their transport properties in terms of an energy flux that is coupled to the Alfv\'en-wave dynamics. 

\subsection{Eddington approximation}

A similarly complex problem is the radiative transfer (RT) equation with its two phase space coordinates photon propagation direction $\bs{n}$ and photon frequency. Powerful methods describing the transport of comoving radiation energy were pioneered by \citet{BookMihalas} and \citet{BookCastor}.

In the case of an optically thick medium, the Eddington approximation is a valuable tool to model the transport of radiation energy. In this approximation, the RT equation is expanded up to first order in $\bs{n}$ while assuming that the contribution from higher-order moments of the radiation distribution can be neglected. This assumption is justified in the optically thick medium because rapid scattering quickly damps any anisotropy.

A more accurate approximation of RT problems with a preferred direction is the assumption of plane-parallel or slab geometry. In this case, all quantities of the medium are taken to be constant on planes perpendicular to this particular direction $\bs{n}$. The RT equation can then be expressed in terms of the coordinate along $\bs{n}$ and the direction cosine $\mu$ between the orientation of a ray and $\bs{n}$. In this setting, the Eddington approximation for the radiation intensity $I$ simplifies to
\begin{align}
	I(\mu) = I_0 + I_1 \mu,
\end{align}
where we suppress the spatial dependence of the first- and second-order moments $I_0$ and $I_1$ in our notation. However, this simplified slab geometry is of limited use because it often does not apply to astrophysical problems at hand.

This is different for CR transport where the mean magnetic field is a priori known as a preferred direction of (gyrophase averaged) motion. Thus, CR transport is locally akin to plane-parallel RT. To model CR transport with such an RT methodology, we have to account for the spatially and temporarily varying plane and translate the corresponding terminologies.

The direction cosine $\mu$ in RT is equivalent to the pitch-angle cosine $\mu$ in CR transport. Thus, we expand equation~\eqref{eq:fpe_skilling} in moments of the pitch angle. This expansion has a long history in CR transport and is frequently revisited \citep[see e.g.][]{1971Klimas, 1973Earl,1987Webb, 2000Zank,2006Snodin,2013Litvinenko,2018Rodrigues}. For completeness, we recall the derivation to introduce our notation. 

In general, any complete basis of functions could be used to expand $f$ in  pitch-angle. Particularly useful are the Legendre polynomials, because of their geometric relationship to the pitch angle.\footnote{The Legendre polynomials are eigenfunctions of the pitch-angle Laplace operator $\upartial_t f \vert_{\rm scatt} = \upartial_\mu [\nu (1-\mu^2)/2 \,  \upartial_\mu f]$. This operator describes pitch-angle diffusion and $\nu$ denotes the scattering frequency. Note that this simple Laplacian resembles the actual scattering operator as discussed in equation~\eqref{eq:mu_scatt}.} Carrying out the complete expansion using these basis functions results in an infinite set of coupled differential equations. Even though this system captures the full dynamics of equation~\eqref{eq:fpe_skilling}, it is not practicable because of the high degree of coupling between the transport terms \citep{2000Zank}.

Similar to RT, we circumvent problems arising from this coupling by truncating the expansion. Because CRs are subject to rapid scattering, anisotropies of their distribution are efficiently damped. We can thus assume that all moments larger than the first are negligibly small and proceed with
\begin{equation}
	f = f_0 + 3 \mu f_1,
   \label{eq:f_expansion}
\end{equation}
while requiring that $f_0 \gg f_1$. Otherwise, higher-order moments could become dynamically important as a result of coupling and the truncated expansion would not converge. This quasi-linear approximation is valid in cases of self-confined CR transport, where sufficiently energetic Alfv\'en waves are generated by CRs. We will explicitly show this later on in Section~\ref{sec:streaming}.

Inserting the expansion \eqref{eq:f_expansion} into equation~\eqref{eq:fpe_skilling} and taking the pitch-angle average results in
\begin{equation}
\frac{\upartial f_0}{\upartial t} +\bs{u} \bs{\cdot} \bs{\nabla} f_0 +  \bs{\nabla} \bs{\cdot} \left( \varv \bs{b} f_1 \right) - \frac{1}{3} (\bs{\nabla} \bs{\cdot} \bs{u})  p \frac{\upartial f_0}{\upartial p} = 
\left.\frac{\upartial f_0}{\upartial t} \right|_{\rm scatt}.
\label{eq:dot_f0}
\end{equation}	
Analogously, taking the $\mu$-moment of equation~\eqref{eq:fpe_skilling} yields:
\begin{align}
\frac{\upartial f_1}{\upartial t} + \frac{\varv}{3} \bs{b} \bs{\cdot} \bs{\nabla} f_0 &+ \bs{u} \bs{\cdot} \bs{\nabla} f_1 + 
\left[ -\frac{2}{5} (\bs{b} \bs{\cdot} \bs{\nabla} \bs{u} \bs{\cdot} \bs{b}) -\frac{1}{5} \bs{\nabla} \bs{\cdot} \bs{u} \right] p \frac{\upartial f_1}{\upartial p}
\nonumber \\ 
&\hspace{-10pt}+ \left[ \frac{1}{5} \bs{\nabla} \bs{\cdot} \bs{u} - \frac{3}{5} (\bs{b} \bs{\cdot} \nabla \bs{u} \bs{\cdot} \bs{b})\right]  f_1 =  
\left.\frac{\upartial f_1}{\upartial t} \right|_{\rm scatt}.
\label{eq:dot_f1}
\end{align}
The scattering terms on the right-hand side of equations \eqref{eq:dot_f0} and \eqref{eq:dot_f1} are calculated in Section~\ref{sec:streaming}.

A more complex expansion would use eigenfunctions of the scattering operator with a pitch-angle dependent scattering rate $\nu(\mu)$. These eigenfunctions exist and form a orthogonal set of functions by virtue of the Sturm-Liouville theory. In general, this would yield a different set of basis functions that differ from the Legendre polynomials. While this approach would render pitch-angle averaging of the scattering coefficient unnecessary, this more rigorous treatment would obfuscate the derivation and make our results inherently dependent on the actual form of $\nu(\mu)$. Since we truncate the expansion after the first order and assume small anisotropies, we do not expect any change of the presented theory. Hence, our choice of a pitch-angle-averaged scattering rate represents a compromise between physical clarity and mathematical rigour.

\subsection{Fluid equations}

The CR energy density is given by
\begin{align}
\ecr = \int \dif{^3p} E(p) f(p, \mu) = \int_0^\infty\!{\rm d}p\,4 \pi p^2 E(p) f_0(p).
\end{align}
where $E(p) = \sqrt{p^2 c^2 + m^2 c^4}$ is the total energy of CR particles. Combining the truncation in the pitch-angle expansion and assuming approximate gyrotropy of the CR distribution yields an isotropic CR pressure tensor:
\begin{align} 
    \mat{P}_{\rm cr} =  \int \dif{^3p} \bs{\varv} \bs{p} f(p, \mu) =  \int \dif{^3p} \bs{\varv} \bs{p} f_0(p) = \pcr \mat{1}.
\end{align} 
where the isotropic CR pressure is given by:
\begin{align}
\pcr = \int \dif{^3p} \frac{p\varv}{3} f(p, \mu) = \int_0^\infty\!{\rm d}p\,4 \pi p^2 \frac{p \varv}{3} f_0(p).
\end{align}
Only the isotropic component $f_0$ of the CR distribution contributes to both quantities because any anisotropy vanishes as a result of pitch-angle integration and higher moments are neglected in our approximation.
Pressure and energy density are coupled via the equation of state
\begin{align}
	\pcr = (\gamma_{\rm cr} - 1) \ecr,
    \label{eq:eos_ecr}
\end{align}
where the adiabatic index $\gamma_{\rm cr} = 4/3$ holds in the ultra-relativistic regime  that we are focusing on.\footnote{There are different definitions for the CR energy density in the literature: while some authors define $\ecr$ as the kinetic energy moment \citep[e.g.,][]{2007Ensslin,2017Pfrommer}, others use the total particle energy moment \citep[as done here or in e.g.,][]{2017Zweibel}. Both definitions of $\ecr$ vary by the rest mass energy density $c^2 \rho_{\rm cr}$, where $\rho_{\rm cr}$ is the mass density of CRs. However, this difference is negligible in the ultra-relativistic limit adopted here.}

Similarly, we define the CR energy flux density ($\bs{\fcr}$) and the CR pressure flux ($\bs{\kcr}$):
\begin{align}
\bs{\fcr} &= c^2\int{\rm d}^3p\, \bs{p} f(p,\mu) = \int{\rm d}^3p\, E(p) \bs{\varv} f(p,\mu),   \\
\bs{\kcr} &= \int{\rm d}^3p\, \frac{p \varv}{3} \bs{\varv} f(p,\mu).
\end{align}
Due to the assumed gyrotropy, both vectors point along the mean magnetic field. This allows us to use the magnitude of $\bs{\fcr}$ and $\bs{\kcr}$ instead of vector quantities to track the energy flux density and pressure flux. We define
\begin{align}
\fcr &= \bs{b} \bs{\cdot} \bs{\fcr} = \int_0^\infty\!{\rm d}p~4 \pi p^2 E(p) \varv f_1(p), \label{def:fcr} \\
\kcr &= \bs{b} \bs{\cdot} \bs{\kcr} = \int_0^\infty\!{\rm d}p~4 \pi p^2 \frac{p \varv}{3} \varv f_1(p), \label{def:kcr}
\end{align}
where we adopted the truncation in the pitch-angle expansion of equation~\eqref{eq:f_expansion} in the last step. Algebraically, the same equation of state holds as for the CR energy density and pressure:
\begin{align}
	\kcr = (\gamma_{\rm cr} - 1) \fcr.
    \label{eq:eos_fcr}
\end{align}

The interpretation of $\fcr$ becomes apparent after multiplying equation~\eqref{eq:dot_f0} by $E(p)$ and successively integrating the equation over momentum space, which yields
\begin{align}
\frac{\upartial \ecr}{\upartial t} + \bs{\nabla} \bs{\cdot} (\bs{u} (\ecr + \pcr) + \bs{b} \fcr) = \bs{u} \bs{\cdot} \bs{\nabla} \pcr + \left.\frac{\upartial \ecr}{\upartial t} \right|_{\rm scatt}. 
\label{eq:dot_ecr}
\end{align}
Hence, $\fcr$ is the flux density of CR energy along the magnetic field. By analogy, $\kcr$ is the corresponding (anisotropic) flux of CR pressure. The interpretation of the remaining terms in equation~\eqref{eq:dot_ecr} is straightforward: the CR energy density is advected with the gas at velocity $\bs{u}$ and subject to adiabatic changes.

We derive the transport equation for the flux density of CR energy, $\fcr$, in the ultra-relativistic limit ($\varv\to c$) and show in Section~\ref{sec:spectrum} how to generalize this simplification to account for the transport of CR energy across the full momentum spectrum. Multiplying equation~\eqref{eq:dot_f1} by $\varv E(p)$ and integrating over momentum space yields
\begin{align}
\frac{\upartial \fcr}{\upartial t} + \bs{\nabla} \bs{\cdot} \left( \bs{u} \fcr  \right) + \frac{c^2}{3} \bs{b} \bs{\cdot} \nabla \ecr &= \nonumber \\ &\hspace{-16pt} - ( \bs{b} \bs{\cdot} \bs{\nabla} \bs{u}) \bs{\cdot} (\bs{b} \fcr) + \left.\frac{\upartial \fcr}{\upartial t} \right|_{\rm scatt}.
\label{eq:dot_fcr}
\end{align}
Here, we use equation~\eqref{eq:eos_fcr} to cast the result in this compact form. The third term on the left-hand side corresponds to the Eddington term in RT. However, it differs from its original appearance since it is projected onto the magnetic field that guides the anisotropic CR transport. This term can be interpreted as a source term: any spatial anisotropy as manifested by a gradient in $\ecr$ gives rise to a change of the local anisotropy and hence to a flux of CR energy. The first term on the right-hand side accounts for the change of the local direction of reference and is thus equivalent to a pseudo force term. This is explicitly demonstrated by deriving the evolution equations~\eqref{eq:dot_ecr} and \eqref{eq:dot_fcr} for $\ecr$ and $\fcr$ in the semi-relativistic limit of the fully covariant conservation equations in Appendix~\ref{app:derivation}.

Both equations fully describe the evolution of $\ecr$ and $\fcr$ in our chosen geometry, i.e. along the local direction of the magnetic field. However, these equations are incomplete without specifying the scattering terms on the right-hand side.

\section{CR scattering by magnetic turbulence}
\label{sec:scattering}

In this section, we compute the scattering terms for the CR energy density and flux density while accounting for the Fokker-Planck coefficients of pitch-angle and momentum diffusion.

\subsection{Pitch-angle scattering}

In our derivation so far, we adopted the essential assumption of rapid CR scattering with Alfv\'en waves. In general this interaction is described by a non-linear stochastic process. If the magnetic perturbations $\delta B$ in the magnetic turbulence are small, $\delta B / B \sim 10^{-3}$ or less, this stochastic scattering process can be simplified and treated analytically. This is conventionally adopted within quasi-linear theory (QLT), where Boltzmann's and Maxwell's equation are evaluated up to linear order \citep{BookKulsrud}. 

The wave-particle scattering can be provided by self-generated Alfv\'en waves through the gyroresonant instability \citep{1969KulsrudPearce}: any residual anisotropy of the CR distribution can excite resonant Alfv\'en waves as a collective interaction. In turn, these Alfv\'en waves scatter wave-generating CRs in pitch angle, eventually leading to (partial) isotropization of the distribution function as we will see. This mechanism is thought to be the principle contributor to all scattering processes and affects CRs at low to intermediate energies \citep[$E\lesssim200$~GeV,][]{2006Lazarian}. 

\begin{figure}
	
\begingroup%
\providecommand\color[2][]{%
	\errmessage{(Inkscape) Color is used for the text in Inkscape, but the package 'color.sty' is not loaded}%
	\renewcommand\color[2][]{}%
}%
\providecommand\transparent[1]{%
	\errmessage{(Inkscape) Transparency is used (non-zero) for the text in Inkscape, but the package 'transparent.sty' is not loaded}%
	\renewcommand\transparent[1]{}%
}%
\providecommand\rotatebox[2]{#2}%
\ifx\svgwidth\undefined%
\setlength{\unitlength}{209.55541132bp}%
\ifx\svgscale\undefined%
\relax%
\else%
\setlength{\unitlength}{\unitlength * \real{\svgscale}}%
\fi%
\else%
\setlength{\unitlength}{\svgwidth}%
\fi%
\global\let\svgwidth\undefined%
\global\let\svgscale\undefined%
\makeatother%
\begin{picture}(1,0.76001702)%
\put(0,0){\includegraphics[width=\unitlength,page=1]{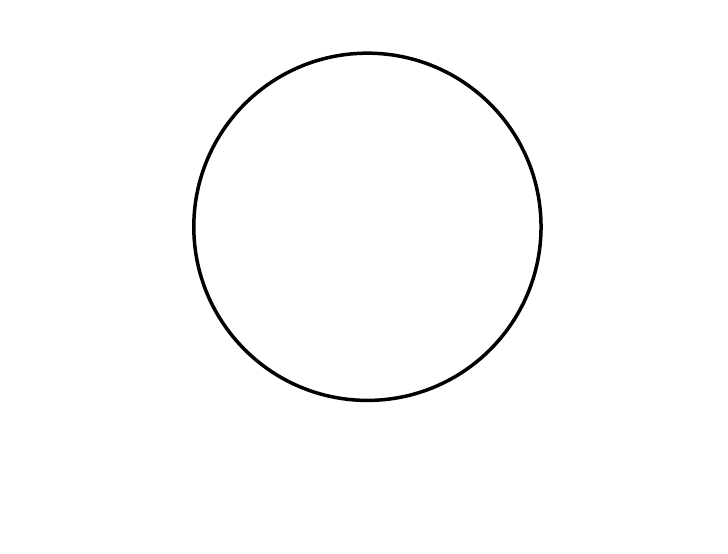}}%
\put(0.30810252,0.44080098){\color[rgb]{0,0,0}\makebox(0,0)[lb]{\smash{$I^{\rm L}_{+}$}}}%
\put(0.67023927,0.44334203){\color[rgb]{0,0,0}\makebox(0,0)[lb]{\smash{$I^{\rm R}_{+}$}}}%
\put(0,0){\includegraphics[width=\unitlength,page=2]{figure-1.pdf}}%
\put(0.17662323,0.06288417){\color[rgb]{0,0,0}\makebox(0,0)[rb]{\smash{$\mu$}}}%
\put(0,0){\includegraphics[width=\unitlength,page=3]{figure-1.pdf}}%
\put(0.79291622,0.01009529){\color[rgb]{0,0,0}\makebox(0,0)[b]{\smash{$+1$}}}%
\put(0.21669839,0.01009529){\color[rgb]{0,0,0}\makebox(0,0)[b]{\smash{$-1$}}}%
\put(0,0){\includegraphics[width=\unitlength,page=4]{figure-1.pdf}}%
\put(0.56221686,0.01037559){\color[rgb]{0,0,0}\makebox(0,0)[lb]{\smash{$\frac{\varv_{\rm a}}{\varv}$}}}%
\put(0.50100914,0.01062472){\color[rgb]{0,0,0}\makebox(0,0)[b]{\smash{$0$}}}%
\put(0,0){\includegraphics[width=\unitlength,page=5]{figure-1.pdf}}%
\put(0.50478914,0.71308865){\color[rgb]{0,0,0}\makebox(0,0)[b]{\smash{$k=0$}}}%
\put(0,0){\includegraphics[width=\unitlength,page=6]{figure-1.pdf}}%
\put(0.50478914,0.23727875){\color[rgb]{0,0,0}\makebox(0,0)[b]{\smash{$k=\infty$}}}%
\end{picture}%
\endgroup%
	
    \caption{The resonance condition for co-propagating Alfv\'en waves. For any given $\mu$ there is only one resonant wave polarization state of left-(L) or right-(R) handedness. At $\mu=\varv_{\rm a}/\varv$ the resonant wave number $k_{{\rm res}, +}$ becomes infinite and switches sign. This corresponds to a pitch angle of $90^\circ$ in the wave frame. By moving the pitch angle across this point, the type of wave polarization state that a CR can resonate with also changes. Thus, the point $k=\infty$ connects both wave spectra in terms of their resonant property of CR scattering. This connection enables us to compactify $k$-space to a circle onto which the $\mu$-axis can be mapped (via an Alexandroff compactification).}
    \label{fig:resonance}
\end{figure}

CRs scatter resonantly off of Alfv\'en waves when, in the wave frame, they gyrate around the mean magnetic field in the same direction as the magnetic field of the circulary polarized Alfv\'en waves. Formally, this requirment is captured by the resonance condition:
\begin{align}
	\omega - k_{\parallel} \varv \mu + \sigma \Omega = 0, && \sigma \in \lbrace +1, -1 \rbrace
    \label{eq:resonance}
\end{align}
where $\omega$ is the wave frequency of the wave, $\sigma = +1$ if CRs scatter with a right-hand polarized wave, and $\sigma = -1$ for scattering with a left-hand polarized wave. Provided that the dielectric contribution to the dispersion relation of Alfv\'en waves is small, the wave frequency is given by
\begin{align}
	\omega = 
\begin{cases} 
+ k_{\parallel} \varv_{\rm a} & \text{for co-propagating waves, and} \\
- k_{\parallel} \varv_{\rm a} & \text{for counter-propagating waves.}
\end{cases}
\label{eq:DR}
\end{align}
A CR particle can always interact with two types of Alfv\'en waves: if the CR co-propagates with the wave, the mode needs to be right-hand polarized, if it counter-propagates, the wave mode needs to be left-hand polarized. From now on, we identify $k\equiv{}k_\parallel$, i.e., we drop the subscript on the wave number but retain its meaning. Combining the dispersion relation~\eqref{eq:DR} and the resonance condition~\eqref{eq:resonance}, we can derive a wave number for CRs that resonantly interact with Alfv\'en waves:
\begin{align}
\label{eq:resonance_condition}
k_{\rm res, \pm} = \frac{\Omega}{\mu \varv \mp \varv_{\rm a} }, 
\end{align}
where we suppress the polarization sign that we encapsulate in the next definition: the energy contained in waves at this wave number is given by the resonant wave power spectrum:
\begin{align}
    R_{\pm}(k_{\rm res, \pm}) = I^{\rm L}_{\pm}(-k_{\rm res, \pm}) + I^{\rm R}_{\pm}(k_{\rm res, \pm}).
    \label{def:resonant_wave_spectrum}
\end{align}
Here, $I^{\rm L, R }_{\pm}$ are the intensities of co-/counter-propagating Alfv\'en waves of each polarization state. The resonant wave intensity $I^{\rm L, R }_{\pm}(k_{\rm res, \pm})=0$ for negative arguments, $k_{\rm res, \pm}<0$. In Fig.~\ref{fig:resonance}, we illustrate this definition together with the resonance condition. Through this definition $R_{\pm}$ identifies the correct polarization state of Alfv\'en waves that are resonant with a particular wave number $k_{\rm res, \pm}$ of our CR particle.

We define a total wave power spectrum that contains all power carried by co- and counter-propagating waves:
\begin{align}
    E_{\pm}(k) = I^{\rm L}_{\pm}(k) + I^{\rm R}_{\pm}(k).
\end{align}
This enables us to define the total Alfv\'en wave energy density:
\begin{equation}
	\ewpm = \int_0^\infty\!{\rm d}k\,E_\pm(k).
    \label{eq:total_wave_energy_def}
\end{equation}
    
Because Alfv\'en waves are purely magnetic perturbations, there are no electric fields in their own frames. Hence, the interaction between Alfv\'en waves and CRs preserves their kinetic energies but changes their pitch angles. Mathematically, this scattering can be described as a diffusion process in phase space (for the general case, see \citet{1989Schlickeiser}; and \citet{2002Teufel} for our specific case). Thus, we have for pure pitch-angle scattering \citep{1971Skilling}:
\begin{equation}
\label{eq:mu_scatt}
	\left.\frac{\upartial f}{\upartial t} \right|_{\rm scatt, wave} = \frac{\upartial}{\upartial \mu} \left. \left( \frac{1-\mu^2}{2} \nu(p, \mu) \frac{\upartial f}{\upartial \mu}\right) \right\rvert_{\rm wave}
\end{equation}
where time and pitch angle derivatives have to be evaluated in the wave frame. The scattering frequencies for forward and backward propagating Alfv\'en waves are given by \citet{1989Schlickeiser}:
\begin{align}
\nu_\pm(p, \mu) = \pi \Omega \frac{\lvert k_{\rm res, \pm} \rvert \, R_\pm(k_{\rm res, \pm})}{\varepsilon_B}.
\label{def:scattering_coeff}
\end{align}
Pitch-angle scattering thus damps the CR anisotropy in the wave frame.

In the comoving frame, propagating waves excite magnetic and electric fields. Accordingly, a scattering event implies an energy transfer between CRs and waves. \citet{1989Schlickeiser} accounted for both pitch-angle and momentum diffusion in slab Alfv\'enic turbulence and found: 
\begin{align}
	\left. \frac{\upartial f}{\upartial t} \right\rvert_{\rm scatt} &= \frac{\upartial}{\upartial \mu} \left( D_{\mu \mu} \frac{\upartial f}{\upartial \mu} + D_{\mu p} \frac{\upartial f}{\upartial p} \right) \nonumber \\&\hspace{6em}+ \frac{1}{p^2} \frac{\upartial}{\upartial p} p^2 \left( D_{\mu p} \frac{\upartial f}{\upartial \mu} + D_{p p} \frac{\upartial f}{\upartial p} \right). \label{eq:ps-scattering}
\end{align}
The diffusion coefficients are given by \citep{1989Schlickeiser,1990Dung}:
\begin{align}
	D_{\mu \mu} &= \frac{1 - \mu^2}{2} \left[ \left(1 - \mu \frac{\varv_{\rm a}}{\varv}\right)^2 \nu_+ + \left(1 + \mu \frac{\varv_{\rm a}}{\varv}\right)^2 \nu_-\right], \\
	D_{\mu p}   &= \frac{1 - \mu^2}{2} p \frac{\varv_{\rm a}}{\varv} \left[\left(1 - \mu \frac{\varv_{\rm a}}{\varv}\right)\nu_+ - \left(1 + \mu \frac{\varv_{\rm a}}{\varv}\right)\nu_-\right], \\
	D_{p p}     &= \frac{1 - \mu^2}{2} p^2 \frac{\varv_{\rm a}^2}{\varv^2} (\nu_+ + \nu_-),
\end{align}
where $D_{\mu \mu}$ is the pitch-angle diffusion coefficient provided by magnetic fluctuations and $D_{p p}$ is the momentum diffusion coefficient as a result of particle acceleration by fluctuating electric fields. The mixed coefficient $D_{\mu p}$ contains elements of both scattering processes and formally derives as a result of cross-correlations between electric and magnetic turbulence.

All coefficients are correct to any order in $\mathcal{O}(\varv_{\rm a} / \varv)$ and completely describe the phase-space diffusion of CRs induced by scattering with parallel propagating Alfv\'en waves in the QLT approximation \citep{1989Schlickeiser}.

\subsection{CR streaming}
\label{sec:streaming}

Evaluating equation~\eqref{eq:ps-scattering} in terms of its moments is difficult, even in the ultra-relativistic limit. The fact that the scattering frequency is unknown precludes a direct calculation of the corresponding scattering terms. 

This situation is reminiscent of RT. The analogue to the scattering by waves is the absorption and scattering of radiation by the gas. Our wave-scattering frequency is related to the absorption coefficient in RT.  This coefficient has an intrinsic dependence on the photon frequency, as different absorption processes (i) operate in different frequency regimes and (ii) have a frequency dependence due to the underlying physical processes. In the context of RT, the absorption coefficient is often assumed to be constant. This strong assumption can be practically justified in cases where the dynamically interesting frequencies are confined to narrow bands. The resulting theory is called {\it grey} RT.

Here, we use a related approximation for CRs and define a reference energy $E'$ of \emph{typical} CRs. These CRs resonate with Alfv\'en waves of wave numbers larger than $k'_{{\rm min}, \pm} = \Omega' / (\varv'\pm \varv_{\rm a})$, where $\Omega'$ and $\varv'$ are the reference gyrofrequency and velocity at energy $E'$. In the following argument, we identify all occurring gyrofrequencies with $\Omega'$.

We further confine our analysis to isospectral Alfv\'en-wave intensities:
\begin{align}
I^{L}_\pm(k) &= H(k - k'_{{\rm min}, \pm}) \, C_{\pm}\frac{1}{k^q}, \\
I^{R}_\pm(k) &= H(k - k'_{{\rm min}, \mp}) \, C_{\pm}\frac{1}{k^q},
\end{align}
where $C_{\pm}$ are normalisation constants, $q$ is the spectral index and $H$ is the Heaviside function. Using equation~\eqref{eq:total_wave_energy_def}, we determine these constants to
\begin{align}
	C_{\pm} = (q - 1) \frac{\ewpm \Omega'^{q-1}}{(\varv' + \varv_{\rm a})^{q - 1} + (\varv' - \varv_{\rm a})^{q - 1}}.
\end{align}
Inserting this into equation~\eqref{def:scattering_coeff} yields
\begin{align}
	\nu_\pm = \pi \Omega' \frac{\ewpm}{\varepsilon_B} (q-1)\frac{\lvert \mu \varv' \mp \varv_{\rm a} \rvert^{q - 1}}{(\varv' + \varv_{\rm a})^{q - 1} + (\varv' - \varv_{a})^{q - 1}}.
    \label{eq:iso_scatt_coeff}
\end{align}

This equation shows that it is impossible to fully embrace the idea of a grey transport theory that becomes trivially independent of pitch angle cosine $\mu$. This would correspond to the case $q=1$, for which the wave spectra $I_\pm^{\rmn{L,R}}\propto k^{-1}$ become degenerate as equation~\eqref{eq:total_wave_energy_def} diverges. For $q>1$, the isospectral scattering rate $\nu_\pm$ is physically well defined and converges. However, in general different moments of the scattering rate of equation~\eqref{eq:ps-scattering} cannot be solved in closed form except for the algebraically convenient choice of $q=2$, which we adopt here. It coincides with the upper limit of theoretically inferred spectral indices of $0.8$ to $2.0$ for the bulk of resonant wave numbers \citep{2006Lazarian,2011Yan}. Assuming $q=2$ in equation~\eqref{eq:iso_scatt_coeff}, the pitch-angle averaged scattering frequencies are given by:
\begin{align}
	\bar{\nu}_\pm
    =\frac{3}{2}\int_{-1}^1\dd\mu\frac{1-\mu^2}{2}\,\nu_\pm
    =\frac{3 \pi}{8} \Omega' \frac{\ewpm/2}{\varepsilon_B}  \left( 1 + \frac{2\varv_{\rm a}^2}{\varv'^2} \right).
    \label{eq:closure_scatt_coeff}
\end{align}
Here, $(1-\mu^2)/2$ is a geometric factor connected to the pitch-angle gradient of equation~\eqref{eq:ps-scattering}. We checked that any different choice for $1<q\leq2$ yields the exact same result for the different moments up to order $\mathcal{O}(\bar{\nu} \varv_{\rm a}^2 / \varv'^2)$.

With every choice $q\neq1$ we encounter a well-known problem of QLT: for CRs with $\mu=\pm \varv_{\rm a} / \varv'$ the scattering coefficient vanishes identically. Formally, these CRs cannot resonate with any wave. As this $\mu$ corresponds to gyration nearly perpendicular to the large-scale magnetic field, this absence of scattering is commonly referred to as the $90^\circ$-problem. This problem can be resolved by two different arguments: (i) in the presence of dielectric effects the sharp resonance is broadened and CRs with wave vectors $k_{\rm res} = \infty$ in our definition are able to resonate with waves of finite wave number and (ii) a second-order treatment of the particle trajectories in small-scale turbulence, which includes a description of perturbed trajectories, introduces further resonance broadening.

As shown by theory and checked by simulations, diffusion coefficients in QLT underestimate their correct values even for $\mu$ near $\pm \varv_{\rm a} / \varv'$ \citep{2005ShalchiII}. Nevertheless the bulk of CRs are scattered with diffusion coefficients in accordance with expectation of QLT. Hence, we expect the impact of second-order QLT to only marginally change the presented result (if at all). 

Equipped with this approximation, we now evaluate moments of equation~\eqref{eq:ps-scattering}. Multiplying this equation by $E(p)$ and $\mu E(p)$, respectively, and integrating over momentum space results in
\begin{align}
		\left.\frac{\upartial \ecr}{\upartial t} \right\rvert_{\rm scatt} &= -3 \frac{\varv_{\rm a}}{c^2} (\bar{\nu}_+ - \bar{\nu}_-) \kcr + 4 \frac{\varv_{\rm a}^2}{c^2} (\bar{\nu}_+ + \bar{\nu}_-) \pcr, \\
		\left.\frac{\upartial \fcr}{\upartial t} \right\rvert_{\rm scatt} &= -(\bar{\nu}_+ + \bar{\nu}_-) \fcr + \varv_{\rm a}(\bar{\nu}_+ - \bar{\nu}_-) (\ecr + \pcr),
\end{align}
where we used the ultra-relativistic approximation $\varv \to c$ again. The symmetry in those terms can be restored by using the equations of state linking energy density and pressure as well as their corresponding anisotropic fluxes. Thus, eliminating the CR pressure via equation~\eqref{eq:eos_ecr} and the corresponding flux via equation~\eqref{eq:eos_fcr}, we arrive at
\begin{align}		
\label{eq:dot_ecr_streaming}
\left.\frac{\upartial \ecr}{\upartial t} \right\rvert_{\rm scatt} &= -\frac{\varv_{\rm a}}{3\kappa_+} \left[ \fcr - \varv_{\rm a} (\ecr + \pcr) \right] \nonumber \\&\hspace{7em}+ \frac{\varv_{\rm a}}{3\kappa_-} \left[ \fcr + \varv_{\rm a} (\ecr + \pcr) \right], \\
		\left.\frac{\upartial \fcr}{\upartial t} \right\rvert_{\rm scatt} &= -\frac{c^2}{3\kappa_+} \left[ \fcr - \varv_{\rm a} (\ecr + \pcr) \right] \nonumber \\&\hspace{7em}- \frac{c^2}{3\kappa_-} \left[ \fcr + \varv_{\rm a} (\ecr + \pcr) \right],
		\label{eq:dot_fcr_streaming}
\end{align}
where the diffusion coefficients associated with either wave are given by (see also Appendix~\ref{app:diffusion})
\begin{equation}
\label{eq:kappa}
\kappa_\pm = \frac{c^2}{3 \bar{\nu}_\pm}.
\end{equation}
The derivation of these equations concludes the proof of equations~\eqref{eq:ecr} and \eqref{eq:fcr}.

In deriving equations~\eqref{eq:dot_ecr_streaming} and \eqref{eq:dot_fcr_streaming} we neglected every boundary term resulting from partial integrations in $p$. Formally, this imposes mathematical constraints on the functional form of the CR proton distribution function that we locally approximate with a power law in momentum, $f\propto{}p^{\alpha_{\rmn{p}}}$. To justify the neglect of boundary terms at low momenta, we require a low-momentum spectral index $\alpha_{\rmn{p}}>-1$, as the phase space volume element scales as $p^2\dd p\dd\mu\dd\varphi$. In practice, a realistic CR distribution fulfills this constraint since at low particle energies, CRs suffer fast Coulomb interactions with the thermal plasma. Hence, the CR population quickly establishes a nearly constant low-momentum spectral index $\alpha_{\rmn{p}}\to 0$ \citep{2007Ensslin}. On the opposite side, our regularization constraint translates to a requirement for the high-momentum spectral index of $\alpha_{\rmn{p}}<-4$. Diffusive shock acceleration at strong shocks generates CRs with a spectral slope of $\alpha_{\rmn{p}}\approx -4.1$  and weaker shocks inject progressively softer spectra, thus meeting our requirement also holds in the high-energy regime \citep{2006Amato}. Moreover, the CR distribution exhibits an exponential cut-off at the maximum proton energy ($\sim10^{15}$~eV for supernova remnants and $\sim10^{20}$~eV for ultra high-energy CRs), which implies that there is no restricting mathematical precondition of our theory due to the spectral form of the CR distribution.

\subsection{Galilean-invariant CR streaming}

This form of equations~\eqref{eq:dot_ecr_streaming} and \eqref{eq:dot_fcr_streaming} highlights the limit of purely Alfv\'enic transport: if one of both waves dominates, CRs constantly lose energy and get scattered  until their flux approaches the Alfv\'enic limit:
\begin{align}
	\fcr \rightarrow \pm \varv_{\rm a} (\ecr + \pcr). \label{eq:alfvenic_limit}
\end{align}
We can understand this process in the wave frame: if the dominant wave scatters CRs, it isotropizes the CRs in its own frame. After the distribution reaches isotropy in the wave frame, the flux density of CR energy vanishes there by definition.  A Galilean transformation into the comoving frame demonstrates that the CR flux density is given by the limit \eqref{eq:alfvenic_limit}. Hence CRs and their energy are transported with $\pm \varv_{a}$ with respect to the gas velocity. This transport mode is called {\it streaming} of CRs and is enforced in modern transport theories through a steady-state assumption \citep{2013Zweibel,2017Pfrommer}.

The above calculation had to be carried out to order $\mathcal{O}(\bar{\nu} \varv_{\rm a}^2 / c^2)$ in order to obtain a consistent result, namely a Galilean invariant expression for scattering.  As can be inferred from equations~\eqref{eq:dot_ecr_streaming} and \eqref{eq:dot_fcr_streaming}, efficient scattering in the wave frame is necessary for a vanishing 
CR energy transfer and flux, which is the case of an isotropic CR distribution in one of the wave frames.

Calculations to lower order in the scattering terms fail to correctly account for the frame change and are thus incompatible with any Galilean invariant theory of CR transport. In Appendix~\ref{app:diffusion} we explicitly demonstrate why a lower-order calculation up to $\mathcal{O}(\bar{\nu})$, which describes pure CR diffusion, is inconsistent.

In Appendix~\ref{sec:alternative_derivation}, we provide an
alternative derivation of the scattering terms, which clarifies the physical origin of the high accuracy order $\mathcal{O}(\bar{\nu} \varv_{\rm a}^2 / c^2)$ that is needed to fully account for Galilean invariant transport. We start by evaluating pure CR pitch-angle scattering in the wave frame, which is free of electric fields. The resulting space-like component of the four-force density that is oriented along the magnetic field is given by $-\bar{\nu}_\pm\fcr/c$ and is formally of order $\mathcal{O}(\varv_{\rm a}/c)$. Performing a Lorentz transformation to lowest order $\mathcal{O}(\varv_{\rm a} / c)$ into the comoving frame picks up another factor of $\varv_{\rm a}/c$, thus explaining the puzzling result.

\subsection{Flux-limited transport}

In moment-based RT, there exists a simple physical constraint for the energy flux. Since photons travel with the speed of light $c$, the speed of the entire photon population is also limited to $c$:
\begin{align}
\left\lvert \frac{f_{\rm rad}}{c (\varepsilon_{\rm rad} + P_{\rm rad})} \right\rvert \leq 1,
\label{eq:flux_limit}
\end{align}
where $\varepsilon_{\rm rad}$, $P_{\rm rad}$ and $f_{\rm rad}$ are the radiation energy density, pressure and energy flux density. Both quantities are defined by analogy with their corresponding CR quantities. 

A similar constraint must also hold for CRs. Consider an isolated population of CRs that carries a super-Alfv\'enic flux, $\lvert \fcr \rvert > \varv_{\rm a} \ecr$. By means of equation~\eqref{eq:dot_ecr_streaming} this flux density induces a strong energy transfer from CRs to Alfv\'en waves via the gyroresonant instability. This possible mode of CR transport is unstable and rapidly decays to the Alfv\'enic streaming limit on the growth timescale of the gyroresonant instability (see Section~\ref{sec:gri}). More formally, the presented argument states  that CRs drift according to
\begin{align}
\left\lvert \frac{\fcr}{\varv_{\rm a} (\ecr+\pcr)} \right\rvert \lesssim 1.
\label{eq:flux_limit_cr}
\end{align}
 This is a posterior justification of our initial assumption that $f_1 / f_0  \ll 1$ as equation~\eqref{eq:flux_limit_cr} implies
\begin{align}
\left\lvert \frac{f_1}{f_0} \right\rvert \lesssim \frac{\varv_{\rm a}}{\varv} \ll 1.
\end{align}
Note that both constraints are not enforced by physical limitations, as in the case of radiation, but due to the assumed self-confinement of CRs.

From the microscopic point of view, this argument holds for CRs at low to intermediate energies, which are indeed self-confided. For externally-confined CRs at energies $E\gtrsim200$~GeV, this Alfv\'enic constraint needs to be replaced by equation~\eqref{eq:flux_limit}. Since low- and intermediate-energy CRs dominate the CR energy density for normal momentum spectral indexes $\alpha_{\rmn{p}}\lesssim-4.2$ (assuming that the distribution function scales as $f\propto p^{\alpha_{\rmn{p}}}$), we conclude that equation~\eqref{eq:flux_limit_cr} is valid for momentum-integrated quantities. 

If the Alfv\'en-wave energy is rapidly damped so that the damping overcomes the growth of waves, then the premise of this argument does not hold. In this situation CRs are insufficiently scattered and can indeed move with bulk velocities that exceed $\varv_{\rm a}$. None the less, the energy transfer of the streaming instability increases by a factor $\fcr / \varv_{\rm a} (\ecr + \pcr)$ provided that the bulk velocities are greater than $\varv_{\rm a}$. This increased growth rate is still able to balance the larger damping rate and a dynamical equilibrium emerges. The complexity of this case prohibits a general discussion and the question whether an equilibrium state can be reached on hydrodynamical or on kinetic time-scales needs to be addressed for the specific scenario at hand.

\section{Alfv\'en wave dynamics}
\label{sec:aflven-waves}

In this section, we embrace the connection between CR and Alfv\'en-wave transport by deriving the energy equation for Alfv\'en waves in our framework. So far, there is only a limited literature on coupled transport of CRs and Alfv\'en waves  available \citep[e.g.,][]{1992Ko,1993Jones,2016Recchia,2017Zweibel}. Hence, we discuss different damping mechanisms and calculate the corresponding energy moments, to cast our treatment of the waves into a hydrodynamical picture. We furthermore show that the gyroresonant instability acts as a source or sink of wave energy.

\subsection{Alfv\'en waves as a fluid}

To use Alfv\'en waves as a mediator between the thermal and the CR fluid, we seek to describe them by their mean energy and momentum content. Such a hydrodynamical description is justified provided the oscillations characterizing the waves do not affect the large-scale hydrodynamics directly but only their spatial and temporal mean properties, implying that the time-scale of Alfv\'en-wave oscillations has to be much shorter than any hydrodynamical time-scale. This is observed by Alfv\'en waves with typical wavelengths that are equal to the gyroradii of (pressure-carrying GeV-to-TeV) CRs. Indeed, those wavelengths are much smaller and wave-frequencies are much larger than the corresponding characteristic scales of their embedding medium such as the ISM or the ICM.

Each Alfv\'en wave consists of two principle components: a perturbation in the mean motion of the thermal gas and one of the electromagnetic field. Both components contribute to the energy density contained in Alfv\'en waves at a certain wavenumber $k$, which amounts to the sum of kinetic energy and magnetic field density:
\begin{align}
	I^{\rm L, R}_\pm(k) = \rho \frac{\left\langle \delta \bs{u}^{\rm L, R}_\pm(k)^2 \right\rangle}{2} + \frac{\left\langle \delta \bs{B}^{\rm L, R}_\pm(k)^2 \right\rangle}{2}, \label{def:spectral_alfven_energy}
\end{align}
as the incompressibility condition of shear (or pseudo) Alfv\'en waves guarantees that no thermal energy is carried by Alfv\'en waves. The perturbations in the velocity and magnetic fields are linked by the MHD relation for Alfv\'en waves:
\begin{align}
	\delta \bs{u}^{\rm L, R}_\pm = \mp\frac{\varv_{\rm a}}{B} \delta \bs{B}^{\rm L, R}_\pm. \label{eq:aw_def}
\end{align}
Thus, the kinetic and magnetic energy of an Alfv\'en wave are in equipartition and the total energy density is:
\begin{equation}
	I^{\rm L, R}_\pm(k) = \left\langle \delta \bs{B}^{\rm L, R}_\pm(k)^2 \right\rangle. \label{eq:final_spectral_alfven_energy}
\end{equation}

The mean momentum of the waves is solely given by the electromagnetic component: the perturbation in the gas momentum $\rho \bs{u}$ oscillates rapidly in space/time and thus has a zero mean over hydrodynamical scales. Contrary, the electromagnetic momentum given by the Poynting vector $\delta \bs{E}^{\rm L, R}_\pm \bs{\times} \delta \bs{B}^{\rm L, R}_\pm / c$ does not oscillate and thus has a non-zero average. For Alfv\'en waves, the mean momentum density in a single wavemode as measured by an observer in the comoving frame is given by
\begin{align}
	\frac{\left\langle \delta \bs{E}^{\rm L, R}_\pm(k) \bs{\times} \delta \bs{B}^{\rm L, R}_\pm(k) \right\rangle}{c} = \mp \frac{I^{\rm L, R}_\pm(k)}{c} \frac{\bs{\varv}_{\rm a}}{c},
\end{align}
since the electric field of an Alfv\'en wave is $\delta \bs{E}^{\rm L, R}_\pm(k) = \mp \bs{\varv}_{\rm a} \bs{\times} \delta \bs{B}^{\rm L, R}_\pm(k) / c$. 
In the non-relativistic MHD approximation, this electromagnetic momentum is neglected as it is assumed to be vanishingly small to order $\mathcal{O}(\varv_{\rm a} / c)$. However, the result above is instructive because Alfv\'en waves do not carry kinetic momentum, which needs to be considered during the discussion of the acting forces.

We now discuss a minor and rather subtle discrepancy between the ideal MHD assumption and our treatment of CRs. In order to fully describe the interaction between CRs and Alfv\'en waves, we arrived at the momentum diffusion terms of \citet{1989Schlickeiser}. The moment expansion of those terms as manifest in equations~\eqref{eq:dot_ecr_streaming} and \eqref{eq:dot_fcr_streaming} include the full effects of magnetic and electric fields and are formally accurate up to oder $\mathcal{O}(\varv_{\rm a}^2 / c^2)$. Here, the first and second order terms derive from the correlations $\langle \delta \bs{E}^{\rm L, R}_\pm \delta \bs{B}^{\rm L, R}_\pm \rangle $ and $\langle \delta \bs{E}^{\rm L, R}_\pm \delta \bs{E}^{\rm L, R}_\pm \rangle$ with $\delta E \sim \varv_{\rm a} \delta B/c$. Consequently, our description of the transfer of momentum and energy between Alfv\'en waves and CRs exhibits an accuracy with a comparable order, at least up to $\mathcal{O}(\varv_{\rm a} / c)$. Contrarily, the formulation of ideal MHD in a non-relativistic setting disregards contributions of order $\mathcal{O}(\varv_{\rm a} / c)$ and thus neglects contributions of the electric fields to any energy balance. Hence, there is a contradiction: electric fields do work on CRs while we neglect their energy density in equation~\eqref{def:spectral_alfven_energy}. 

To resolve this contradiction, we could account for the Poynting flux and energy density of the electric fields in the unperturbed MHD equations in a semi-relativistic approximation \citep{1970Boris,2002Gombosi}. This would restore the missing energy density of the electric field of the Alfv\'en waves to equation~\eqref{def:spectral_alfven_energy} but would simultaneously change the momentum equation of MHD, too. The latter step alters the dispersion relation of Alfv\'en waves to $\omega(k) = \pm k \varv_{\rm a} / \sqrt{1 + \varv_a^2 / c^2}$ and we would need to reevaluate the right-hand side of equation~\eqref{def:spectral_alfven_energy}. However, due to cancellations of Lorentz factors, expressing the total energy density $I^{\rm L, R}_\pm(k)$ in terms of $\langle \delta \bs{B}^{\rm L, R}_\pm(k)^2 \rangle$ again results in equation~\eqref{eq:final_spectral_alfven_energy}. In the end, the equipartition between kinetic and electromagnetic energy of an Alfv\'en wave remains unchanged. 

\subsection{Alfv\'en waves on an inhomogeneous background}

To account for the inhomogeneous background of Alfv\'en waves, we perform a WKB (Wentzel-Kramers-Brillouin) approximation for the defining properties of Alfv\'en waves. In the following, we suppress the super- and sub-scripts of perturbations indicating their propagation directions and polarization states ($\pm$ respectively L, R) for simplicity. We shall therefore assume that the subsequent arguments only hold for a distinct wave of given propagation direction and polarization state. For example, we decompose the magnetic field into plain waves:
\begin{align}
\delta \bs{B}(\bs{x}) = \int {\rm d}k \, \delta \bs{B}(k, \bs{x}) \exp\left[ i (k \bs{b} \bs{\cdot} \bs{x} - \omega(\bs{k}) t ) \right],
\label{eq:wkb_magnetic_field_perturb}
\end{align}
where $\delta \bs{B}(k, \bs{x})$ is slowly varying in space and time. The quantity $\delta \bs{B}(k, \bs{x})$ can be interpreted as the perturbation of a single Alfv\'en wave with kinetic wavelength $\bs{k} = k \bs{b}$ located at $\bs{x}$. On hydrodynamical timescales the turbulent motions of these waves can be described by a few statistical parameters which are defined as time averages over the high-frequency wave oscillations. We can exchange this time average by an ensemble average assuming the validity of the ergodic theorem. For parallel propagating Alfv\'en waves in slab turbulence the wave statistics is given by the mean and second-order correlation \citep{BookSchlickeiser,BookZank}:
\begin{align}
\langle \delta \bs{B}(k, \bs{x}) \rangle &= 0, \\
\langle \delta \bs{B}(k, \bs{x}) \, \delta \bs{B}(k', \bs{x}')^* \rangle &= \frac{I^{\rm L, R}_\pm(k)}{2} \, \delta_{\rmn{d}}(\bs{x} - \bs{x}') \, \delta_{\rmn{d}}(k - k') \, (\mat{1} - \bs{b} \bs{b}), \label{eq:alfven_statistics}
\end{align}
where $\delta_{\rmn{d}}$ is Dirac's delta distribution. Both delta distributions reflect that two different Alfv\'en waves at different localizations in configuration and wave-number space are uncorrelated. The tensor $\mat{1} - \bs{b} \bs{b}$ accounts for the specific directions of the magnetic perturbations: they must be perpendicular to the mean magnetic field for parallelly propagating Alfv\'en waves.

The inhomogeneous wave background has further consequences: as the waves exert both magnetic and kinetic pressure (here in the form of ram pressure) on their surroundings, a spatially varying distribution of Alfv\'en waves induces a current that counteracts these imbalances. This slightly changes Amp\`ere's law which reads in the MHD approximation as:
\begin{align}
\bs{\nabla} \bs{\times} \delta \bs{B} = \frac{1}{c} \delta \bs{j}
\end{align}
where $\delta \bs{j} = \delta \bs{j}_{\rm gas} + \delta \bs{j}_{\rm cr}$ denotes the total current induced by Alfv\'en waves, which is carried by the thermal gas \citep{1981Achterberg}. Inserting equation~\eqref{eq:wkb_magnetic_field_perturb} into Amp\`ere's law yields:
\begin{align}
\bs{\nabla} \bs{\times} \delta \bs{B}(k, \bs{x}) + i \bs{k} \bs{\times} \delta \bs{B}(k, \bs{x}) = \frac{1}{c} \delta \bs{j}(k, \bs{x}).
\end{align}
We assume that the gas current can be decomposed into two contributions: one that is inherent to the oscillatory motion of the waves, $\delta \bs{j}_{\rm a}$, and one that is a direct consequence of the inhomogeneities, $\delta \bs{j}_{\rm inh}$, with $\delta \bs{j}_{\rm inh}=\bs{0}$ if the background is homogeneous. Thus we find in the absence of such inhomogeneities  
\begin{align}
i \bs{k} \bs{\times} \delta \bs{B}(k, \bs{x}) = \frac{1}{c} [\delta \bs{j}_{\rm a}(k, \bs{x}) + \delta \bs{j}_{\rm cr}(k, \bs{x})].
\end{align}
This is in accordance to the usual expression of Amp\`ere's law for Fourier-components in plasma physics.
This decomposition allows us to directly use the results obtained in our local analysis where the WKB approximation is not applied and hence the Fourier components are assumed to be spatially invariant. This is particularly useful, since plasma kinetic effects are usually investigated on a homogeneous background.

\subsection{Macroscopic energy transport}
\label{sec:marco}

The transport equations for MHD waves can be derived with the action principle and Whitham's  (\citeyear{1961Whitham}) transport theory for waves 
\citep{1970Dewar, 1977Jacques}. Here, we rederive their results for parallel propagating Alfv\'en waves following a different approach. This allows us to accurately identify the exerted forces and their associated work done on thermal and CR fluids.

We start with the Euler equation in the lab frame, which describes the thermal gas subject to a Lorentz force in its Lagrangian form (${\rm d}/{\rm d}t = \upartial / \upartial t + \bs{u} \bs{\cdot} \bs{\nabla}$):
\begin{align}
	\rho \frac{{\rm d} \bs{u}}{{\rm d} t} = - \bs{\nabla} P_{\rm th} + \frac{\bs{j}_{\rm gas} \bs{\times} \bs{B}}{c}.
\end{align}
Introducing perturbations $q \to q + \delta q$ where $q \in \lbrace \bs{u}, \bs{B}, \bs{j}\rbrace$ in a Reynolds decomposition results in  
\begin{align}
	\rho &\left[\frac{{\rm d} \bs{u}}{{\rm d} t} + \frac{{\rm d} \delta \bs{u}}{{\rm d} t}  + (\delta \bs{u} \bs{\cdot} \bs{\nabla}) \bs{u} + (\delta \bs{u} \bs{\cdot} \bs{\nabla}) \delta \bs{u} \right] = - \bs{\nabla} P_{\rm th}  \nonumber \\
    &\hspace{1em} + \frac{1}{c}\left(\bs{j}_{\rm gas} \bs{\times} \bs{B} + \delta \bs{j}_{\rm gas} \bs{\times} \delta \bs{B} + \delta \bs{j}_{\rm gas} \bs{\times} \bs{B} + \bs{j}_{\rm gas} \bs{\times}  \delta \bs{B}\right).
\end{align}
We separate mean and fluctuating components by taking the ensemble average with $\langle \delta q \rangle$ and subtract the averaged from the unaveraged equations to arrive at:
\begin{align}
\rho \frac{{\rm d} \bs{u}}{{\rm d} t} &= -\bs{\nabla} P_{\rm th} + \frac{\bs{j}_{\rm gas} \bs{\times} \bs{B}}{c} + \left\langle \frac{\delta \bs{j}_{\rm gas} \bs{\times} \delta \bs{B}}{c} - (\rho \delta \bs{u} \bs{\cdot} \bs{\nabla}) \delta \bs{u} \right\rangle \label{eq:unperturb_euler}\\
\rho \frac{{\rm d} \delta \bs{u}}{{\rm d} t} &= -(\rho \delta \bs{u} \bs{\cdot} \bs{\nabla}) \bs{u} +\frac{\delta \bs{j}_{\rm gas} \bs{\times} \bs{B}}{c} + \frac{\bs{j}_{\rm gas} \bs{\times}  \delta \bs{B}}{c} \nonumber\\&\phantom{=}+ \left[\left\langle \frac{\delta \bs{j}_{\rm gas} \bs{\times} \delta \bs{B}}{c} - (\rho \delta \bs{u} \bs{\cdot}\bs{\nabla}) \delta \bs{u} \right\rangle\right], \label{eq:perturb_euler}
\end{align}
where we introduced the abbreviation $[\langle q \rangle] = q - \langle q \rangle$. We identify the forces acting on the mean motion as the Reynolds stress and the pondermotive Lorentz force. Multiplying equation~\eqref{eq:perturb_euler} by $\delta \bs{u}$, adding the continuity equation times $\bs{u}^2/2$, and ignoring terms that are third order in fluctuations results in an evolution equation for the kinetic energy in Alfv\'en waves:
\begin{align}
	\frac{\upartial}{\upartial t}\left(\rho \frac{\delta \bs{u}^2}{2}\right) + \bs{\nabla} \bs{\cdot} \left( \bs{u} \rho \frac{\delta \bs{u}^2}{2} \right) + (\rho \delta \bs{u} \delta \bs{u}) \bs{:}\bs{\nabla} \bs{u} = \pm \bs{\varv}_{\rm a} \bs{\cdot} \frac{\delta \bs{j}_{\rm gas} \bs{\times} \delta \bs{B}}{c}.
\end{align}
The same procedure results in Poynting's theorem for magnetic energy in fluctuations:
\begin{align}
	\frac{\upartial}{\upartial t} \left(\frac{\delta \bs{B}^2}{2}\right) + \bs{\nabla} \bs{\cdot} \left[ (\bs{u} \pm \bs{\varv}_{\rm a}) \delta \bs{B}^2 \right] - (\delta \bs{B} \delta \bs{B})\bs{:\nabla} \bs{u} &= \nonumber \\
    &\hspace{-10em} (\delta \bs{B} \bs{u}) \bs{:} \bs{\nabla} \delta \bs{B} - (\bs{u} \pm \bs{\varv}_{\rm a}) \bs{\cdot} \frac{\delta \bs{j} \bs{\times} \delta \bs{B}}{c}.
\end{align}
Here, we used the MHD relation for Alfv\'en waves in the lab frame, $\delta\bs{E}=-(\bs{u}\pm\bs{\varv}_\rmn{a})\bs{\times} \delta\bs{B}/c$, and the relation $\delta\bs{B}\bs{\cdot}\bs{\varv}_\rmn{a}=0$ for parallel propagating Alfv\'en waves. Adding both equations for magnetic and kinetic energy of both  polarizations states together with equations~\eqref{eq:wkb_magnetic_field_perturb} and \eqref{eq:alfven_statistics}, we obtain a conservation law for the energy contained in Alfv\'en waves:
\begin{align}
	\frac{\upartial E_\pm(k)}{\upartial t} + \bs{\nabla} \bs{\cdot} \left[ \left( \bs{u} \pm \varv_{\rm a} \bs{b} \right)  E_\pm(k) \right] + \frac{1}{2} (\bs{\nabla} \bs{\cdot} \bs{u}) E_\pm(k) &= \nonumber \\
    &\hspace{-12em} \mp \bs{\varv}_{\rm a} \bs{\cdot} \frac{\langle \delta \bs{j}_{\rm cr}(k) \bs{\times} [\delta \bs{B}^{\rm L}_\pm(k) + \delta \bs{B}^{\rm R}_\pm(k)] \rangle}{c},
    \label{eq:dot_p_alf}
\end{align}
which coincides with the result obtained by \citet{1970Dewar} except for the work done by CRs. Following the arguments of \citet{1981Achterberg} we express this work using the growth rate Alf\'ven waves by:
\begin{align}
\mp \bs{\varv}_{\rm a} \bs{\cdot} \frac{\langle \delta \bs{j}_{\rm cr}(k) \bs{\times} [\delta \bs{B}^{\rm L}_\pm(k) + \delta \bs{B}^{\rm R}_\pm(k)] \rangle}{c} = \Gamma_{\rm gri, \pm}(k) R_\pm(k),
\label{eq:pdv_alfven}
\end{align}
where $\Gamma_{\rm gri, \pm}(k)$ is the growth rate of Alfv\'en waves caused by the gyroresonant instability ({\it gri} for short). There are additional physical loss-processes that act on Alfv\'en waves and convert kinetic and magnetic energy into heat. We model those processes by effective growth rates $\Gamma_{\rm loss,\pm}(k)$, such that we finally obtain
\begin{align}
\frac{\upartial E_\pm(k)}{\upartial t} + \bs{\nabla} \bs{\cdot} \left[ \left( \bs{u} \pm \varv_{\rm a} \bs{b} \right)  E_\pm(k) \right] + \frac{1}{2} (\bs{\nabla} \bs{\cdot} \bs{u}) E_\pm(k) &= \nonumber \\
    &\hspace{-12em} \Gamma_{\rm gri, \pm}(k) R_\pm(k) - \Gamma_{\rm loss,\pm}(k) E_\pm(k).
    \label{eq:dot_p_alf_2}
\end{align}

The interpretation of the left-hand side is straightforward: wave energy is transported with the Alfv\'en speed relative to the gas and experiences adiabatic changes due to the spectral wave pressure $E_\pm(k)/2$. The total energy contained in Alfv\'en waves is
\begin{equation}
	\ewpm = \int_0^\infty\!{\rm d}k\,E_\pm(k).
\end{equation}
The total wave pressure obeys the equation of state
\begin{align}
  	\pwpm = (\gamma_{\rm a} - 1) \ewpm,
\end{align}
with an adiabatic index of $\gamma_{\rm a} = 3/2$. We can readily integrate equation~\eqref{eq:dot_p_alf_2} over wave number space to obtain
\begin{align} 
\label{eq:wave_energy}
\frac{\upartial \ewpm}{\upartial t} + \bs{\nabla} \bs{\cdot} \left[ \left( \bs{u} \pm \varv_{\rm a} \bs{b} \right)  \ewpm \right] + \frac{1}{2} (\bs{\nabla} \bs{\cdot} \bs{u}) \ewpm = S_{\rmn{gri}, \pm} - Q_{\pm},
\end{align}
where the Fourier integrated source terms for energy gains and losses are given by
\begin{align}
S_{\rmn{gri}, \pm} &= \int_0^\infty\!{\rm d}k\,\Gamma_{{\rm gri}, \pm}(k)R_\pm(k),\mbox{ and}
\label{eq:Sgri}\\
Q_{\pm} &= \int_0^\infty\!{\rm d}k\,\Gamma_{{\rm loss}, \pm}(k)E_\pm(k).
    \label{eq:Saw}
\end{align}
In the following, we discuss different wave creation and annihilation processes, which are known to operate in ISM or ICM conditions and provide expressions for $Q_{\pm}$.

\subsection{Gyroresonant instability}
\label{sec:gri}

As CRs drift with an anisotropy in the Alfv\'en frame and gyrate around the mean magnetic field, collectively they excite Alfv\'en waves in resonance with their gyromotion. This effect is intimately related to CR scattering: any CR distribution with a residual anisotropy of pitch angles  transfers energy to or extracts energy from the waves via scattering. For instance, if CRs are moving in the same direction as an Alfv\'en wave packet, but the CR streaming velocity $\fcr / (\ecr + \pcr)$ exceeds the Alfv\'en velocity $\varv_{\rm a}$, then these waves gain energy while the CR distribution loses energy. The growth rate of this process is \citep{2017Zweibel,1969KulsrudPearce}:\footnote{The growth rate in equation (4) of \citet{2017Zweibel} is calculated for electric fields. In order to obtain a growth rate for the resonant wave energy of Alfv\'en waves in equation~\eqref{eq:Sgri}, we multiply equation (4) of \citet{2017Zweibel} by a factor of 2.}
\begin{align}
\Gamma_{\rmn{gri},\pm} = 
    &\pm \int {\rm d}^3 p \, \frac{\pi \Omega^2 \varv_{\rm a}}{\varepsilon_B} \frac{1-\mu^2}{2} p 
    \label{eq:gri_rate} \\ 
    & \times \left[\left(1\mp\mu\frac{\varv_\rmn{a}}{\varv}\right)\frac{\upartial f}{\upartial \mu} \pm \frac{\varv_{\rm a}}{\varv} p \frac{\upartial f}{\upartial p} \right] \delta_{\rmn{d}} \left((\mu \varv \mp \varv_{\rm a}) k  - \Omega\right). \nonumber
\end{align}
Dirac's $\delta$ distribution is the formal consequence of the gyroresonance condition of equation~\eqref{eq:resonance_condition}. Again, we account for the polarization dependence of the resonance by the definition of the resonant energy in equation~\eqref{def:resonant_wave_spectrum}. If we directly evaluate the ${\rm d}^3 p$-integral of equation~\eqref{eq:gri_rate}, this definition and Dirac's $\delta$ distribution select the correct CR momenta and pitch angles, which are scattered by waves with a given $k$.

To obtain the source function of Alfv\'en wave energy in equation~\eqref{eq:Sgri}, we integrate over $k$-space and evaluate $R_\pm(k)$ at the zero of the argument of the $\delta$ distribution. Accounting for the approximation of isospectral wave intensities as discussed in Section~\ref{sec:streaming}, we find in the ultra-relativistic limit:
\begin{align}
S_{\rmn{gri},\pm} =  \pm \frac{\varv_{\rm a}}{3 \kappa_\pm} \left[\fcr \mp \varv_{\rm a} (\ecr + \pcr)  \right].
\label{eq:gri_source_term}
\end{align}
Comparing this result to the CR energy loss term on the right-hand side of equation~\eqref{eq:dot_ecr_streaming} we find that the sum of CR and wave energy is exactly conserved during gyroresonant scattering. 

We can directly infer the acting forces from the growth and decay of Alfv\'en wave energy by integrating both sides of equation~\eqref{eq:pdv_alfven}:
\begin{align}
	S_{\rmn{gri},\pm} = \pm \bs{\varv}_{\rm a} \bs{\cdot} \bs{g}_{\rm gri, \pm},
\end{align}
where $\bs{g}_{\rm gri, \pm}$ is the corresponding force density. Because the momentum of an Alfv\'en wave is aligned with the propagation direction of the Alfv\'en wave itself, each exerted force on or by Alfv\'en waves must be aligned with $\bs{b}$, too. We can finally conclude that
\begin{align}
	\bs{g}_{\rm gri, \pm} = \pm \frac{S_{\rmn{gri},\pm}}{\varv_{\rm a}} \bs{b} = \frac{\bs{b}}{3 \kappa_\pm} [\fcr \mp(\ecr + \pcr)].
    \label{eq:gri_forces_final}
\end{align}
The derivation of equations~\eqref{eq:wave_energy} and \eqref{eq:gri_forces_final} concludes the proof of equation~\eqref{eq:eaw}.

\subsection{Ion-neutral damping}

One of the first damping mechanisms considered was the indirect damping of waves by the friction between ions and neutrals in a partially ionized medium \citep[see Appendix C of][]{1969KulsrudPearce}. The process can be understood as follows: collisions between ions and neutrals maintain near equilibrium so that they share a similar temperature and mean velocity (modified by the square root of the mass ratio). The ions are additionally accelerated by the Lorenz force generated by the Alfv\'en waves. As before,  the waves lose energy due to this acceleration, while the ions gain this as kinetic energy. However, this force can be cancelled by friction between both particle species. In the end, the energy lost by waves is thermalised and heats both ions and neutrals.

We here account for the friction between ions ($\rm i$), neutral hydrogen ($\rm H$) and neutral helium ($\rm He$). The damping rate for this three-component fluid was derived by \cite{2016Soler}, whom we closely follow here. First, we consider the definition of the friction coefficient for collisions between ions and neutrals with small relative drift velocities:
\begin{align}
	\alpha_{\beta \beta'} = n_{\beta} \, n_{\beta '} \,  m_{\beta \beta'} \, \sigma_{\beta \beta'} \frac{4}{3} \sqrt{\frac{8 k_{\rm b} T_{\rm th}}{\pi m_{\beta \beta'}}},
\end{align}
where $\beta, \beta' \in \lbrace {\rm i}, {\rm H}, {\rm He} \rbrace$, $m_{\beta \beta'} = m_{\beta} \,m_{\beta '} / (m_{\beta} + m_{\beta '})$ is the reduced mass of either two species, $n_{\beta}$ and $m_{\beta}$ are the number density and mass of species $\beta$, $T_{\rm th}$ and $k_{\rm B}$ are temperature and Boltzmann's constant, respectively. We implicitly assume that all plasma components share the same temperature. The momentum-transfer cross sections of interest are $\sigma_{\rm i H} = 10^{-18}~{\rm m}^{-2}$ and $\sigma_{\rm i He} = 3 \times 10^{-19}~{\rm m}^{-2}$.
The resulting damping rate is given by 
\begin{align}
\Gamma_{{\rm in}} = \frac{1}{2} \left( \frac{\alpha_{\rm iH}}{\rho_{\rm i}} + \frac{\alpha_{\rm iHe}}{\rho_{\rm i}} \right),
\end{align}
where we neglect terms, which are second order in the collision frequencies $\alpha_{\rm \beta \beta'} / \rho_{\rm i}$ and $\rho_{\rm i}$ is the mass density of ions.

Since $\Gamma_{{\rm in}}$ is independent of wave number, we conclude that the total loss term of Alfv\'en waves by ion-neutral damping is given by
\begin{align}
	Q_{{\rm in},\pm} = \Gamma_{{\rm in}} \, \ewpm.
\end{align}

\subsection{Non-linear Landau damping}
\label{sec:NLL}

The thermal gas can be directly heated via another mechanism. Consider two waves 1 and 2 with wave numbers $k_i$ and wave frequencies $\omega_i$ ($i \in \{1,2\}$) that interact to form a beat wave, which propagates at the group velocity
\begin{align}
	\varv_{\rm beat} = \frac{\omega_1 - \omega_2}{k_1 - k_2}.
\end{align}
Associated with this beat wave is a second-order electric field, which accelerates thermal particles travelling at similar velocities. More formally, the two waves 1 and 2 interact through their beat wave at the Landau resonance with particles around the thermal speed $\varv_{\rm th}$:
\begin{align}
	\varv_{\rm beat} - \varv_{\rm th} = 0.
\end{align}
In a linear perturbation analysis, \citet{1973Lee} calculated the resulting damping of waves in a general setting. In a high-$\beta$ plasma ($\beta_\rmn{plasma}=\varv_{\rm th}^2 / \varv_{\rm a}^2$), where thermal electrons and protons share the same temperature, the non-linear Landau (nll) damping rate can be approximated by \citep{1981Voelk,1991Miller}
\begin{align}
\Gamma_{{\rm nll},\pm}(k) = \frac{\sqrt{\pi}}{8} \frac{\varv_{\rm th}}{\varepsilon_B} k \int_{0}^{k} {\rm d}k' \, E_\pm(k').
\label{eq:wave_gamma_nll}
\end{align}
While this damping rate strictly only applies to waves of the same propagation direction, there can also be non-linear Landau damping between counter-propagating waves. However, this effect is smaller by an order of magnitude for high-$\beta$ plasmas compared to the case of non-linear Landau of co-propagating waves \citep{1981AchterbergII,1991Miller}, hence we neglected this case here.

We can introduce a suitably averaged wave number $\langle k \rangle$ \citep[as in][]{1983McKenzie} so that the hydrodynamic version of equation~\eqref{eq:wave_gamma_nll} can be written as:
\begin{align}
	Q_{{\rm nll}, \pm} = \alpha \ewpm^2,
\end{align}
where the interaction coefficient is given by
\begin{align}
	\alpha = \frac{\sqrt{\pi}}{8} \frac{\varv_{\rm th}}{\varepsilon_B} \langle k \rangle,
\end{align}
with an averaged wave number \citep{1981Voelk}:
\begin{align}
\label{eq:mean_k}
\langle k \rangle = \frac{1}{\ewpm^2} \int_{0}^{\infty} {\rm d}k \, k E_\pm(k) \int_{0}^{k} {\rm d}k' \, E_\pm(k'),
\end{align}
which, to order of magnitude, corresponds to the resonant wave number of CRs. Please note that our particular choice of the algebraic form of $E_\pm(k)\propto k^{-2}$ formally gives raise to an ultra-violet divergence ($k\to\infty$) of wave energy loss by virtue of equations~\eqref{eq:wave_gamma_nll} and \eqref{eq:Saw}. We remind the reader that this profile was an appropriate choice for intermediate wave numbers ($k\sim c/\Omega$), where the turbulence is driven by the bulk of CRs. At larger wave numbers, i.e., in the inertial range and in the dissipation regime of the CR-driven turbulence, this spectrum is not applicable and would have to be modified to account for turbulent cascading and dissipation. This modification also cures the apparent ultra-violet divergence of the integral. 

\subsection{Turbulent and linear Landau damping}

Magnetic turbulence becomes anisotropic through the elongation of wave packets along the mean magnetic field on scales much smaller than the injection scale \citep{1995Goldreich}. Two interacting wave packets shear each other and cause field-line wandering. As the two counter-propagating wave packets follow the perturbed field lines of their corresponding collision partner, they are distorted transverse to the mean magnetic field \citep{2001Lithwick}. This process operates on the eddy turnover time and results in a cascade of energy to higher wave numbers $k_{\parallel}$ \citep{2004Farmer}.

It also acts as a damping process because it removes energy from scales where it was injected. The damping rate is minimized at the largest scale where waves are driven that obey the gyroresonance condition $\lambda_{\parallel,\rm max} \sim k_{\parallel, {\rm min}}^{-1} \sim r_{\rm L}$, and can be estimated as \citep{2004Farmer,2013Zweibel}:
\begin{align}
	\Gamma_{\rm turb} \approx \varv_{\rm a} k_{\parallel, {\rm min}} \sqrt{ \frac{k_{\rm mhd, turb}}{ k_{\parallel, {\rm min}}} },
\end{align}
where $k_{\rm mhd, turb}$ is the wave number at which the large scale MHD turbulence is driven. 

A related process is linear Landau damping of oblique waves \citep{2017Zweibel}. Here the electric field of a single wave can interact with the gas through the Landau resonance. Since Alfv\'en waves constantly change their propagation angle relative to the mean magnetic field, this effect is directly linked to large-scale magnetic turbulence and the anisotropic cascade. The corresponding damping rate can be estimated as
\begin{align}
\Gamma_{\rm ll} \approx \varv_{\rm a} \frac{\sqrt{\pi}}{4} k_{\parallel, {\rm min}} \sqrt{\beta_{\rm plasma} \frac{k_{\rm MHD, turb}}{ k_{\parallel, {\rm min}}}}.
\end{align}

Combining both damping rates, the loss of total energy density by processes related to turbulence is
\begin{align}
	Q_{{\rm turb + ll},\pm} = (\Gamma_{\rm turb} + \Gamma_{\rm ll}) \ewpm.
\end{align}

\section{Coupling to the thermal gas}
\label{sec:coupling}

After deriving the CR-Alfv\'enic subsystem, which describes the hydrodynamics of Alfv\'en wave-mediated CR transport, we are now coupling the forces and work done by this subsystem to the MHD equations and address energy and momentum conservation of this new theory. First, we review the evolution equations of kinetic, thermal, and magnetic energy.

In the preceding section we have derived the Euler equation~\eqref{eq:unperturb_euler} for the mean motion of the thermal gas, which can be written in its conservation form as:
\begin{align}
	\frac{\upartial (\rho \bs{u})}{\upartial t} + \bs{\nabla} \bs{\cdot} (\rho \bs{u} \bs{u} &+ \mat{1} P_{\rm th})
    =  \frac{\bs{j}_{\rm gas} \bs{\times} \bs{B}}{c} \nonumber\\
    &\hspace{1em}+ \left\langle \frac{\delta \bs{j}_{\rm gas} \bs{\times} \delta \bs{B}}{c} - (\rho \delta \bs{u} \bs{\cdot}\bs{\nabla}) \delta \bs{u} \right\rangle.
    \label{eq:euler}
\end{align}
To derive the evolution equation for the mean kinetic energy, we multiply equation~\eqref{eq:unperturb_euler} with $\bs{u}$, the continuity equation with $\bs{u}^2/2$ and add both results to obtain:
\begin{align}
	\frac{\upartial \varepsilon_{\rm kin}}{\upartial t} + \bs{\nabla} \bs{\cdot} (\bs{u} \varepsilon_{\rm kin}) &= - \bs{u} \bs{\cdot} \bs{\nabla} P_\rmn{th} + \bs{u} \bs{\cdot} \frac{\bs{j}_{\rm gas} \bs{\times} \bs{B}}{c} \nonumber \\
    &\phantom{=}+\bs{u} \bs{\cdot} \left\langle \frac{\delta \bs{j}_{\rm gas} \bs{\times} \delta \bs{B}}{c} - (\rho \delta \bs{u}\bs{\cdot} \bs{\nabla}) \delta \bs{u} \right\rangle \label{eq:pre_kin_energy},
\end{align}
where $\varepsilon_{\rm kin}=\rho \bs{u}^2/2$ is the mean kinetic energy density. The thermal or internal energy equation for a gas with the equation of state $P_{\rm th} = (\gamma_{\rm th} - 1) \varepsilon_{\rm th}$, is given by
\begin{align}
\frac{\upartial \varepsilon_{\rm th}}{\upartial t} + \bs{\nabla} \bs{\cdot} [ \bs{u} (\varepsilon_{\rm th} + P_{\rm th})] = \bs{u} \bs{\cdot} \bs{\nabla} P_{\rm th} + Q_{+} + Q_{-} \label{eq:pre_th_energy},
\end{align}
where we added the heating contributions of the Alfv\'en wave damping processes. The magnetic energy of the large-scale fields is given in the MHD-approximation by
\begin{align}
	\frac{\upartial \varepsilon_{\rm mag}}{\upartial t} + \bs{\nabla} \bs{\cdot} [\bs{u} (\varepsilon_{\rm mag} + \bs{B}^2/2) - \bs{u} \bs{\cdot} \bs{B} \bs{B}] &= -\bs{j} \bs{\cdot} \bs{E} \\
    &= -\bs{u} \bs{\cdot} \frac{\bs{j} \bs{\times} \bs{B}}{c} \label{eq:pre_mag_energy},
\end{align}
where $\varepsilon_{\rm mag}=\bs{B}^2/2$. We are now going to discuss the forces and their associated works exerted by the large- and small-scale electromagnetic fields.

\subsection{Perpendicular forces}
\label{sec:perpendicular_forces}
We first focus on the large-scale Lorentz force. The mean current $\bs{j}_{\rm gas}$ is composed of electron and ion currents. It is an unknown quantity of the gas and cannot be expressed in terms of $\rho$, $\bs{u}$ and $\varepsilon_{\rm th}$ in general. In the framework of ideal MHD the dependence of the Euler equation on $\bs{j}_{\rm gas}$ can be directly removed using Amp\`ere's law. This is not possible in the presence of CRs, because the mean motion of CRs also drives a current and hence affects the magnetic field. Accounting for both currents, we obtain for Amp\`ere's law:
\begin{align}
	\bs{\nabla} \bs{\times} \bs{B} = \frac{\bs{j}_{\rm gas} + \bs{j}_{\rm cr}}{c}.
	\label{eq:large_scale_ampere}
\end{align}
Again, the CR current is an unknown quantity but can be inferred by reversing our arguments above for the thermal gas. 

By definition CRs gyrate around the mean magnetic field, and thus on time-scales that are short in comparison to those of hydrodynamics. Furthermore, we expect CRs to be nearly gyrotropic, where deviations from purely parallel motions are induced by the macroscopic motions and the pressure of CRs. This inertia of CRs counteracts the Lorentz force and we expect that both reach dynamical equilibrium on time-scales of a few gyro-orbits, which is much shorter than the hydrodynamical time-scales. We can use this fact for a Chapman--Enskog expansion of the perpendicular components of the total momentum balance to infer a gyroaveraged expression for the large-scale Lorentz force.  For this, we integrate equation~\eqref{eq:vlasov} multiplied with $\bs{p}$ over momentum space, neglect the respective inertial terms that act on gyration time-scales, and project the result perpendicular to the mean magnetic field to obtain (see also equation~\ref{eq:rel_fcr_dot3}):\footnote{This argument can be easily seen for non-relativistic/low-energy CRs in the lab frame: for those, we can neglect all velocity components of the CR mean velocity $\bs{u}_\rmn{cr}$ that are perpendicular after a gyroaverage for an approximately gyrotropic CR distribution and directly deduce equation~\eqref{eq:final_cr_lorentz_force} from the CR Euler equation $\upartial_t(\rho_\rmn{cr} \bs{u}_\rmn{cr})+ \bs{\nabla}\bs{\cdot}(\rho_\rmn{cr} \bs{u}_\rmn{cr} \bs{u}_\rmn{cr} + \pcr \mat{1})=(\bs{j}_\rmn{cr} \bs{\times} \bs{B}) / c + \bs{g}_{\rmn{gri}, +} + \bs{g}_{\rmn{gri},-}$.}
\begin{equation}
\frac{\bs{j}_{\rm cr} \bs{\times} \bs{B}}{c} = \bs{\nabla}_{\perp} \pcr.
    \label{eq:final_cr_lorentz_force}
\end{equation}
This implies that any Lorentz force originating from a large-scale CR current is identically balanced by the pressure of CRs. Combining this equation with Amp\`ere's law~\eqref{eq:large_scale_ampere} yields an expression for the Lorentz force exerted on the gas,
\begin{align}
	\frac{\bs{j}_{\rm gas} \bs{\times} \bs{B}}{c} &= (\bs{\nabla} \bs{\times} \bs{B}) \bs{\times} \bs{B} - \frac{\bs{j}_{\rm cr} \bs{\times} \bs{B}}{c} \\
    &= -\bs{\nabla} \bs{\cdot} \left(\frac{\bs{B}^2}{2} \mat{1}- \bs{B} \bs{B}\right) - \bs{\nabla}_{\perp} \pcr.
    \label{eq:final_large_lorentz_force}
\end{align}
One consequence of the MHD assumption is that the electromagnetic field does not carry momentum. Consequently, equation~\eqref{eq:final_large_lorentz_force} is already the momentum balance of the electromagnetic field, where the individual Lorentz forces are identically balanced by the magnetic pressure and stress. We arrive at a combined momentum balance of the thermal gas and electromagnetic fields by inserting equation~\eqref{eq:final_large_lorentz_force} into equation~\eqref{eq:euler}. This would be the form of the momentum equation known from textbooks when the contributions of CRs would vanish. But equation~\eqref{eq:final_large_lorentz_force} introduces the perpendicular CR pressure gradient into the Euler equation as an additional term that apparently acts as an additional force.

\subsection{Parallel forces}

CRs do not directly interact with the thermal gas in the direction parallel to the mean magnetic field. Their influence is indirect, with the Alfv\'en waves acting as a mediator. Once the streaming instability causes a substantial growth of Alfv\'en waves, CRs lose energy and momentum. Various damping mechanisms subsequently transfer this energy and momentum from the waves to the thermal gas. Only in this scenario all three participants are tightly coupled. We have already discussed the tight correspondence between the energy loss rate of Alfv\'en waves and the momentum transfer through the associated forces in Section~\ref{sec:marco}. Using similar steps as presented therein, we will derive a relation for the ponderomotive force acting on the thermal gas. Taking the cross-product of Amp\`ere's law for Alfv\'en waves with $\delta \bs{B}$ results in (see equation~\ref{eq:final_large_lorentz_force}):
\begin{align}
\frac{\delta \bs{j}_{\rm gas} \bs{\times} \delta\bs{B}}{c} &= (\bs{\nabla} \bs{\times}  \delta\bs{B}) \bs{\times}  \delta\bs{B} - \frac{ \delta \bs{j}_{\rm cr} \bs{\times}  \delta \bs{B}}{c} \\
    &= -\bs{\nabla} \bs{\cdot} \left(  \frac{\delta \bs{B}^2}{2} \mat{1}-  \delta\bs{B}  \delta\bs{B}\right) - \frac{ \delta \bs{j}_{\rm cr} \bs{\times}  \delta \bs{B}}{c}.
\end{align}
Combining this result with the perturbed continuity equation in the comoving frame, $\bs{\nabla\cdot} (\delta\bs{u} \rho)=0$, and equation~\eqref{eq:aw_def}, we can simplify the ensemble average of the perturbed quantities on the right-hand side of Euler's equation~\eqref{eq:euler}:
\begin{align}
\left\langle \frac{\delta \bs{j}_{\rm gas} \bs{\times} \delta \bs{B}}{c} - (\rho \delta \bs{u} \bs{\cdot}\bs{\nabla}) \delta \bs{u} \right\rangle &\\&\hspace{-30pt}= -\bs{\nabla} \bs{\cdot} \left(  \frac{\left\langle \delta \bs{B}^2 \right\rangle}{2} \mat{1} \right) - \left\langle\frac{ \delta \bs{j}_{\rm cr} \bs{\times}  \delta \bs{B}}{c} \right\rangle
\\&\hspace{-30pt}=- \bs{\nabla} (\pwp + \pwm) + \bs{g}_{{\rm gri},+} + \bs{g}_{{\rm gri},-} \label{eq:final_small_lorentz_force},
\end{align}
after inserting $\delta \bs{B} = \delta \bs{B}^{L}_+ + \delta \bs{B}^{R}_+ + \delta \bs{B}^{L}_- + \delta \bs{B}^{R}_-$ and separating the different contributions of co- and counter-propagating waves. Again, CR-associated forces appear in the mean momentum equation solely through the assumption of vanishing electromagnetic momenta of Alfv\'en waves.

We can combine those results to obtain our final equations for the mean gas momentum in equation~\eqref{eq:final_gas_euler} by inserting equations~\eqref{eq:final_large_lorentz_force} and \eqref{eq:final_small_lorentz_force} into equation~\eqref{eq:unperturb_euler}. Similarly, we obtain the evolution equation for the combined kinetic and thermal gas energies and large scale electromagnetic fields in equation~\eqref{eq:final_gas_energy} by inserting equations~\eqref{eq:final_large_lorentz_force} and \eqref{eq:final_small_lorentz_force} into equation~\eqref{eq:pre_kin_energy} and adding the three equations~\eqref{eq:pre_kin_energy}, \eqref{eq:pre_th_energy} and \eqref{eq:pre_mag_energy}. This completes our derivation of the MHD equations augmented by the presence of CRs and small-scale Alfv\'en waves and finally our full CR--MHD system.

\subsection{Conservation laws}

We now discuss the conservation of the total energy and momentum of the MHD-CR system. Formally, the total energy and the momentum vector as measured by an observer residing in an inertial frame has to be conserved in the Newtonian limit. Mathematically, this requirement implies that the total momentum and energy densities obey a conservation equation of the type:
\begin{align}
\frac{\upartial q}{\upartial t} + \bs{\nabla} \bs{\cdot} \bs{f} = 0,
\end{align}
where $q$ is the volume density of the conserved quantity and $\bs{f}$ is its flux. The relativistic generalization of energy and momentum conservation requires the covariant derivative of the energy-momentum tensor to identically vanish, $T^{\alpha\beta}{}_{;\alpha}\equiv0$ (assuming Einstein's sum convention). Because the total momentum and energy contains contributions from (relativistic) CRs, we have to derive their conservation laws in the relativistic framework, taking appropriate approximations that are aligned with the non-relativistic formulation of ideal MHD. To proceed, we need to transform the CR energy and momentum densities from the comoving frame to the lab frame. To first order, the Lorentz-transformation of these quantities is given by (\citealt{BookMihalas}, see also Appendix~\ref{app:lab}): 
\begin{align}
\left.\ecr\right\rvert_{\rm lab} &= \ecr + 2 \frac{\bs{u} \bs{\cdot} (\fcr \bs{b})}{c^2} + \mathcal{O}\left(u^2 / c^2\right), \label{eq:transform_ecr}\\
\left.\bs{\fcr}\right\rvert_{\rm lab} &= \fcr \bs{b} + \bs{u} (\ecr +\pcr ) + \mathcal{O}\left(u^2 / c^2\right). \label{eq:transform_fcr}
\end{align}
The truncation of the Lorentz transformation after terms of order $\mathcal{O}(u/c)$ can lead to non-conservation of energy and momentum: the evolution equations of $\left.\ecr\right\rvert_{\rm lab}$ and $\left.\fcr\right\rvert_{\rm lab}$ contain terms that cannot be expressed as pure flux terms. This can be seen by taking the time derivative of equations~\eqref{eq:transform_ecr} and \eqref{eq:transform_fcr} and by inserting the respective evolution equations for the comoving quantities. In principle, we could circumvent this problem by including increasingly higher-order terms. However, we would have to simultaneously increase the order of the Lorentz-transformation and the evolution equations for $\ecr$ and $\fcr$ to obtain a consistent result.

Nevertheless, the underlying problem of the above procedure is our initial semi-relativistic approximation of ideal MHD and macroscopic CR transport, which in general do not satisfy the fully relativistic conservation equations. We instead downgrade our approximations to the Newtonian limit and demonstrate energy and momentum conservation therein. In the Newtonian limit, the relativistic CR population can only be transported with a non-relativistic mean velocity. Such a situation is realized, e.g., when CRs are streaming with $\rvert \fcr / (\ecr + \pcr) \lvert \sim \varv_{\rm a} \ll c$. In this case, the CR momentum density is negligible, $\fcr / c^2 \sim 0$, and the introduced degeneracy between the two frames vanishes. Neglecting terms containing the CR momentum, equation~\eqref{eq:fcr} reduces to
\begin{align}
\bs{\nabla}_\parallel \pcr = - \bs{g}_{\rm gri, +} - \bs{g}_{\rm gri, -}, \label{eq:force_balance}
\end{align}
where $\bs{g}_{\rm gri, \pm}$ is defined in equation~\eqref{eq:gri_forces_final}. Combining equations~\eqref{eq:final_large_lorentz_force} and \eqref{eq:force_balance} implies that all acting external forces are identically balanced by the CR pressure or vice versa when their momentum is negligible.

In this Newtonian case, equation~\eqref{eq:transform_ecr} shows that the CR energy density in lab and the comoving frame coincide. We can hence derive a total energy equation by adding the energy equations of the thermal gas, CRs and Alfv\'en waves of equations \eqref{eq:final_gas_energy}, \eqref{eq:ecr}, and \eqref{eq:eaw}, which results in 
\begin{align}
	\frac{\upartial \varepsilon_{\rm tot}}{\upartial t} + \bs{\nabla} \bs{\cdot} \bs{f}_{\rm tot} &= \bs{u} \bs{\cdot} \left( \bs{\nabla}_\parallel \pcr + \bs{g}_{\rm gri, +} + \bs{g}_{\rm gri, -} \right) \\
    &= 0,
\end{align}
where the right-hand side vanishes by equation~\eqref{eq:force_balance}. Furthermore, in the Newtonian limit neither CRs nor the electromagnetic field carry momentum, which leaves the mean motion of the thermal gas as the sole contributor to the total momentum balance. Thus, by inserting equation~\eqref{eq:force_balance} into the Euler equation~\eqref{eq:final_gas_euler}, we arrive at the total momentum balance in the Newtonian limit,
\begin{align}
	\frac{\upartial \rho \bs{u}}{\upartial t} + \bs{\nabla} \bs{\cdot} (\rho \bs{u} \bs{u} + P_{\rm tot}\mat{1} - \bs{BB}) &= \bs{0}.
\end{align}
To summarize, in the Newtonian limit the total momentum and energy is conserved, while our semi-relativistic approximation prohibits such a statement in the relativistic case in the comoving frame. In contrast to this, we can ensure energy and momentum conservation in the semi-relativistic limit if we formulate all CRs transport equation in the lab frame from the beginning. However, this comes with the serious drawback that we would need to resolve the gyroscale dynamics of our CR distribution, which is impossible for astrophysical simulations on the macroscale, see Appendix~\ref{app:lab} for details.

\section{Discussion}
\label{sec:discussion}

After the derivation of the Alfv\'en wave-mediated CR transport equation, here we show how it relates to the classical streaming-diffusion equation that was previously used to model CR transport. We close by showing how to generalize our simplified picture of grey CR transport to include spectral information of CR momentum space and Alfv\'en wave-number space.

\subsection{Relation to the streaming-diffusion equation}
\label{sec:streaming-diffusion}

The Newtonian limit in the preceding section can be realized for rapid scattering. In this case the scattering times of CRs are fast compared to the slow time-scales of the macroscopic evolution of the CR--gas fluid and terms associated with the fast time-scale can be evaluated in their steady state limit. This argument is similar to the Chapman--Enskog expansion previously applied at the kinetic level \citep{1975SkillingI,1989Schlickeiser}, but here adapted for the macroscopic description. The only terms in the CR transport equations that are assumed to be small in this expansion are those containing the CR mean momentum density $\fcr / c^2$ in equation~\eqref{eq:fcr}. Interestingly, exactly these terms have also been neglected in the Newtonian limit to derive equation~\eqref{eq:force_balance}. Thus, the results obtained in the Newtonian limit are indistinguishable to the Chapman--Enskog expansion of the CR equations. Indeed, by rearranging and expanding equation~\eqref{eq:force_balance} using the definition of $\bs{g}_{{\rm gri}, \pm}$ of equation~\eqref{eq:gri_forces_final}, we obtain Fick's law for CRs:
\begin{align}
	\bs{b} \bs{\cdot} \bs{\nabla} \ecr = - \frac{1}{\kappa} \left[ \fcr - u_{\rm st} (\ecr + \pcr) \right],
    \label{eq:steady_fcr}
\end{align}where the streaming velocity with respect to the fluid is given by
\begin{align}
\label{eq:ust}
u_{\rm st} = \varv_{\rm a} \frac{\bar{\nu}_+ - \bar{\nu}_-}{\bar{\nu}_+ + \bar{\nu}_-}
\end{align}
and the total diffusion coefficient is
\begin{align}
	\kappa = \frac{c^2}{3 (\bar{\nu}_+ + \bar{\nu}_-)}.
\end{align}
Comparing equation~\eqref{eq:steady_fcr} to its original and complete evolution equation~\eqref{eq:fcr}, it becomes clear that the Chapman--Enskog expansion approximates the flux in steady state. Inserting equation~\eqref{eq:steady_fcr} into equation~\eqref{eq:ecr} results in:
\begin{align}
\frac{\upartial \ecr}{\upartial t} + \bs{\nabla} \bs{\cdot} \left[ (\bs{u} + u_{\rm st} \bs{b}) (\ecr + \pcr) - \kappa \bs{b} \bs{b} \bs{\cdot} \bs{\nabla} \ecr \right] = \nonumber \\
+(\bs{u} + u_{\rm st} \bs{b}) \bs{\cdot} \bs{\nabla} \pcr + 4 \frac{\bar{\nu}_+ \bar{\nu}_-}{\bar{\nu}_+ + \bar{\nu}_-} \frac{\varv_{\rm a}^2}{c^2} (\ecr + \pcr),\label{eq:old_transport}
\end{align}
which coincides with the streaming-diffusion equation, modified by the inclusion of last term \citep{1992Ko,2017Zweibel,2017Pfrommer}. This term represents the second-order Fermi process, which accelerates CRs via electro-magnetic interactions with Alfv\'en waves. Since both $\bar{\nu}_+$ and $\bar{\nu}_-$ are  positive, this process always transfers energy from Alfv\'en waves to CRs. Note that this term only takes this simple form if one assumes that the collision frequencies $\nu_{\pm}$ are independent of CR momentum. If this were not the case, we would obtain a formal integral over momentum space of the CR distribution.

To correctly account for the second-order Fermi process to order $\mathcal{O}(\bar{\nu}\varv_{\rmn{a}}^2/c^2)$, two aspects of our final set of equations are necessary: (i) the Galilean invariance of the scattering terms and (ii) the inclusion of the pressure and energy density of Alfv\'en waves to accurately estimate the scattering coefficient. It further guarantees that the total energy, $E_{\rm tot}$, in the system is conserved by this process. 

\citet{1975SkillingI} derives similar expressions for the streaming velocity $u_{\rm st}$, the total diffusion coefficient $\kappa$ and the time-scale of the second-order Fermi process, as given by equation~\eqref{eq:old_transport}, see the expressions below his equation~(9). His results are formulated and valid in the kinetic framework. Our results for these three quantities can be obtained by replacing the scattering frequencies $\nu_\pm$ with their pitch-angle averaged counterparts $\bar{\nu}_\pm$ and moving into the fluid picture by taking the energy moment of his equation (9).

\subsection{Spectral CR hydrodynamics}
\label{sec:spectrum}

Here, we outline how to extend the presented theory to model the propagation of the CR momentum spectrum from the non-relativistic to the ultra-relativistic regime, which would be equivalent to dropping our grey approximation of CR transport. This extension would provide a more accurate description of CR transport at the expense of being more complicated algebraically and numerically. 

The fundamental assumption when evaluating moments of the focused transport equation~\eqref{eq:fpe_skilling} is the validity of the ultra-relativistic limit for the intrinsic CR speed. While this is certainly true for high-energy CRs, it fails for CR protons with a kinetic energy around their rest mass energy. This issue could be addressed by describing CR transport in a multi-spectral approach: instead of using the total energy of the entire CR population as the fundamental quantity, we could define spectral CR energy densities, i.e., integrated over a finite momentum range instead of the full momentum space. This would enable us to define a typical CR velocity of that spectral momentum range (thereafter called {\em bin}), which could be used instead of $c$ in the transport equation.

The spectral bins could also be used to better account for the inherent momentum dependence of CR scattering. We adopted the approximation of replacing the momentum-dependent gyrofrequency by a typical value in order to obtain very compact expressions. Instead of choosing a reference gyrofrequency, we would be able to more accurately capture the typical momenta and associated gyrofrequencies in the multi-spectral approach.

So far, we adopted an isospectral ansatz for the Alfv\'en wave intensities to account for their inherent scale dependence. As discussed, this assumption has some drawbacks. In conjunction to or separate from the multi-spectral description for the CRs, it would be possible to also drop this approximation. The overall procedure would be the same: first, we would define wave number bins for the Alfv\'en-wave intensities and describe the emerging dynamics in those bins. This directly enables a more accurate description of the gyroresonant scattering process and non-linear Landau damping, as both strongly depend on wave number.

However, this extension would come with a price: all of our derivations rely on partial integrations and each spectral bin introduces new boundary terms for every partial integration. This inevitably would expand our evolution equations. Aiming for transparency in this work, we decided in favour of describing the CR distribution by only two independent thermodynamical quantities, namely an energy density and its corresponding flux. For this choice every boundary term vanishes identically and we obtain our compact results. 

\section{Numerical demonstration}
\label{sec:numerics}

In this section, we demonstrate the feasibility of our presented approach in one dimension that is oriented along a magnetic flux tube. We showcase the interplay of CR transport mediated by Alfv\'en wave dynamics in a few selected idealized cases and discuss the strengths and weaknesses of our approach in comparison to other approaches used in the literature.

\subsection{Methods}

Here, we solely focus our attention to the dynamics of the new CR-Alfv\'en wave subsystem of the full set of hydrodynamical equations and leave a three-dimensional implementation and study of the dynamical impact of CRs to future work. Hence, we assume that the background gas is at rest and all MHD quantities are constant ($\rho,B_0={\rm const.}$, $\bs{B} = B_0\bs{e}_x$, $\bs{u}={\mathbf 0}$). With this reduction, the CR transport and Alfv\'en wave equations~\eqref{eq:ecr}, \eqref{eq:fcr}, and \eqref{eq:eaw} simplify to the numerical standard form:
\begin{align}
	\frac{\upartial \bs{Q}}{\upartial t} + \frac{\upartial \bs{F}(\bs{Q})}{\upartial x} = \bs{S}(\bs{Q}),
    \label{eq:num_full_eq_set}
\end{align}
where the state and flux vectors are
\begin{align}
\bs{Q} = \left[ \begin{array}{c} \ecr \\ \fcr \\ \ewp \\ \ewm \end{array} \right], 
\quad\bs{F}(\bs{Q}) = \left[ \begin{array}{c} \fcr \\ c^2\ecr / 3 \\ +\varv_{\rm a}\ewp \\ -\varv_{\rm a}\ewm \end{array} \right],
\end{align} 
while the sources are given by
\begin{align}
\bs{S}(\bs{Q}) = \left[ \begin{array}{c} -\frac{\dps\varv_{\rm a}}{\dps3\kappa_+} \left( \fcr - \varv_{\rm a} \gamma_{\rm cr} \ecr \right)  + \frac{\dps\varv_{\rm a}}{\dps3\kappa_-} \left( \fcr + \varv_{\rm a} \gamma_{\rm cr} \ecr \right) \\[.5em] 
-\frac{\dps{}c^2}{\dps3\kappa_+} \left( \fcr - \varv_{\rm a} \gamma_{\rm cr} \ecr \right)  - \frac{\dps{}c^2}{\dps3\kappa_-} \left( \fcr + \varv_{\rm a} \gamma_{\rm cr} \ecr \right) \\[.5em]  
+\frac{\dps\varv_{\rm a}}{\dps3\kappa_+} \left( \fcr - \varv_{\rm a} \gamma_{\rm cr} \ecr \right) - \alpha \ewp^2 + S_{\rmn{inj}} \\[.5em]  
-\frac{\dps\varv_{\rm a}}{\dps3\kappa_-} \left( \fcr + \varv_{\rm a} \gamma_{\rm cr} \ecr \right) - \alpha \ewm^2 + S_{\rmn{inj}}
\end{array} \right],
\end{align}
where $S_{\rmn{inj}}$ accounts for unresolved sources of Alfv\'en-wave energy. We only account for non-linear Landau damping (Section~\ref{sec:NLL}) and neglect other damping processes. Throughout this section we use internal code units ($\varv_{\rm a}=1$) and write for the diffusion coefficients:
\begin{align}
\label{eq:def_chi}
\frac{1}{3\kappa_\pm} = \chi \ewpm,
\end{align}
using equation~\eqref{eq:closure_scatt_coeff}.

We solve this equation with a finite volume scheme, which is second order by design. The hyperbolic eigenvalues of equation~\eqref{eq:num_full_eq_set} have characteristic velocities $\pm c / \sqrt{3}$ and $\pm \varv_{\rm a}$. To avoid excessive numerical diffusion we separately calculate numerical fluxes for the CR subsystem ($\ecr$, $\fcr$) and for the Alfv\'en-wave system ($\ewp$, $\ewm$) respectively. The fluxes at the cell boundaries are determined via the localized Lax-Friedrichs approximate Riemann solver, which calculates fluxes for the left and right states \citep{BookLeVeque}. For these states we use a space-time predictor, which approximates the left and right boundary values at a half-step akin to the MUSCL-Hancock method, which implicitly includes source terms \citep{1979vanleer}. This predictor uses reconstructed state gradients, which are limited in characteristic variables by a minmod limiter \citep{BookToro}. The full time-step results from the divergence of the flux and a calculation of the source term. To resolve the small time-scales of scattering, we subcycle the source terms and implicitly update the state-vector in each cycle. A necessary condition for numerical convergence of a hyperbolic partial differential equation is a Courant-Friedrichs-Lewy (CFL) number less than unity; we adopt $0.3$. Whether our scheme achieves its convergence order in practice remains to be seen. No analytic solution of the full set of equations is known to the authors, which precludes a formal convergence study.

In astrophysical environments under our consideration (ISM, CGM, ICM) and thus in simulations of those systems the light speed is $10^2$ to $10^4$ times larger than any MHD velocity. Thus, in order to follow CR dynamics that propagates information with the light speed, the resulting timestep is $3\times10^2$ to $3\times10^4$ times smaller by virtue of the CFL condition. Furthermore, this high signal velocity entails larger numerical diffusion for any Riemann solver of the same order. To reduce the diffusivity of the solution, we are either forced to increase the numerical order of our scheme or increase the spatial resolution, which would render most simulations unfeasible due to the increase in computational time.

Both problems can be addressed simultaneously by the reduced-speed-of-light approximation, which replaces the physical speed of light by an ad-hoc choice of a reduced value:
\begin{align}
 c \rightarrow c_{\rm red} < c.
\end{align}
However, to ensure physical validity of this approximation, the characteristic signal speed $c_{\rm red} / \sqrt{3}$ of the CR subsystem has to be larger than any MHD velocity, which guarantees the correct propagation of information when coupled to MHD. This can be motivated by looking at the opposite case: if the CR signal speed is equal or smaller than the largest MHD velocity, then CRs are unable to outrun advection by the gas and hence, information contained in the CR distribution is transported differently in the numerical scheme in comparison to Nature.   

\begin{table}
	\centering
    \caption{Adopted numerical parameters for simulations with our method.}
    \label{tab:our_num_param}
	\begin{tabular}{l c c c c c}
		\hline
		name & ICs & $c_{\rm red}$ & $\chi$ & $\alpha$ & $S_{\rm inj}$ \\
		\hline
        {\tt tp\_A\_c100} & A & $100$ & $5\times10^7$ & $5\times10^{10}$ & $1\times10^{-8}$\\
        {\tt tp\_A\_c100\_ld} & A & $100$ & $1\times10^6$ & $1\times10^{11}$ & $5\times10^{-6}$\\
        {\tt tp\_A\_c100\_id} & A & $100$ & $1\times10^6$ & $5\times10^{11}$ & $5\times10^{-6}$\\
        {\tt tp\_A\_c100\_hd} & A & $100$ & $1\times10^6$ & $1\times10^{12}$ & $5\times10^{-6}$\\
        {\tt tp\_A\_c10} & A & \phantom{0}$10$ & $5\times10^7$ & $5\times10^{10}$ & $1\times10^{-8}$\\
        {\tt tp\_B\_c100} & B & $100$ & $5\times10^7$ & $5\times10^{10}$ & $1\times10^{-8}$\\
        {\tt tp\_C\_c10} & C & \phantom{0}$10$  & $5\times10^7$ & $5\times10^{10}$ & $1\times10^{-8}$\\
        {\tt tp\_C\_c100} & C & $100$ & $5\times10^7$ & $5\times10^{10}$ & $1\times10^{-8}$\\
		\hline
	\end{tabular}
\begin{quote}
\ \\
(i) We employ three different initial conditions (ICs).\\
(ii) Here, $c_{\rm red}$ the reduce speed of light, $\chi$ is a numerical factor that describes the diffusion coefficient (see equations~\eqref{eq:def_chi} and \eqref{eq:closure_scatt_coeff}), $\alpha$ is the wave damping coefficient due to non-linear Landau damping for which we distinguish three cases: low, intermediate and high damping rates (labelled with {\tt ld}, {\tt id} and {\tt hd}, respectively); $S_{\rm inj}$ accounts for unresolved sources of Alfv\'en-wave energy.
\\
\end{quote}
\end{table}

\subsection{Set-up and simulations}

To demonstrate the emerging CR dynamics, we simulate three different cases, each of which probes a specific characteristics of CR transport. In Table~\ref{tab:our_num_param} we summarize the adopted numerical parameters for our method. 

We always choose the source term for Alfv\'en-wave energy $S_{\rm inj}$ in such a way that $\ewpm=4\times10^{-10}\ldots7\times10^{-9}$ would be the equilibrium wave energy density if only injection and non-linear Landau damping changed the wave energy. At this level of wave energy, CR scattering by the waves would have no effect on the time-scales considered here. Gyroresonant scattering introduces an additional source or sink term in the balance of wave energy. Hence, the injection of waves via $S_{\rm inj}$ serves as a dynamical process to avoid numerically degenerate solutions. The values of $\chi$ and $\alpha$ are chosen so that the typical wave energy  $\ewpm\sim10^{-6}$ in the regime of CRs streaming with Alfv\'en velocity. We use 1024 grid cells that are uniformly distributed in the computational domain, which we state together with the initial conditions. We use outflowing boundary conditions after \citet{1990Thompson}.

All these models are chosen to highlight different aspects of CR dynamics and to emphasize differences between the methods. They are all pathological, as we assume some functional forms of the physical quantities $\ecr$, $\fcr$ and $\ewpm$ which may not have a realization in reality. Now, we successively introduce the initial conditions for the simulations shown in this work. 

\subsubsection{Initial conditions A: isolated Gaussian}

Our canonical example is a Gaussian distribution of CR energy density. We set up the energy flux density so that CRs stream initially with at the Alfv\'en velocity down their gradient. We also assume that there is a constant pool of Alfv\'en waves of both propagation directions initially present while the Gaussian contains additional wave energy. The specific initial conditions are given by
\begin{align}
		g(x) &= \exp(-40 x^2), \\
        \ecr(x) &= g(x), \\
        \fcr(x) &= \gamma_{\rm cr} {\rm sgn}(x) \ecr(x), \\
        \ewpm(x) &= (1 + g(x)) \times 10^{-6}.
\end{align}
We use $[-1, 1]$ as our simulation domain.

\begin{figure*} 
	\centering
	\includegraphics[width=\linewidth]{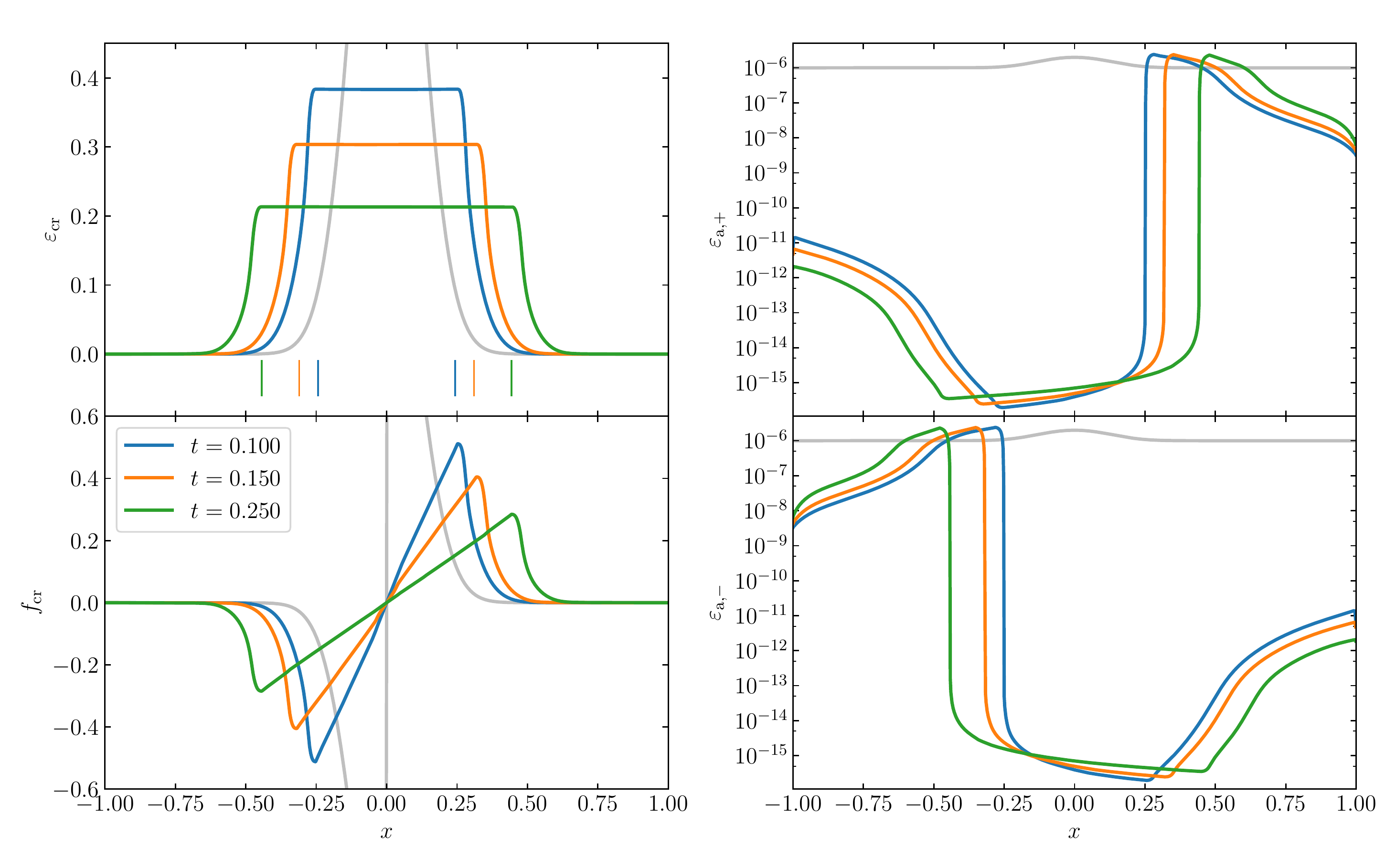}
    \caption{Temporal evolution of the Gaussian model {\tt tp\_A\_c100} using our new description for Alfv\'en-wave-mediated CR transport. We show the CR energy density $\ecr$ and its flux $\fcr$ (left-hand panels) as well as the energy density of co- and counter propagating Alfv\'en waves (right-hand panels) at three different times. The grey graphs show the initial conditions of each quantity, respectively. Vertical coloured lines in the top-left panel indicate the spatial extent of the Alfv\'enic characteristics of the initial Gaussian standard deviation that expand with velocity $\gamma_{\rm cr} \varv_{\rm a}$ in both directions.}
    \label{fig:full_panel_box}
\end{figure*}

\begin{figure*} 
	\centering
	\includegraphics[width=\linewidth]{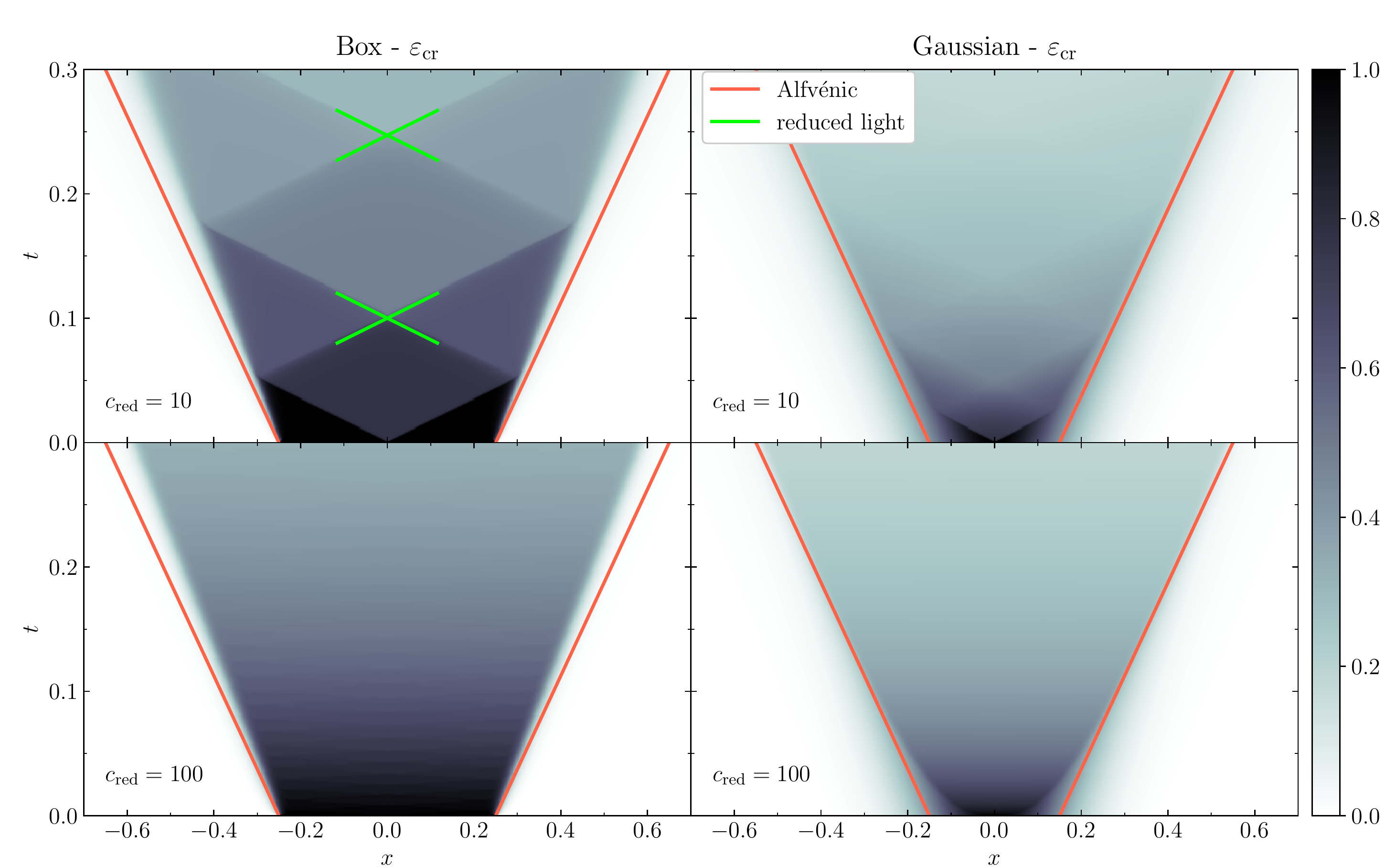}
    \caption{Time evolution of initial box (left column, models {\tt tp\_A\_c10} and {\tt tp\_A\_c100}) and Gaussian (right column, models {\tt tp\_C\_c10}  and {\tt tp\_C\_c100}) CR distributions at different reduced speeds of light, $c_{\rm red}$. While light-like characteristics are clearly visible for the evolving box distribution at low values of $c_{\rm red}$, those disappear for larger values ($c_{\rm red} = 100$) or for more realistic smooth initial CR distributions. These bird's-eye views of the evolving CR distributions show that CRs are self-confined in all cases and stream at the adiabatic Alfv\'en speed $\pm\gamma_{\rm cr} \varv_{\rm a}=\pm4/3$. We colour-code the corresponding \emph{Alfv\'enic} characteristics, which enclose the spatial extent of the CR distribution. }
    \label{fig:bird_characteristics}
\end{figure*}

\subsubsection{Initial conditions B: Gaussian with background}

The second set of initial conditions are given by
\begin{align}
		g(x) &= \exp(-40 x^2) \\
        \ecr(x) &= 10 + g(x) \\
        \fcr(x) &= \gamma_{\rm cr} {\rm sgn}(x) g(x) \ecr(x) \\
        \ewpm(x) &= (1 + g(x)) \times 10^{-6}
\end{align}
and follow the same reasoning as for initial conditions A, except that we place the Gaussian CRs distribution on top of a constant background of CR energy density, which has a 10 times larger amplitude in comparison to the Gaussian distribution. Setting up the flux this way ensures that only the Gaussian is streaming in the beginning while the background is kept at rest. Using this example, we can asses how the different numerical methods react to CR energy sources in the presence of an existing CR background. Here we simulate the domain $[-4, 4]$.

\subsubsection{Initial conditions C: isolated box}

The last set of initial conditions considered here is an isolated compact box of CR energy density, which is defined by
\begin{align}
		g(x) &= 1_{[-1/4,1/4]}(x), \\
        \ecr(x) &= g(x), \\
        \fcr(x) &= \gamma_{\rm cr} {\rm sgn}(x) \ecr(x), \\
        \ewpm(x) &= (1 + g(x)) \times 10^{-6},
\end{align}
where $1_A$ is the characteristic function of a set $A$. Again, we add an initial background of Alfv\'en waves which is enhanced in the region containing CRs. Here, the computational domain is given by $[-1, 1]$.

These initial conditions serve as a formal example to investigate the characteristics of the hyperbolic part of our differential equations. It is unlikely that this extremely sharp transition between the CR plateau and the region outside is realized in nature as the flat plateau would have to be communicated instantaneously and initial CR confinement would have to be perfect.

\subsection{CR streaming and diffusion}
\label{sec:sim_CRstreaming}

In Fig.~\ref{fig:full_panel_box} we show the temporal evolution of the isolated  Gaussian initial conditions for $\ecr$ (model {\tt tp\_A\_c100}). We adopt a reduced speed of light of $c_{\rm red}=100$ and thus begin with one of the more natural set-ups.

The most prominent feature of the solution is the expanding plateau in the CR distribution, which propagates with the adiabatic Alfv\'en velocity $\pm\gamma_{\rm cr} \varv_{\rm a}=\pm4/3$. The effective CR streaming velocity $u_{\rmn{st}}\equiv\fcr/(\ecr+\pcr)$ is sub-Alfv\'enic in the plateau region, as can be inferred from the bottom-left panel of Fig.~\ref{fig:full_panel_box}. The plateau is flat since the CR energy density flux approximately scales as $\fcr \sim x$ in this region, which yields $\upartial \fcr / \upartial x\sim {\rm const}$. Hence, there is a coherent local CR energy loss which results in a decreasing energy level of the entire plateau. Co- and counter-propagating wave energy densities, $\ewpm$, are strongly damped as sub-Alfv\'enic streaming corresponds to a transfer of both wave energies to CRs because both wave types attempt to scatter CRs into their propagation direction. The injection of wave energy balances wave loses due to non-linear Landau damping and second-order Fermi processes at a low level of $\ewpm \sim 10^{-16}$.

The outer wings of the initial Gaussian CR population are spread out by CR diffusion because there is less wave energy available to efficiently scatter CRs. In these regions, the gyroresonant instability decelerates CRs and transfers their kinetic energy to Alfv\'en waves. This results in an increase of wave energy of the outwards propagating mode. Exactly at the transition between plateau and wings, there are spikes in $\ewpm$. These correspond to fronts at which CRs are scattered most efficiently and hence, stream almost perfectly with $\fcr \approx \pm \varv_{\rm a} (\ecr + \pcr)$ so that residual growth of wave energy prevails over non-linear Landau damping.

As described, in the plateau region a large fraction of wave energy is damped. Hypothetical CR perturbations introduced there would not be efficiently scattered, because in order to do so, the waves would have to grow for approximately ten e-folding times to a level where the wave energy density would be large enough to affect the CR evolution. Hence these CR perturbations would propagate ballistically, an effect which we investigate now.

Using Fig.~\ref{fig:bird_characteristics} we investigate how the reduced-speed-of-light approximation affects the overall solution. We accomplish this by simulating the time evolution of the idealized initial box and Gaussian CR distributions, each with two values of $c_{\rm red}$ (models {\tt tp\_A\_c100} and {\tt tp\_A\_c10} for the Gaussian as well as {\tt tp\_C\_c100} and {\tt tp\_C\_c10}  for the box simulations, see Table~\ref{tab:our_num_param}). Because the numerical scattering time scales as $3 \kappa / c_{\rm red}^2$, lowering $c_{\rm red}$ from 100 to 10 enables us to gain information about processes that usually happen at very small time-scales. 

\begin{figure*} 
	\centering
	\includegraphics[width=\textwidth]{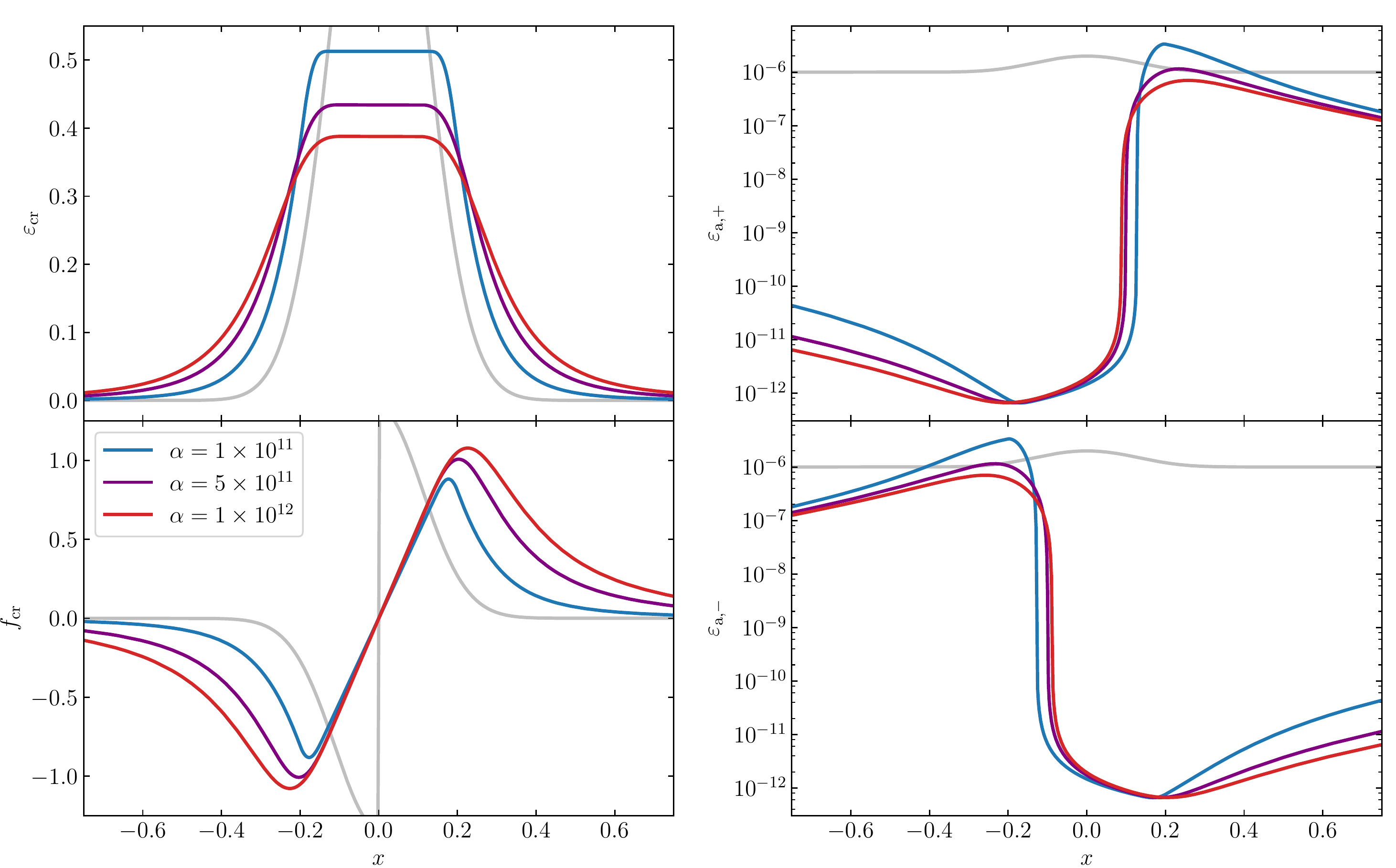}
    \caption{Effects of increasing the wave damping coefficient due to non-linear Landau damping, $\alpha$. As $\alpha$ increases, the initial Gaussian becomes broader due to lack of efficient scattering, which is accompanied by an increasing CR energy flux density $\fcr$. We use the same initial conditions for all three displayed values of $\alpha$ (shown in grey), adopt a reduced speed of light of $c_{\rm red} = 100$, and choose a snapshot at $t=0.04$.}
    \label{fig:vary_diffusion}
\end{figure*}

In the left column of Fig.~\ref{fig:bird_characteristics} we display our solution for the box initial conditions in an $x-t$ diagram. In this plot every straight line corresponds to a characteristic velocity $u_{\rm char}$, as $x = u_{\rm char} t$. The most prominent characteristics is the adiabatic Alfv\'en velocity  $\pm\gamma_{\rm cr} \varv_{\rmn{a}}$, which encloses the extent of the evolved box. Visually, the true velocity appears to be somewhat smaller, which results from the onset of diffusion at the box edges, causing them to spread apart.

In the pathological case of $c_{\rm red}=10$ (box initial conditions) we observe strong light-like characteristics propagating with velocity $\pm c_{\rm red} / \sqrt{3}$. The initial sharp transition between the CR plateau and the region outside rapidly introduces an anisotropy via the geometric contribution of the Eddington term. As most of the initial wave energy has been used up to accelerate CRs via the second-order Fermi process, there is only a small amount of waves available to scatter CRs. Thus the anisotropically moving CRs cannot be efficiently scattered into one of the wave frames. 

Furthermore, even though the CR gradient introduces anisotropy and should promote wave growth via the Eddington term, the growth rate is too small in order to efficiently reproduce waves. There are waves generated, but since the characteristics is a feature of small spatial extent travelling at large velocity its transition time is smaller than the wave growth time. The combination of both effects leads to incomplete scattering of these light-like characteristics, so that they propagate ballistically until they encounter one of the Alfv\'en characteristics. However, wave energy deposited by the light-like characteristics smoothes its wake. As the light-like characteristics interacts with the Alfv\'en characteristics, there is an evanescent wave transmitted and a reflected wave generated. While the evanescent wave damps instantaneously (the Alfv\'en mode prevails in presence of sufficiently energetic scattering waves), the reflected light-like characteristics propagates with a smaller amplitude in opposite direction.

The corresponding time-evolution of an isolated Gaussian for $c_{\rm red}=10$ and $c_{\rm red}=100$ is displayed in the right column of Fig.~\ref{fig:bird_characteristics}. Again, the entire CR population is enclosed by adiabatic \emph{Alfv\'enic} characteristics that propagate at speed $\pm\gamma_{\rm cr} \varv_{\rm a}=\pm4/3$. Here, the light-like characteristics are only present in the case $c_{\rm red}=10$ for an initial transient after which they quickly diffuse and vanish almost entirely. In the case of $c_{\rm red}=100$ there are no residual light-like characteristics visible and the evolution is completely smooth.

\subsection{Impact of damping}

In Fig.~\ref{fig:vary_diffusion} we compare the influence of the damping coefficient $\alpha$ on the solution of isolated Gaussian simulations (initial conditions~A). We show the results for simulations with $\alpha=1\times10^{11}$, $5\times10^{11}$ and $1\times10^{12}$ at $t=0.04$ and $c_{\rm red} = 100$ (models {\tt tp\_A\_c100\_ld}, {\tt tp\_A\_c100\_id}  and {\tt tp\_A\_c100\_hd}). Here, we use a smaller CR-Alfv\'en wave coupling constant $\chi=10^6$ to increase the relative impact of damping.

Corresponding to the notion of stronger damping, the maximum wave energy decreases for increasing damping coefficients. The overall shape of $\ecr$ remains similar while increasing damping coefficients yield broadened solutions of $\ecr$. This behaviour is expected: as less wave energy is available to scatter the CRs into their frame, the mode of ballistic transport starts to influence the solution. Hence, CRs get less efficiently scattered in the direction opposing their current propagation direction. As a result, an increasing damping rate yields an increasing CR flux density and consequently a broader, more diffusive solution of $\ecr$.

The particular numerical solutions presented here are clearly influenced by our choice of non-linear Landau damping. However, the overall trend remains the same for all damping processes. Consider two situations that start with the same CR distribution but exhibit varying damping strengths. The case of stronger damping implies a more evolved CR distribution with a larger spatial support in comparison to the situation with the weaker damping process. 

Lowering the imposed dynamical Alfv\'en wave energy threshold by altering the injection rate does not change the presented qualitative results in terms of $\ecr$ and $\fcr$. Doing so results in lower overall levels of wave energy density.

\subsection{Comparison to previous approaches}

Here, we compare our approach to two other approaches for CR transport in the literature: \citet{2010Sharma} model equilibrium CR streaming that is augmented with numerical diffusion to ensure numerical stability and \citet{2018Jiang} employ an ansatz inspired by RT. In Table~\ref{tab:sharma_num_param} we summarize the adopted numerical parameters for each of their methods. 

\begin{table}
	\centering
    \caption{Adopted numerical parameters for simulations with the methods of \citet{2010Sharma} and \citet{2018Jiang}.}
    \label{tab:sharma_num_param}
	\begin{tabular}{l c c c c c}
		\hline
        \multicolumn{5}{l}{\it{Method of \citet{2010Sharma}:}}\\
		name & ICs & $\delta$ & $c_{\rm red}$ & $\chi$ & $\alpha$ \\
		\hline
        {\tt sc\_A} & A & $100$ & & $5 \times 10^{7}$ & $5 \times 10^{10}$ \\
        {\tt sc\_B} & B & $100$ & & $5 \times 10^{7}$ & $5 \times 10^{10}$ \\
		\hline
		\hline
        \multicolumn{5}{l}{\it{Method of \citet{2018Jiang}:}}\\
		name & ICs & $\delta$ & $c_{\rm red}$ & $\chi$ & $\alpha$ \\
		\hline
        {\tt jo\_A} & A & & $100$ & $5 \times 10^{7}$ & $5 \times 10^{10}$ \\
        {\tt jo\_B} & B & & $100$ & $5 \times 10^{7}$ & $5 \times 10^{10}$ \\
	\end{tabular}
\begin{quote}
\ \\
(i) We employ two different initial conditions (ICs).\\
(ii) Here, $\delta$ is the regularization parameter of the streaming velocity in equation~\eqref{eq:tanh_streaming_velocity}, $c_{\rm red}$ the reduced speed of light that enters equation~\eqref{eq:jiangs_transport2}, $\chi$ is a numerical factor that describes the diffusion coefficient, and $\alpha$ is the wave damping coefficient due to non-linear Landau damping.
\end{quote}
\end{table}

We compare simulations of two set-ups: the evolution of an isolated Gaussian of $\ecr$ (initial conditions~A, see Fig.~\ref{fig:gauss_without_background}) and of a Gaussian CR distribution on a homogeneous background (initial conditions~B, see Fig.~\ref{fig:gauss_with_background}). While we have discussed the evolution of an isolated Gaussian with our theory in Section~\ref{sec:sim_CRstreaming}, here we briefly comment on the additional features that the solution assumes when we consider the Gaussian on a homogeneous background.

As CRs are streaming away from the extremum, the wings of the Gaussian expand and the central extremum decreases. As a result, the background $\ecr$ needs to respond to this change because the available volume for background CRs decreases since CRs cannot stream upwards their gradient. If this change were communicated with infinite signal speed, the level of CR background would steadily rise by analogy with the water level of a basin in which the walls are moving together. However, in reality this change needs to be communicated with the fastest signal speed. Indeed, our numerical solution shows two propagating bow waves ahead of the streaming Gaussian wings into the background medium that travel with the fastest characteristics $\pm{}c/\sqrt{3}$ (see Fig.~\ref{fig:gauss_with_background}). The wave front induces a small anisotropy which distorts the local equilibrium. With this anisotropy, a local preferred direction exists and CRs are scattered with different strengths by co- and counter-propagating waves. Due to this imbalance, the CRs begin to stream with the Alfv\'en speed in the preferred direction (see bottom-left panel of Fig.~\ref{fig:gauss_with_background}).

\subsubsection{Method of \citet{2010Sharma}}

\begin{figure*} 
	\centering
	\includegraphics[width=\textwidth]{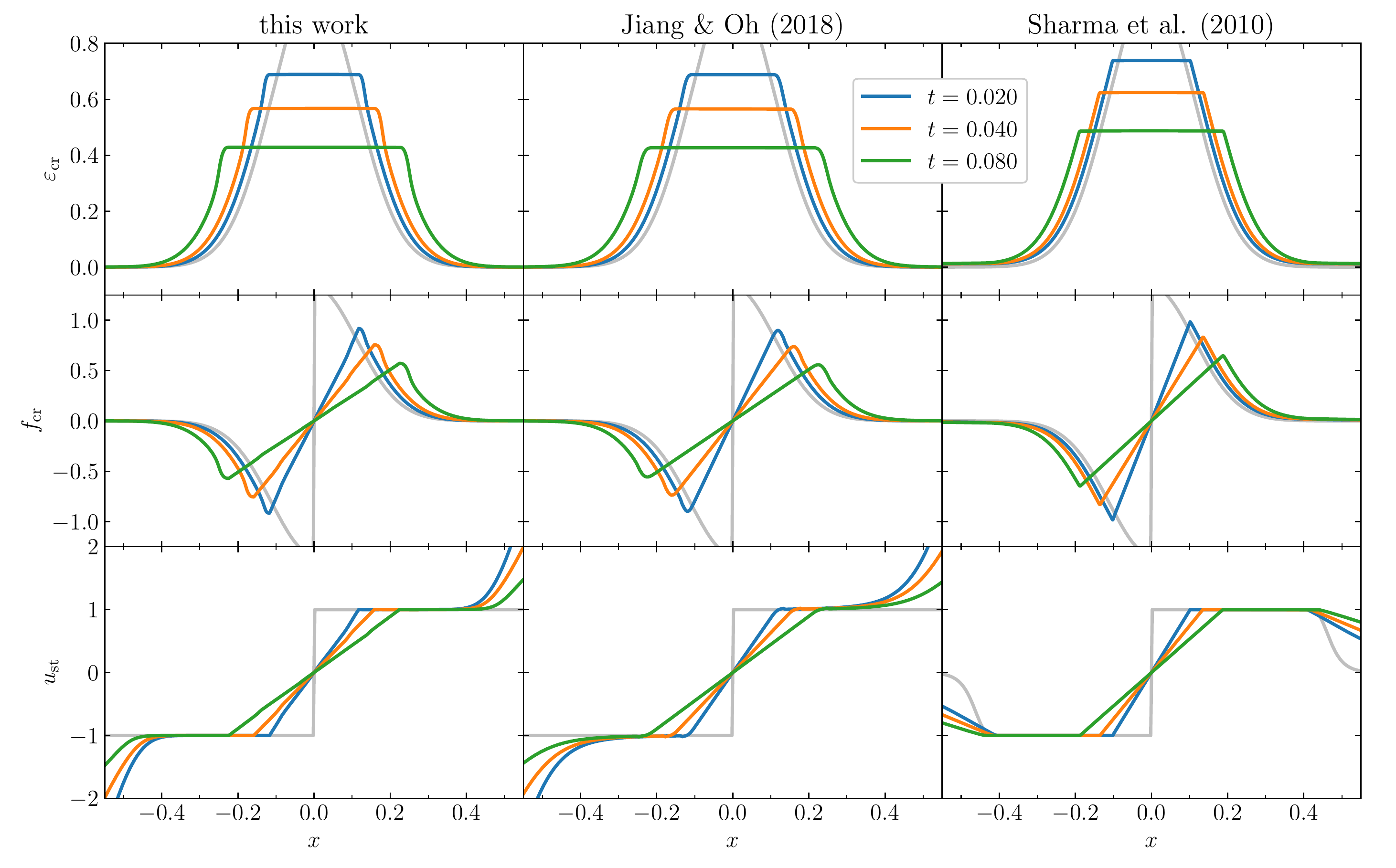}
    \caption{Comparison of the time evolution of an isolated Gaussian of $\ecr$ using three numerical methods. From left to right we compare the models {\tt tp\_A\_c100}, {\tt jo\_A}, and {\tt sc\_A}. From top to bottom, we show CR energy density $\ecr$, its flux density $\fcr$ and the CR streaming velocity $u_\rmn{cr}$ at three different times (colour-coded). In the model of \citet{2010Sharma}, we use the regularized streaming velocity of equation~\eqref{eq:tanh_streaming_velocity}, regularized CR energy flux of equation~\eqref{eq:sharma_energy_flux} and define $u_{\rmn{st}}\equiv\fcr/(\ecr+\pcr)$ for the other two approaches.}
\label{fig:gauss_without_background}
\end{figure*}

\begin{figure*} 
	\centering
	\includegraphics[width=\textwidth]{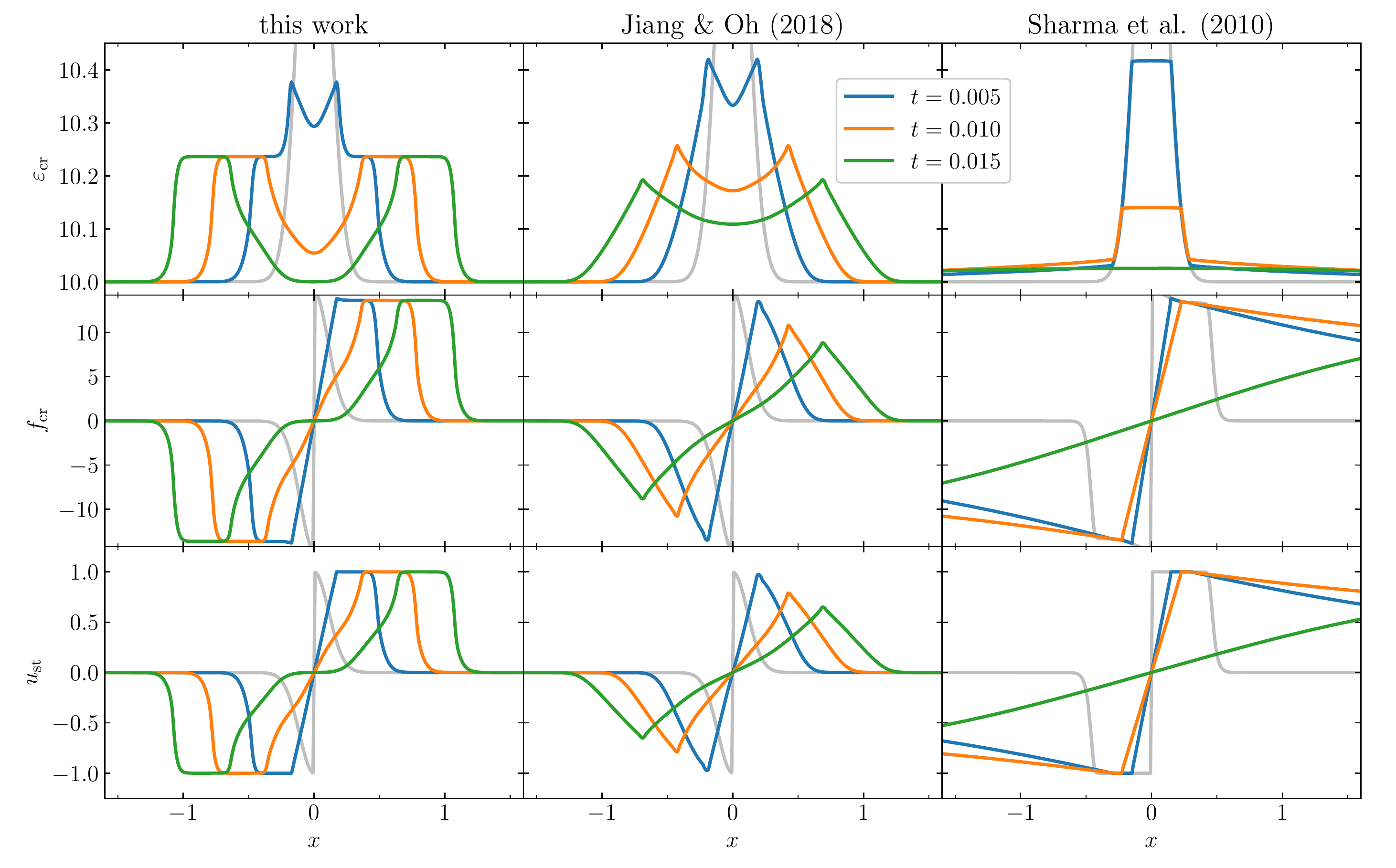}
    \caption{Same as in Fig.~\ref{fig:gauss_without_background}, but for the Gaussian on top of a homogeneous distribution of CR energy density. From left to right we compare the models {\tt tp\_B\_c100}, {\tt jo\_B}, and {\tt sc\_B}.}
    \label{fig:gauss_with_background}
\end{figure*}

Neglecting contributions from second-order Fermi processes, the steady-state version of the streaming-diffusion equation~\eqref{eq:old_transport} reads in our simplified setting:
\begin{align}
\frac{\upartial \ecr}{\upartial t} +  \frac{\upartial }{\upartial x} \left[ u_{\rm st} (\ecr + \pcr) - \kappa_0 \frac{\upartial \ecr}{\upartial x}  \right] = u_{\rm st} \frac{\upartial \pcr}{\upartial x},
\label{eq:sharma_streaming}
\end{align}
where the streaming velocity is given by
\begin{align}
u_{\rm st} = -\varv_{\rm a} \rmn{sgn}\left(\frac{\upartial \ecr}{\upartial x}\right),
\label{eq:sign_streaming_velocity}
\end{align}
and $\kappa_{0}$ is a diffusion coefficient valid in equilibrium, where gains of Alfv\'en-wave energy via the gyroresonant instability are exactly canceled by wave-loss processes.

This equation cannot be integrated using conventional finite-volume numerical methods because the functional form of this streaming velocity implies that the equation is a highly non-linear diffusion equation. This problem was first analysed by \citet{2010Sharma}, who suggested to regularize the streaming velocity of equation~\eqref{eq:sign_streaming_velocity} via
\begin{align}
\tilde{u}_{\rm st} = -\varv_{\rm a} \tanh \left(\frac{1}{\delta}\frac{\upartial \ecr}{\upartial x}\right),
\label{eq:tanh_streaming_velocity}
\end{align}
where $\delta=\rmn{const.}$ is a (small) regularization parameter. By analogy with our theory applied to steady state, we define the regularized CR energy flux density:
\begin{align}
\tilde{f}_{\rm cr} = \tilde{u}_{\rm st} (\ecr + \pcr).
\label{eq:sharma_energy_flux}
\end{align}
For $\delta\rightarrow 0$ the regularized streaming velocity matches its analytic counterpart. Inserting expression~\eqref{eq:tanh_streaming_velocity} for $\tilde{u}_{\rm st}$ into equation~\eqref{eq:sharma_streaming} yields for the advection term

\begin{align}
\label{eq:sharma_streaming_regularization}
\frac{\upartial}{\upartial x} \left[ \tilde{u}_{\rm st} (\ecr + \pcr) \right] = \tilde{u}_{\rm st} \frac{\upartial}{\upartial x}(\ecr + \pcr) -\kappa_{\rmn{reg}}\frac{\upartial^2\ecr}{\upartial x^2},
\end{align}
with
\begin{align}
\kappa_{\rmn{reg}}=\varv_{\rmn{a}}\gamma_\rmn{cr}\ecr\frac{1}{\delta}\rmn{sech}^2\left(\frac{1}{\delta}\frac{\upartial \ecr}{\upartial x}\right).
\label{eq:kappa_reg}
\end{align}
Here, $\kappa_\rmn{reg}$ is a numerical diffusion coefficient that depends on the regularization parameter $\delta$, the CR energy density, and its gradient. For weak CR energy gradients ($\upartial\ecr/\upartial x\ll\delta$) numerical diffusion dominates the solution while steady-state CR streaming emerges for steep CR energy density gradients ($\upartial\ecr/\upartial x\gg\delta$). 

With this choice, the streaming-diffusion equation is classified as a non-linear diffusion equation, even in the limit of negligible physical diffusion ($\kappa_0=0$), and can be numerically integrated. In practical terms, this regularization attempts to emulate the steady state by reconstructing local CR streaming based on the energy gradient. The extension of equation~\eqref{eq:sign_streaming_velocity} to values in between $\pm\varv_{\rmn{a}}$ is justified, as this corresponds to a smooth transition between the two limiting stationary cases. The fact that we observe this behaviour in simulations of our new non-equilibrium CR transport model provides additional physical justification for this regularization.

To compare the results of this method to ours, we include the effects of finite physical diffusion modelled by $\kappa_0$. Performing a simultaneous Chapman--Enskog expansion of the equations for wave energy and CR energy flux (equation~\ref{eq:num_full_eq_set}) we can estimate the energy density of the dominant wave species as:
\begin{equation}
\varepsilon_{\rm a} = \sqrt{ \frac{\varv_{\rm a}}{3 \alpha} \left\lvert\frac{\partial \ecr}{\partial x} \right\rvert}.
\end{equation}
The associated diffusion coefficient $\kappa_0$ is given by
\begin{equation}
\label{eq:kappa0}
\frac{1}{\kappa_0} = \chi \sqrt{\frac{3 \varv_{\rm a}}{\alpha} \left\lvert\frac{\partial \ecr}{\partial x} \right\rvert}
\end{equation}
using the closure relation of equation~\eqref{eq:def_chi}. This argument for the evaluation of the diffusion coefficient $\kappa_{0}$ corresponds to that given before by \citet{1981Voelk, 2018Jiang} and numerically implemented by \cite{2013Wiener}.

We implement this regularized scheme by evaluating the gradient in equation~\eqref{eq:tanh_streaming_velocity} on cell faces while all gradients in equation~\eqref{eq:sharma_streaming} are discretized using central differences. We integrate the time derivative using an explicit super-timestepping Runge-Kutta method \citep{2012Meyer}, which permits us to treat the non-linear parabolic terms robustly and with large, second-order accurate time-steps, thereby circumventing the restrictive parabolic von-Neumann criterion. In particular, we adopt a regularization parameter of $\delta=0.005$ and use 120 super-timesteps.

In the right column of Fig.~\ref{fig:gauss_without_background} we show the time evolution of the isolated Gaussian of $\ecr$ for the model by \citet{2010Sharma} (using model {\tt sc\_A}) and compare it to our numerical method. Both wings of the Gaussian are correctly transported with $\pm\gamma_{\rmn{cr}}\varv_{\rm a}=\pm4/3$. In this regime, the CRs transfer a sufficient amount of energy to resonant Alfv\'en waves in order to balance the damping of this wave type. In the assumed steady-state limit this corresponds to the regime in which one (i.e., the resonant) wave type dominates. Furthermore, the steady-state assumption implies that the CR distribution is isotropic in the frame of the dominant wave. As CRs stream away from the maximum, the spatial support of the Gaussian broadens. Energy conversion smoothes out the initial maximum in $\ecr$ so that it converges onto a plateau distribution. The residual gradient on the plateau continuously connects both wings in terms of energy flux density. This arises as a result of the diffusive nature of the regularization scheme, which would smooth any strong gradient of the energy flux.  

In Fig.~\ref{fig:gauss_with_background} we add a constant background of CR energy density (corresponding to the model {\tt sc\_B}). Now, the initial Gaussian is a small addition to the background, which is quickly erased by the diffusive nature of the regularization scheme. As a result, the CR energy converges to a constant value, losing any information about the initial conditions. Thus, in the picture of \citet{2010Sharma}, the solution of this problem depends entirely on numerical diffusion and the specific choice of the regularization parameter.

\subsubsection{Method of \citet{2018Jiang}:}

The fundamental idea of \citet{2018Jiang} was to describe CR transport with the equations of RT while modifying the scattering terms in order to restore the strongly-coupled limit of CR transport. For a medium at rest their one-dimensional equations read:
\begin{align}
\phantom{\frac{1}{c^2}} \frac{\upartial \ecr}{\upartial t} +  \frac{\upartial \fcr}{\upartial x} &= u_{\rm st} \frac{\upartial \pcr}{\upartial x}, \mbox{ and}
\label{eq:jiangs_transport1}\\
\frac{1}{c_{\rmn{red}}^2} \frac{\upartial \fcr}{\upartial t} +  \frac{\upartial \pcr}{\upartial x} &= - \frac{1}{\kappa} \fcr,
\label{eq:jiangs_transport2}
\end{align}
where all quantities retain the same meaning as in the preceding sections. The resemblance to the transport part of our equations~\eqref{eq:dot_ecr} and \eqref{eq:dot_fcr} is not incidental since both descriptions root in the same ideas that originate from RT. In multiple dimension, however, both transport theories differ fundamentally as \citet{2018Jiang} model the transport of CR energy in terms of a three-dimensional flux while our theory is based on its projection onto the direction of the local magnetic field. 
The system of equations~\eqref{eq:jiangs_transport1} and \eqref{eq:jiangs_transport2} is closed by 
\begin{align}
u_{\rm st} &= - \varv_{\rm a} \, {\rm sgn}\left(\frac{\upartial \pcr}{\upartial x}\right),\mbox{ and}\\
\kappa &= \kappa_{0} +  \left \vert \frac{\upartial \pcr}{\upartial x} \right\vert^{-1} \varv_{\rm a} \left(\ecr + \pcr\right), \label{eq:jiang_kappa}
\end{align}
where the first equation encodes the steady-state limit of the streaming velocity. The second equation contains a diffusion coefficient $\kappa_{0}$ and a second term that is proportional to the Alfv\'en speed times the CR gradient length. We adopt the same approximation for the physical diffusion as laid down in equation~\eqref{eq:kappa0}. Taking the steady state limit with $\kappa_0\rightarrow 0$ of equation~\eqref{eq:jiangs_transport2} results in
\begin{align}
\fcr = u_{\rm st} (\ecr + \pcr),
\end{align}
which corresponds to the correct energy flux in the limit of streaming CRs. Thus, the equations of \citet{2018Jiang} can be regarded as a compromise between our theory and that of \citet{2010Sharma}. 

This closure of the diffusion coefficient in equation~\eqref{eq:jiang_kappa} is the distinguishing feature of the formalism by \citet{2018Jiang}, as it aims at reconstructing an effective diffusion coefficient from local information of $\ecr$ only. Its particular choice roots in the idea that any CR pressure gradient excites Alfv\'en waves, which efficiently scatter CRs so that they are primarily transported via streaming. Because the scattering coefficient scales as $\bar{\nu} \propto \varepsilon_{\rm a} \propto \kappa^{-1}$, a large gradient of $\ecr$ and consequently of steady-state wave energy implies a small diffusion coefficient. Conversely, CRs experience a large diffusion coefficient if the wave energy and hence the CR pressure gradient is negligible.

The numerical problems arising when evaluating the derivative of the discontinuity of the streaming velocity (i.e., ${\rm sgn}(\upartial \ecr / \upartial x)$) are naturally circumvented here, as the problematic term in equation~\eqref{eq:sharma_streaming} is replaced by the flux density $\fcr$. Nevertheless, equation~\eqref{eq:jiangs_transport2} depends non-linearly on $\upartial \pcr / \upartial x$ through equation~\eqref{eq:jiang_kappa} and is thus classified as a Hamilton-Jacobi equation.\footnote{A Hamilton Jacobi equation in the theory of partial differential equations has the form $\upartial q / \upartial t = H(x, q, p=\upartial  q / \upartial x),$ where in general $q$ is a state vector and $H(x,q,p)$ is the Hamiltonian. Matching our physical intuition, the signal velocity $s$ of the system is given by $s=\upartial H(x,q,p) / \upartial p$.} As a consequence of this model, the closure alters the signal propagation speed of the entire system. 

We implement a numerical solver similar to that described by \citet{2018Jiang}. To this end we use the modified two-stage Runge-Kutta scheme to advance all quantities. The hyperbolic part of the equations is calculated using slope-limited piece-wise linear extrapolation to cell faces and the Lax--Friedrichs Riemann solver while the sources terms are treated implicitly. We calculate the diffusion coefficient of equation~\eqref{eq:jiang_kappa} before each stage as the mean of the left and right gradients of a cell. For this method, we use the same values of $\chi$ and $\alpha$ as for our method. 

We show the results of this scheme for the isolated Gaussian (defined by model {\tt jo\_A}) in the middle column of Fig.~\ref{fig:gauss_without_background}. All displayed schemes show a flattening of the initial maximum of the CR pressure to a plateau where the gradient of $\pcr$ approaches zero. This implies a large CR diffusion coefficient according to equation~\eqref{eq:jiang_kappa}. The wings of the Gaussian are characterized by a sizeable CR gradient, which limits the diffusivity onto the diffusion coefficient $\kappa_{0}$. Here, the advective aspect of the scheme dominates, which results in a streaming CR distribution. In the transition zone CRs stream with Alfv\'en speed because the scattering coefficient is increased so that the flux rapidly converges towards $\fcr = \pm \varv_{\rm a} (\ecr + \pcr)$. This corresponds to the equilibrium flux in equation~\eqref{eq:jiangs_transport2} for a negligible diffusion mediated by $\kappa_0$. Overall the results with the method of \citet{2018Jiang} for this test problem are almost identical to those obtained with our theory. 

If we add a constant background to the Gaussian (model {\tt jo\_B}) and evolve the equations of \citet{2018Jiang}, the results are qualitatively comparable to that obtained by our new formulation but there are notable quantitative differences. The Gaussian is broadened, while maintaining a clear spatial separation to the background, which responds to the expanding enhancement of CR energy density. This scheme captures the undershoot at the position of the initial maximum of $\ecr$, however not to its full extent in comparison to the results of our new theory. The travelling plateaus seen in the evolution with our method are erased in the scheme of \citet{2018Jiang} so that only a spike remains, which marks the transition between the moving wings and undershoot.

In conclusion, the transport equations by \citet{2018Jiang} qualitatively agree better with our results and also show clear differences to the regularization approach of \citet{2010Sharma}. This is the result of the adopted closure of \citet{2018Jiang} for the diffusion coefficient, which emulates the actual CR dynamics more accurately. 

\section{Conclusions}
\label{sec:conclusion}

We succeeded in developing a new macroscopic transport theory for CR transport, which includes both CR diffusion and streaming along magnetic field lines in the self-confinement picture: as CRs stream super-Alfv\'enically along the magnetic field, they resonantly excite Alfv\'en waves through the gyroresonant instability. Scattering off of this wave field modulates the macroscopic mode of CR transport in interesting and non-trivial ways.

For the first time, we provide a mathematically rigorous derivation of the equations of CR hydrodynamics that are coupled to the evolution of Alfv\'en waves in the Eddington approximation of RT. We accomplish this by evaluating the zeroth- and first-order CR pitch-angle moment of the gyro-averaged CR transport equation and successive integration over CR-momentum space. As a result, we obtain two coupled evolution equations for the CR energy density $\ecr$ and its flux density $\fcr$, which resemble the equations of classical radiation hydrodynamics. 

However, both equations depend on CR scattering terms, which need to be specified to close this set of equations. Our key insight for evaluating CR scattering at magnetic turbulence consists in considering a reference energy of typical CRs, similar to the grey approximation of RT. This yields a pitch-angle-averaged scattering frequency that depends on the energy level of co- and counter-propagating waves, $\bar{\nu}_\pm\propto\ewpm$, and is not a constant value as often assumed in the literature. We explicitly demonstrate that CR-wave scattering terms to order $\mathcal{O}(\bar{\nu}\varv_{\rm a}^2 / c^2)$ need to be considered in order to provide a Galilean invariant and flux-limited CR transport. A Chapman--Enskog expansion of this new set of equations (i.e., filtering out fast time-scales associated with non-equilibrium transients) or equivalently the Newtonian limit enables us to recover the classical streaming-diffusion equation of CR transport in the steady-state flux limit.

The dependence of the scattering rate on $\ewpm$ immediately exemplifies the need to dynamically also evolve the Alfv\'en wave equations for self-consistency. To this end, we derive the transport of wave energy and cast it into our new picture. We provide a complete review of all available wave damping processes such as sub-Alfv\'enically streaming CRs, non-linear Landau damping, ion-neutral damping, turbulent and linear Landau damping, and show how their contributions change $\ewpm$. Most importantly, we explicitly demonstrate that the energy lost by CRs owing to the gyroresonant instability exactly matches the energy gained by Alfv\'en waves only if the calculation is done at least to order $\mathcal{O}(\bar{\nu}\varv_{\rm a}^2 / c^2)$.

We couple the new CR-Alfv\'enic subsystem to ideal MHD, ensuring energy and momentum conservation in the non-relativistic limit of MHD. A covariant derivation of our CR hydrodynamical equations in the semi-relativistic limit in the Appendix demonstrates the validity of these conservation laws. In particular, this calculation reveals that the adiabatic CR source term in the Newtonian limit can be understood as a non-vanishing metric connection (Christoffel symbol), i.e., it has the meaning of a geometric term that results from the transformation into the non-inertial comoving frame and is not equivalent to the work associated with a force.

We finally show numerical solutions of our new CR-Alfv\'en wave subsystem in one dimension that is oriented along a magnetic flux tube. Our first-principle approach significantly advances over previous steady-state approaches because it enables us for the first time to include non-equilibrium kinetic effects such as non-linear Landau damping, second-order Fermi acceleration or energy transfer via the gyroresonant instability in hydrodynamical settings. In particular, our numerical implementation enables to quantify the relative impact of these kinetic effects on CR transport and on how CR and wave-pressure gradients impact the dynamics of thermal plasma. 

We provide a first parameter study of our CR transport theory and asses how it reacts to variations in 
\begin{itemize}
\item the reduced speed of light, $c_{\rmn{red}}$: smaller values give access to processes that act on faster time-scales but also promote (unphysical) ballistic CR transport;
\item the wave damping coefficient $\alpha$ (due to the non-linear Landau process): larger values damp the peak wave energy, increase the CR flux density, and make CR transport more diffusive;
\item the unresolved sources of Alfv\'en-wave energy, $S_{\rmn{inj}}$: as long as there is some initial wave energy, the solution for the peak wave energy (that determines the mode of CR transport) is independent on the exact wave amplitude.
\end{itemize}
We emphasize that our theory has no tunable free parameters: $c_{\rmn{red}}$ is chosen so that the solution does not depend on its specific value; the wave damping coefficient $\alpha$ and the inverse CR diffusion coefficient $\chi$ are given by MHD quantities and the characteristic gyrofrequency of our CR population in the grey approximation; and $S_{\rmn{inj}}$ does not impact on the solution as long as it does not become dynamically important. This should enable us to accurately capture momentum and energy deposition of propagating CRs in future simulations of galaxy formation.

Our numerical simulations recover CR streaming and diffusion at self-generated waves. Comparing our solutions to two previously suggested approaches, we highlight similarities and differences. Our approach recovers the equation of \citet{2010Sharma} in the physical steady-state limit (i.e., in the presence of sufficient Alfv\'en wave energy). However, their approach has the weakness of excessive numerical diffusion for small CR gradients. This problem is reinforced in the presence of highly stratified CR energy densities, which are inevitably encountered in simulations of galaxies and galaxy clusters. This renders the solution problematic if CRs are injected into an already pre-existing CR background as this represents the weak-gradient regime.

\citet{2018Jiang} describe CR transport with the equations of RT while modifying the scattering terms. Overall, the results obtained with their method are almost identical to those obtained with our theory for the problems studied here. This is due to a similar two-moment treatment of CR transport in those approaches.

Summarizing, our novel derivations of CR hydrodynamics holds the promise to provide a sustainable framework to assess the importance of CR momentum and energy feedback for galaxy formation and the cosmological evolution of cool core galaxy clusters.

\section*{Acknowledgements}
We acknowledge Avery Broderick, Yan-Fei Jiang and Peng Oh for valuable discussions, and an anonymous referee for a constructive report that greatly helped to improve this work. TT and CP acknowledge support by the European Research Council under ERC-CoG grant CRAGSMAN-646955.




\bibliographystyle{mnras}
\bibliography{main} 



\appendix
\onecolumn
\section{CR diffusion}
\label{app:diffusion}

Following up on our discussion in Section~\ref{sec:streaming-diffusion} we saw that the diffusion term in the streaming-diffusion equation is a combination of $\mathcal{O}(\bar{\nu}_{\pm})$ terms in the Taylor expansion with respect to the small variable $\varv_{\rm a} / \varv$. Hence, the diffusion term is present in any expansion. As we will show in the following, against common belief a purely diffusing CR population is physically inconsistent in case of self-confined CRs. 
 
We start by expanding equation~\eqref{eq:ps-scattering} to order $\mathcal{O}(\nu_{\pm})$ in $\varv_{\rm a} / \varv$, which yields
\begin{align}
	\left. \frac{\upartial f}{\upartial t} \right\rvert_{\rm scatt} = \frac{\upartial}{\upartial \mu} \left[ \frac{1-\mu^2}{2} (\nu_+ + \nu_-) \frac{\upartial f}{\upartial \mu} \right].
\end{align}
In physical terms this expansion neglects any contribution from electric fields. Consequently, the interaction between waves and CRs is fully described by pitch-angle scattering (even in the comoving frame). Taking the appropriate moments as in Section~\ref{sec:streaming} results in
\begin{align}
	\left. \frac{\upartial \ecr}{\upartial t} \right\rvert_{\rm scatt} &= 0, \label{eq:diff_dot_ecr_scatt}\\
	\left. \frac{\upartial \fcr}{\upartial t} \right\rvert_{\rm scatt} &= - (\bar{\nu}_+ + \bar{\nu}_-) \fcr \label{eq:diff_dot_fcr_scatt}.
\end{align}
The physical interpretation is straightforward: as CRs scatter solely off of magnetic fields, there is no energy transfer between waves and CRs. Furthermore, according to this equation, the CR population will evolve towards an isotropic distribution in the comoving frame because the scattering term of equation~\eqref{eq:ps-scattering} is formulated in the comoving frame. This is in direct contrast to our finding that the appropriate frame of isotropization is the frame moving with Alfv\'en waves, which precludes an accurate formulation of CR transport to lowest (zeroth) order in the expansion variable $\varv_{\rm a} / \varv$. In fact, in order to account for the Doppler-shift between gas and propagating waves, terms of order $\varv_{\rm a} / \varv$ are necessary.

Inserting these scattering terms in our fluid equations \eqref{eq:dot_ecr} and \eqref{eq:dot_fcr}, we obtain a compact set of transport equations:
\begin{align}
	\frac{\upartial \ecr}{\upartial t} + \bs{\nabla} \bs{\cdot} \left[ \bs{u} (\ecr + \pcr) + \bs{b} \fcr \right] &= \bs{u} \bs{\cdot} \bs{\nabla} \pcr \label{eq:app_tele_dot_ecr} \\
    \frac{\upartial \fcr}{\upartial t} + \bs{\nabla} \bs{\cdot} \left( \bs{u} \fcr \right) + \frac{c^2}{3} \bs{b} \bs{\cdot} \bs{\nabla} \ecr &= - ( \bs{b} \bs{\cdot} \bs{\nabla} \bs{u}) \bs{\cdot} (\bs{b} \fcr) - (\bar{\nu}_+ + \bar{\nu}_-) \fcr \label{eq:app_tele_dot_fcr}
\end{align}
In the following, we restrict ourselves to the special case of a background gas at rest and a homogeneous magnetic field (i.e., $\bs{u}={\bf 0}$ and $\bs{b}={\rm const.}$). Differentiating equation~\eqref{eq:app_tele_dot_ecr} with respect to $t$, taking the gradient along the magnetic field direction of equation~\eqref{eq:app_tele_dot_fcr}, and combining the resulting equations, we obtain the \emph{telegraph equation}:
\begin{align}
\frac{1}{\bar{\nu}} \frac{\upartial^2 \ecr}{\upartial t^2} + \frac{\upartial \ecr}{\upartial t}  = \bs{\nabla} \bs{\cdot} \left( \kappa \bs{b} \bs{b} \bs{\cdot} \bs{\nabla} \ecr \right),\label{eq:telegrapher}
\end{align}
where again $\bar{\nu}=\bar{\nu}_+ + \bar{\nu}_-$ and $\kappa=c^2/(3\bar{\nu})$. This equation is a hyperbolic expansion of the usual (parabolic) diffusion equation, which is obtained by dropping the first term in the Chapman--Enskog expansion. 

 There has been a longstanding discussion concerning the validity of the telegraph equation to describe a diffusion process. The earliest work known to the authors which addresses this problem discusses the related process of heat transfer \citep{1958Vernotte}. The same discussion has recently resurfaced in the context of CRs.

The fundamental solution of equation~\eqref{eq:telegrapher} contains two singular wave fronts travelling at signal speed $\pm c / \sqrt{3}$ which decay at a typical rate of $\bar{\nu}/2$ \citep{2015Malkov}. The existence of these characteristics casts doubt on the validity of the telegraph equation because direct numerical solutions of the underlying Boltzmann equation do not show these wave solutions \citep{2013Litvinenko,2016Litvinenko} not even at early times. At times greater at $2/\bar{\nu}$ the solutions to both, the telegraph and diffusion equation qualitatively match those of the Boltzmann equation. Thus, at times $t  \lesssim \, 2 / \bar{\nu}$ in the ballistic regime of transport, the telegraph and the diffusion equation fail to correctly describe CR transport, while at later times, both equations reproduce the \emph{physical} diffusion of CRs. 

The telegraph equation has the mathematical appeal that it is a hyperbolic equation that contains a finite signal speed, whereas the diffusion equation \emph{apparently} has an infinitely fast signal speed. Moreover, if a physical system has an intrinsic anisotropy, the telegraph equation preserves these anisotropic properties. On the contrary, those features are smeared out in the diffusive solution \citep{2016Litvinenko, 2016Tautz}.

In conclusion, \emph{physical} diffusion can be well described by the \emph{mathematical} diffusion equation as well as the telegraph equation after a few scattering times, while the latter preserves more physical properties. However, either equation is not suited to model CR transport in the self-confinement picture.

\section{Alternative derivation of the scattering terms}
\label{sec:alternative_derivation}

The derivation of the scattering terms in Section~\ref{sec:streaming} includes all effects up to the order $\mathcal{O}(\varv_{\rm a}^2 / c^2)$ and hence includes terms that should be negligible
as $\varv_{\rm a} / c \sim 10^{-5}$ to $10^{-3}$. Here, we present an alternative derivation that is accurate to order $\mathcal{O} (\varv_{\rm a} / c)$ which yields essentially equivalent results and discuss the origin of this apparent contradiction.

We consider the CRs to be represented by a non-degenerate relativistic gas in Minkowski space with a metric tensor $\eta_{\alpha\beta}=\rmn{diag} (-1,1,1,1)$, where a CR of rest mass $m$ is characterized by the space-time coordinates $(x^\alpha) = (x^0 = ct,\bs{x})$ and by the particle momentum four-vector $(p^\alpha) = (p^0,\bs{p})$. The momentum four-vector has a constant length $p^\alpha p_\alpha=-m^2 c^2$, which implies that the total particle energy $E=c p^0$ can be expressed in terms of $\bs{p}$ by $E(p) = \sqrt{\bs{p}^2 + m^2 c^2}$. Accounting for the Lorentz invariant $\rmn{d}^3p/p^0$, the energy-momentum tensor is given in terms of the CR distribution function $f(\bs{x}, \bs{p}, t)$ by
\begin{equation}
\label{eq:Tcr}
	T^{\alpha \beta}_{\rm cr} = c \int \frac{{\rm d}^3p}{p^0} \, p^\alpha p^\beta f(\bs{x}, \bs{p}, t).
\end{equation}
Evaluating the momentum integral yields
\begin{equation}
	\left(T^{\alpha \beta}_{\rm cr}\right) = \left( \begin{matrix} \ecr & \frac{\dps1}{\dps c} \fcr \bs{b} \\ \frac{\dps1}{\dps c} \fcr \bs{b}  & \pcr \mat{1}\end{matrix} \right). \label{eq:rel_ene_mom_tensor_crs}
\end{equation}
The force density four-vector $G^{\alpha}$ is defined to be the covariant divergence of the energy-momentum tensor (implying Einstein's sum convention),
\begin{equation}
	\bs{\nabla}_{\beta}T^{\alpha \beta}_{\rm cr} \equiv G^{\alpha}.
\end{equation}
This equation holds in any coordinate system, which becomes evident through its covariant notation.

Because we solely focus on the interaction of CRs with Alfv\'en waves, this momentum and energy exchange is modelled via CR scattering in equation~\eqref{eq:ps-scattering}. We can separate the contributions of the co- and counter-propagating Alfv\'en waves to this interaction and thus to the force density $G^{\alpha}$. In their own frame these waves are purely magnetic and thus only scatter CRs in pitch angle. We already calculated the appropriate energy and flux moments of pure pitch-angle scattering in Appendix~\ref{app:diffusion}. There we purposefully neglected any electric field in the fluid frame to calculate forces exerted and work done by the magnetic fields of Alfv\'en waves. This procedure fails to capture CR streaming but allowed discussing CR propagation to lower order as it is done in the main text. In both wave frames, there are no electric fields of Alfv\'en waves present. Thus, we are able reuse the results given in Appendix~\ref{app:diffusion} without loss of generality here and only need to adjust our definition for the mean scattering frequency. Rather than using the mean scattering frequency $\bar{\nu}_\pm$ as evaluated in the fluid frame, here we use the mean scattering frequencies $\bar{\nu}\rvert_{\rm wave,\pm}$ as evaluated in the respective wave frames. With this definition, the force density four-vector is 
\begin{equation}
	\left(G_\pm^{\phantom{\pm}\alpha}\right)_{\rm wave, \pm} = \left( \begin{matrix} 0 \\  \left.-\frac{\dps\bar{\nu}_{\rm}}{\dps c^2} \fcr\right\rvert_{\rm wave,\pm} \bs{b}\end{matrix} \right).
\end{equation}
These expressions only hold in the wave frames. In order to obtain an expression for this force density in the comoving frame, we apply the Lorentz transformation
\begin{equation}
\left.G^{\alpha}\right|_{\rm cmf} = \Lambda_{+\phantom{\alpha}\beta}^{\phantom{+}\alpha} \left.G_{+}^{\phantom{+}\beta}\right|_{\rm wave,+} + \Lambda_{-\phantom{\alpha}\beta}^{\phantom{-}\alpha} \left.G_{-}^{\phantom{-}\beta}\right|_{\rm wave,-}
\end{equation}
in the $\mathcal{O}(\varv_{\rm a} / c)$ limit:
\begin{align}
	\left(\Lambda^{\phantom{\pm}\alpha}_{\pm\phantom{\alpha}\sigma}\right) &= \left( \begin{matrix} \gamma & \pm\gamma \bs{\beta} \\ \pm\gamma \bs{\beta} & \bs{\mat{1}} + (\gamma - 1) \frac{\dps\bs{\beta} \bs{\beta}}{\dps\beta^2} \end{matrix} \right)_{\rm wave\to cmf} = \left( \begin{matrix} 1\phantom{\Big|} & \pm\frac{\dps\bs{\varv}_{\rm a}}{\dps c} \\ \pm\frac{\dps\bs{\varv}_{\rm a}}{\dps c} & \bs{\mat{1}} \end{matrix} \right) + \mathcal{O}\left({\varv_{\rm a}^2/c^2}\right), \label{eq:rel_transform}
\end{align}
which yields for the four-force density in the comoving frame: 
\begin{align}
	\left(G^\alpha\right)_{\rm cmf} = \left( \begin{matrix} 
    \left.~-\frac{\dps\bar{\nu}\varv_{\rm a}}{\dps c^3} \fcr\right\rvert_{\rm wave, +} \phantom{\bs{b}}&+\left.\frac{\dps\bar{\nu}\varv_{\rm a}}{\dps c^3} \fcr\right\rvert_{\rm wave, -} \phantom{\bs{b}}\\
    \left.~-~\frac{\dps\bar{\nu}}{\dps c^2} \fcr\right\rvert_{\rm wave, +} \bs{b}&- \left.\frac{\dps\bar{\nu}}{\dps c^2}~~ \fcr\right\rvert_{\rm wave, -} \bs{b} \end{matrix} \right).
    \label{eq:appc_gas_force}
\end{align}
This result corresponds to Fermi's (\citeyear{1949Fermi}) original idea: while energy is conserved during the interaction between electromagnetic `clouds' and CRs in the the `cloud' frame, energy is transferred between the two in any other frame.

The CR energy flux evaluated in the wave frame is
\begin{equation}
	\left.\fcr\right|_{\rm wave,\pm}= \fcr \mp \varv_{\rm a} (\ecr + \pcr), \label{eq:rel_flux}
\end{equation}
which can be derived from the Lorentz-transformed momentum four-vector to  order $\mathcal{O}(\varv_{\rm a} / c)$ or simply seen as the Galilean boost of the CR flux. Inserting this into the force density in equation~\eqref{eq:appc_gas_force} results in the scattering terms
\begin{align}
\left. \frac{\partial \ecr}{\partial t} \right\rvert_{\rm scatt} &= -\frac{\varv_{\rm a}\bar{\nu}_{\rm wave,+}}{c^2} \left[ \fcr - \varv_{\rm a} (\ecr + \pcr) \right] +\frac{\varv_{\rm a}\bar{\nu}_{\rm wave,-}}{c^2} \left[ \fcr + \varv_{\rm a} (\ecr + \pcr) \right], \label{eq:rel_ecr_scatter}  \\
\left. \frac{\partial \fcr}{\partial t} \right\rvert_{\rm scatt} &= -\bar{\nu}_{\rm wave, +} \left[ \fcr - \varv_{\rm a} (\ecr + \pcr) \right] -\bar{\nu}_{\rm wave, -} \left[ \fcr + \varv_{\rm a} (\ecr + \pcr) \right].\label{eq:rel_fcr_scatter}
\end{align}
Both equations algebraically coincide with their counterparts in equations~\eqref{eq:dot_ecr_streaming} and \eqref{eq:dot_fcr_streaming}. The sole difference between both equation pairs is the definition of the scattering frequency. While our derivation in the main text uses the scattering frequency as averaged in the comoving frame, we adopt an average in the wave frames for the derivation leading to equations~\eqref{eq:rel_ecr_scatter} and \eqref{eq:rel_fcr_scatter}. 
To summarize, we presented two derivations that yield the same expression for the CR scattering terms. Both derivations differ by the adopted approach: the derivation in Section \ref{sec:streaming} is based on kinetic diffusion theory of CRs in the fluid frame, while the alternative derivation here uses a Lorentz transformation to transform the four-force provided by pich-angle scattering of CRs. In addition, the derivation in the main text was accurate to order $\mathcal{O}(\varv_{\rm a}^2 / c^2)$, whereas here, we only needed an accuracy of order $\mathcal{O}(\varv_{\rm a} / c)$. 

Why do the two derivations that are accurate to different orders in $\varv_{\rm a}/c$ yield the same expression? To answer this question, we first recall the different orders of fundamental quantities appearing in our theory. The thermodynamic quantities $\ecr$ and $\fcr$ share the same $\mathcal{O}(1)$ in $\varv_{\rm a}/c$ since $\fcr \sim \varv_{\rm a} (\ecr + \pcr)$ in a self-confinement scenario. However, the CR energy flux density enters the energy-momentum tensor in equation~\eqref{eq:rel_ene_mom_tensor_crs} in the relativistically appropriate form as $\fcr / c$, which is of order $\mathcal{O}(\varv_{\rm a} / c)$. Transforming this quantity via a semi-relativistic Lorentz-transformation of order $\mathcal{O}(\varv_{\rm a} / c)$ and retaining all terms may then introduce terms of order $\mathcal{O}(\varv_{\rm a}^2/ c^2)$ into the transformed equations. This is exactly the case for the Lorentz-transformed four-force above: the time-like component of equation~\eqref{eq:appc_gas_force} is a transformed space-like component of the Lorentz-force, which contains a factor $\fcr / c$, and is of order $\mathcal{O}(\varv_{\rm a}^2/ c^2)$. 

At first sight, the terms of order $\mathcal{O}(\varv_{\rm a}^2/ c^2)$ could be regarded vanishingly small in a formally rigorous treatment, which aims to be correct up to order $\mathcal{O}(\varv_{\rm a}^2/ c^2)$. On the other side, there is a practical argument that underlines the importance of the $\mathcal{O}(\varv_{\rm a}^2 / c^2)$ terms. All terms in equation~\eqref{eq:rel_ecr_scatter} are of this order. Neglecting them would correspond to a vanishing energy transfer between CRs and Alfv\'en waves. In consequence, there would be no amplified Alfv\'en waves and we would only account for damping processes, which left the CR essentially unscattered. Hence, any theory that builds on the idea of CR self-confinement has to include scattering processes up to order $\mathcal{O}(\varv_{\rm a}^2 / c^2)$ to be consistent.

\section{Comoving Vlasov equation}
\label{app:Vlasov}

Here, we present two different derivations of the Vlasov equation~\eqref{eq:vlasov}. The first and formally correct derivation yields Vlasov's equation for relativistic particles in the comoving frame. To capture all relevant pseudo forces, we need to evaluate the equations of motion in the semi-relativistic limit up to order $\mathcal{O}(u^2 / c^2)$. Surprisingly, the resulting Vlasov equation can also be derived in Newtonian mechanics without any relativistic corrections.

\subsection{Semi-relativistic derivation}

We first recall definitions of special relativity and the geometry of Minikoski space to find approximate expressions in the semi-relativistic limit. The Lorentz transformation from the comoving frame into the lab frame is given by
\begin{align}
\left(\Lambda^\alpha_{\phantom{\alpha}\hat{\alpha}}\right) &= \left( \begin{matrix} \gamma & \gamma \bs{\beta} \\ \gamma \bs{\beta} & \bs{\mat{1}} + (\gamma - 1) \frac{\dps\bs{\beta} \bs{\beta}}{\dps\beta^2} \end{matrix} \right)_{\rm cmf\to lab} = \left( \begin{matrix} 1\phantom{\Big|} & \frac{\dps\bs{u}}{\dps c} \\ \frac{\dps\bs{u}}{\dps c} & \bs{\mat{1}} \end{matrix} \right) + \mathcal{O}\left(u^2/c^2\right),
\end{align}
where $\bs{\beta}=\bs{u}/c$ and we approximated the transformation in the semi-relativistic limit. The motivation for choosing this approximation are our equations of CR hydrodynamics, which derive from the focused transport equation of CRs -- a semi-relativistically transformed variant of the comoving Vlasov equation. The metric connection (Christoffel symbols) associated with this Lorentz transformation can be calculated in an otherwise flat space-time by \citep{BookMTW}
\begin{align}
\Gamma^{\hat{\alpha}}_{\hat{\nu} \hat{\mu}} =  \Lambda^{\hat{\alpha}}_{\phantom{\hat{\alpha}} \alpha} \Lambda^\mu_{\phantom{\mu} \hat{\mu}} \frac{\upartial \Lambda^{\alpha}_{\phantom{\alpha} \hat{\nu}}}{\upartial x^\mu},
\end{align}
where we use Greek letters with hats to denote quantities evaluated in the comoving frame, i.e., $\Lambda^{\mu}_{\phantom{\mu}\hat{\mu}}$ transforms a contravariant vector from the comoving into the lab frame. The connection symbol is not symmetric in its lower indices owing to the torsion introduced by a non-vanishing curl of the flow. The only non-vanishing components of $\Gamma^{\hat{\alpha}}_{\hat{\nu} \hat{\mu}}$ in the semi-relativistic limit are given by
\begin{align}
\Gamma^{\hat{0}}_{\hat{\imath} \hat{\jmath}} &= \Gamma^{\hat{\imath}}_{\hat{0} \hat{\jmath}} = \frac{\nabla_{\hat{\jmath}} u_{\hat{\imath}}}{c}, \label{eq:conn_first_order} \\
\Gamma^{\hat{\imath}}_{\hat{0} \hat{0}} &= \Gamma^{\hat{0}}_{\hat{\imath} \hat{0}} = \frac{1}{c^2} \frac{\rmn{d} u_{\hat{\imath}}}{\rmn{d} t}, \label{eq:conn_second_order}
\end{align}
where Roman indices denote space-like components, $\hat{\imath}, \hat{\jmath}\in\{1,2,3\}$.  The second set of Christoffel symbols, which are of order $\mathcal{O}(u^2/c^2)$, is necessary to restore the Newtonian limit even though the symbols are formally small. This will become apparent at the end of Section~\ref{app:newtonian_argument}. We first consider a particle at position $x^{\hat{\alpha}}$ and velocity $\varv^{\hat{\alpha}}$. Its equations of motion generalise in Minkowski space to
\begin{align}
	\frac{\rmn{d} x^{\hat{\alpha}}}{\rmn{d} \tau} &= \varv^{\hat{\alpha}}, \label{eq:eom_position} \\ 
	\frac{\rmn{d} p^{\hat{\alpha}}}{\rmn{d} \tau} + \Gamma^{\hat{\alpha}}_{\hat{\sigma} \hat{\nu}} \varv^{\hat{\sigma}} p^{\hat{\nu}} &= F^{\hat{\alpha}}, \label{eq:eom_momentum}
\end{align}
where the connection symbols account for local changes of the defining velocity of the comoving frame that occur during the motion of the particle, $\tau$ is the proper time and $(F^{\hat{\alpha}}) = \gamma (\bs{\varv} \bs{\cdot} \bs{F} / c, \bs{F})$ is the Minkowski four-force representing the generalised electromagnetic force. The momentum equation can be simplified using the approximations in equation~\eqref{eq:conn_first_order} and \eqref{eq:conn_second_order} to yield
\begin{align}
    \frac{\rmn{d} p^{\hat{\imath}}}{\rmn{d} \tau} + m \frac{\rmn{d} u^{\hat{\imath}}}{\rmn{d} t} + \bs{p} \bs{\cdot} \bs{\nabla} u^{\hat{\imath}}=F^{\hat{\imath}}. \label{eq:semirel_pseudo_force}
\end{align}
In absence of any particle creation or annihilation processes, the number of CRs is a conserved quantity along any path that is described by the equations of motion. Thus, we have
\begin{align}
	0 &= \frac{\rmn{d} f}{\rmn{d} \tau} = \frac{\rmn{d} x^{\hat{\alpha}}}{\rmn{d} \tau} \frac{\upartial f}{\upartial x^{\hat{\alpha}}} + \frac{\rmn{d} p^{\hat{\alpha}}}{\rmn{d} \tau} \frac{\upartial f}{\upartial p^{\hat{\alpha}}}, \label{eq:relativistic_boltzmann}
\end{align}
which describes the full phase-space evolution of CRs to any order.
The covariant derivative in the semi-relativistic limit is given by
\begin{align}
\left(\upartial_{\hat{\beta}} \right) = \left( \Lambda^{\beta}_{\phantom{\beta}\hat{\beta}} \, \upartial_{\beta} \right) = \left( \frac{\upartial}{c \upartial t} + \frac{\bs{u}}{c} \bs{\cdot} \bs{\nabla}, \bs{\nabla} \right) + \mathcal{O}\left(u^2/c^2\right). \label{eq:comov_covar_deri}
\end{align}
Here, the derivative in the comoving frame is expressed by quantities that are measured in the lab frame, $\bs{u}$ and $t$. This is in preparation of the conventional mixed coordinate system for the  transport equation of CR energy density: while the ambient gas velocity $\bs{u}$ and the direction of the large scale magnetic field $\bs{b} = \bs{B} / B$ are measured in the lab frame, the CR energy and momentum densities, $\ecr$ and $\fcr/c$ as well as the generalised force densities are given with respect to the comoving frame.

The on-shell condition $p^{\hat{\alpha}} p_{\hat{\alpha}} = -m^2 c^2$ is a constraint equation that reduces the set of four independent momentum variables by one. We choose the space-like components $p^{\hat{\imath}}$ as the independent variables so that we have $f\equiv f(x^{\hat{\alpha}},p^{\hat{\imath}})$. Adopting this definition, inserting the covariant derivative of equation~\eqref{eq:comov_covar_deri} and the equations of motion ~\eqref{eq:eom_position} and \eqref{eq:semirel_pseudo_force} into equation~\eqref{eq:relativistic_boltzmann} yields
\begin{align}
	0 &= \frac{\rmn{d} x^{\hat{\alpha}}}{\rmn{d} \tau} \frac{\upartial f}{\upartial x^{\hat{\alpha}}} + \frac{\rmn{d} p^{\hat{\imath}}}{\rmn{d} \tau} \frac{\upartial f}{\upartial p^{\hat{\imath}}} 
   = \varv^{\hat{\alpha}} \frac{\upartial f}{\upartial x^{\hat{\alpha}}} + \left(F^{\hat{\imath}} - \Gamma^{\hat{\imath}}_{\hat{\sigma} \hat{\nu}} \varv^{\hat{\sigma}}   p^{\hat{\nu}} \right) \frac{\upartial f}{\upartial p^{\hat{\imath}}}\\
    &= \frac{\upartial f}{\upartial t} + (\bs{u} + \bs{\varv}) \bs{\cdot}\frac{\upartial f}{\upartial \bs{x}} + \left(\bs{F} - m \frac{\rmn{d} \bs{u}}{\rmn{d} t} - (\bs{p} \bs{\cdot} \bs{\nabla}) \bs{u}\right) \bs{\cdot}\frac{\upartial f}{\upartial \bs{p}},
\end{align}
which coincides with the Vlasov equation~\eqref{eq:vlasov}. 

\subsection{Newtonian derivation}
\label{app:newtonian_argument}

Here, we derive the same result with similar arguments but solely within the framework of Newtonian mechanics. Consider a non-relativistic particle with velocity $\bs{\varv}_\rmn{lab}$ that is measured in the lab frame. Its velocity in the comoving frame $\bs{\varv}$ is defined by
\begin{align}
 \bs{\varv}_\rmn{lab} = \bs{\varv} + \bs{u}(\bs{x}(t), t),
\end{align}
where the background velocity $\bs{u}$ is evaluated at the position of the particle itself. Expressing Newton's equations of motion in comoving quantities yields
\begin{align}
m \frac{\rmn{d}\bs{\varv}_\rmn{lab}}{\rmn{d} t} &= m \frac{\rmn{d}\bs{\varv}}{\rmn{d} t} + m \frac{\rmn{d}}{\rmn{d} t} 
\bs{u}(\bs{x}(t),t) = m \frac{\rmn{d}\bs{\varv}}{\rmn{d} t} + m \left(\frac{\rmn{d} \bs{x}}{\rmn{d} t} \bs{\cdot} \bs{\nabla} \right)\bs{u} + m \frac{\upartial \bs{u}}{\upartial t} = \bs{F},
\end{align}
where $\bs{F}$ is the electromagnetic force. This force is the same in all frames, as long as the Newtonian limit holds. We rearrange this equation to obtain
\begin{align}
\frac{\rmn{d} \bs{p}}{\rmn{d} t} = \bs{F} - m \frac{\rmn{d} \bs{u}}{\rmn{d} t} - (\bs{p} \bs{\cdot} \bs{\nabla}) \bs{u},
\end{align}
which coincides with the semi-relativistic equation~\eqref{eq:semirel_pseudo_force}. Note our different usages of $\rmn{d}/\rmn{d}t$: while the left-hand side is the time derivative of the particle momentum, the time derivative of $\bs{u}$ on the right-hand side of this equation is the convective derivative of the mean gas velocity. 
Although this derivation is valid only for a non-relatvistic gas, it allows us to interpret the result and further highlights the following two points: (i) the comoving velocity can either change due to spatial inhomogeneities or due to acceleration of the background velocity $\bs{u}$. This insight is the basis of our discussion in Section~\ref{sec:focused_CRs}. (ii) Obviously, the pseudo forces arise from the boost into a non-inertial frame. One can show that the total kinetic energy and momentum are conserved based on the equations of comoving momentum and kinetic energy evolution if and only if all terms in the transformation from the comoving frame into the lab frame are retained. 

To derive the Vlasov equation, we had to include factors of order $\mathcal{O}(u^2/c^2)$ in the semi-relativistic derivation while our 
Newtonian derivation is only accurate to order $\mathcal{O}(1)$ in $u/c$. This apparent contradiction is alleviated by inspection of equation~\eqref{eq:eom_momentum} in connection with the Christoffel symbols of equation~\eqref{eq:conn_second_order}. Multiplying the Christoffel symbols by $\varv^{\hat{0}}$ and $p^{\hat{0}}$ yields another factor of $c^2$ that increases the accuracy in the final Vlasov equation from $\mathcal{O}(u^2/c^2)$ to $\mathcal{O}(1)$.

\section{Semi-Relativistic derivation of the CR hydrodynamics equations}
\label{app:derivation}

In this Appendix, we present an alternative derivation of our hydrodynamical equations for CRs based on the conservation of the energy-momentum tensor in special relativity. Those equations include the contribution by Lorentz forces associated with large-scale fields and Alfv\'en waves and read in their covariant form:
\begin{align}
	\nabla_{\beta} T^{\alpha \beta}_{\rm cr} &= j_{\rm cr}^\beta F_{\phantom{\beta}\beta}^\alpha + \left\langle \delta j_{\rm cr}^\beta \left(\delta F_{+\,\beta}^\alpha + \delta F_{-\,\beta}^\alpha\right) \right\rangle, \label{eq:lab_ene_mom}
\end{align}
where the CR energy-momentum tensor $T^{\alpha \beta}_{\rm cr}$ is given by equation~\eqref{eq:rel_ene_mom_tensor_crs}, $F^{\alpha \beta}$ and $\delta F_\pm^{\alpha \beta}$ are the contravariant components of the electromagnetic field tensors for the large-scale and small-scale fields, and $j_{\rm cr}^{\alpha}$ and $\delta j_{\rm cr}^{\alpha}$ are the CR four-currents induced by large-scale fields and Alfv\'en waves, respectively. Note that $F^{\alpha \beta}$ is linear in $\bs{E}$ and $\bs{B}$ such that a Reynolds decomposition in the mean and fluctuating components is straightforward. We identify the small-scale Lorentz forces by extending the result of our discussion in Section~\ref{sec:coupling} to the relativistic case according to:
\begin{align}    
    G_\pm^\alpha&= \left\langle \delta j_{\rm cr}^\beta\, \delta F_{\pm\,\beta}^\alpha \right\rangle,
\end{align}
and write the large-scale Lorentz force as
\begin{align}
G^{\alpha}_{\rm Lorentz} =  j_{\rm cr}^\beta F_{\phantom{\beta}\beta}^\alpha.
\end{align}

We derive the two-moment CR equations in Section~\ref{sec:focused_CRs} for $\ecr$ and $\fcr$ as measured by an observer in the comoving frame. The equations of momentum and energy conservation are transformed as
\begin{align}
	\nabla_{\hat{\beta}} T^{\hat{\alpha} \hat{\beta}}_{\rm cr} = \upartial_{\hat{\beta}} T^{\hat{\alpha} \hat{\beta}}_{\rm cr} + \Gamma^{\hat{\alpha}}_{\hat{\sigma} \hat{\beta}} T^{\hat{\sigma} \hat{\beta}}_{\rm cr} + \Gamma^{\hat{\beta}}_{\hat{\sigma} \hat{\beta}} T^{\hat{\alpha} \hat{\sigma}}_{\rm cr}.\label{eq:transformed_ene_mom}
\end{align}

During the following evaluation of equation~\eqref{eq:transformed_ene_mom}, we assume that the CR mean momentum density is described by the general vector $\bs{\fcr} / c^2$. We will insert  $\bs{\fcr} = \fcr \bs{b}$ after the calculation of the space-like and time-like components of this equation. With the connection symbols from equations \eqref{eq:conn_first_order}-\eqref{eq:conn_second_order} and the derivative from equation~\eqref{eq:comov_covar_deri} in place, the time-like component of equation~\eqref{eq:transformed_ene_mom} reads
\begin{align}
	\nabla_{\hat{\beta}} T^{\hat{0}\, \hat{\beta}}_{\rm cr} &= \upartial_{\hat{\beta}} T^{\hat{0}\, \hat{\beta}}_{\rm cr} + \Gamma^{\hat{0}}_{\hat{\imath}\,\hat{0}} T^{\hat{\imath}\, \hat{0}}_{\rm cr} + \Gamma^{\hat{0}}_{\hat{\imath} \hat{\jmath}} T^{\hat{\imath} \hat{\jmath}}_{\rm cr} + \Gamma^{\hat{0}}_{\hat{\imath} \hat{0}} T^{\hat{0}\, \hat{\imath}}_{\rm cr} + + \Gamma^{\hat{\jmath}}_{\hat{0} \hat{\jmath}} T^{\hat{0} \hat{0}}_{\rm cr} \label{eq:rel_ecr_dot1}\\
    &= \frac{1}{c} \frac{\upartial \ecr}{\upartial t} + \frac{\bs{u}}{c} \bs{\cdot} \bs{\nabla} \ecr + \bs{\nabla} \bs{\cdot} \frac{\bs{\fcr}}{c} + \frac{\bs{\nabla} \bs{u}}{c} \bs{:} (\pcr \mat{1}) + \frac{\bs{\nabla} \bs{\cdot} \bs{u}}{c} \ecr + \frac{2}{c^2} \frac{\rmn{d}\bs{u}}{\rmn{d}t} \bs{\cdot} \frac{\bs{\fcr}}{c} \label{eq:rel_ecr_dot2} \\  
    &= \frac{1}{c} \left\lbrace \frac{\upartial \ecr}{\upartial t} + \bs{\nabla} \bs{\cdot} [\bs{u} (\ecr + \pcr) + \fcr \bs{b}] - \bs{u} \bs{\cdot} \bs{\nabla} \pcr \right\rbrace = G^{\hat{0}}_{+}+G^{\hat{0}}_{-}, \label{eq:rel_ecr_dot3}
\end{align}
which coincides with the combined equations~\eqref{eq:dot_ecr} and \eqref{eq:dot_ecr_streaming} after inserting the scattering terms from equation~\eqref{eq:appc_gas_force}. There is no energy exchange due to the large-scale Lorentz force because the corresponding electric field vanishes identically due to infinite-conductivity assumption of ideal MHD. In the final step in equation~\eqref{eq:rel_ecr_dot3}, we neglect the work done by the pseudo force containing the factor $\rmn{d}\rmn{\bs{u}}/\rmn{d} t$ because it is of order $\mathcal{O}(u^2/c^2)$.

Equivalently, we insert the connection symbols into the space-like components of equation~\eqref{eq:transformed_ene_mom} and obtain:
\begin{align}
	\nabla_{\hat{\beta}} T^{\hat{\imath}\, \hat{\beta}}_{\rm cr} &= \upartial_{\hat{\beta}} T^{\hat{\imath}\, \hat{\beta}}_{\rm cr} + \Gamma^{\hat{\imath}}_{\hat{0}\,\hat{0}} T^{\hat{0}\,\hat{0}}_{\rm cr} + \Gamma^{\hat{\imath}}_{\hat{0}\,\hat{\jmath}} T^{\hat{0}\,\hat{\jmath}}_{\rm cr} + \Gamma^{\hat{0}}_{\hat{\jmath}\, \hat{0}} T^{\hat{\imath}\, \hat{\jmath}}_{\rm cr}  + \Gamma^{\hat{\jmath}}_{\hat{0}\, \hat{\jmath}} T^{\hat{\imath}\, \hat{0}}_{\rm cr} \label{eq:rel_fcr_dot1}\\
    &= \left[\frac{\upartial \bs{\fcr} / c^2}{\upartial t} + \bs{\nabla\cdot} \left( \bs{u} \bs{\fcr} / c^2 + \pcr \mat{1} \right) + \left(\bs{\fcr} / c^2 \bs{\cdot} \bs{\nabla}\right) \bs{u} + \frac{1}{c^2} \frac{\rmn{d}\bs{u}}{\rmn{d}t} (\ecr + \pcr)\right]^{\hat{\imath}} \label{eq:rel_fcr_dot2} \\
    &= \left[\frac{\upartial \bs{\fcr} / c^2}{\upartial t} + \bs{\nabla\cdot} \left( \bs{u} \bs{\fcr} / c^2 + \pcr \mat{1} \right) + \left(\bs{\fcr} / c^2 \bs{\cdot} \bs{\nabla}\right) \bs{u}\right]^{\hat{\imath}} = G^{\hat{\imath}}_{\rm Lorentz}+G^{\hat{\imath}}_{+}+G^{\hat{\imath}}_{-}. \label{eq:rel_fcr_dot3}
\end{align}
Again, we neglect the vanishingly small contribution from the pseudo force containing the factor $\rmn{d}\bs{u}/\rmn{d} t$ in our semi-relativistic limit. If we included that pseudo force, the CR momentum and energy equations would coincide with those obtained by \citet{1979Buchler} for the respective radiation quantities except for the different scattering processes of CRs and radiation, respectively, cf.\ equations (29) and (30) of \citet{1979Buchler}.

After taking the dot product of equation~\eqref{eq:rel_fcr_dot3} with $\bs{b}$, using ${\rm d} (\bs{b} \bs{\cdot} \bs{b}) = 0$ where ${\rm d}$ is some differential and realizing that the Lorentz-force only acts perpendicular to the magnetic field, we arrive at the combined equations~\eqref{eq:dot_fcr} and \eqref{eq:dot_fcr_streaming}. Again the scattering terms $G^{\hat{\imath}}_{\pm}$ are given by equation~\eqref{eq:appc_gas_force}. 

Thus, we can rederive our hydrodynamical CR equations if we treat CRs as a relativistic fluid and approximate its evolution equations in the semi-relativistic limit.  An interesting aspect of this alternative derivation is the clear separation between pseudo forces and the acting pressure and Lorentz forces. For example, the term $\bs{u}\bs{\cdot}\bs{\nabla}\pcr$ in equation~\eqref{eq:rel_ecr_dot3} is commonly attributed to $P\rmn{d}V$ work done on the thermal gas, which is associated with a pressure force $\bs{\nabla} \pcr$ acting in the Newtonian limit (see Section~\ref{sec:coupling}). This is clearly not the case in the semi-relativistic limit for two reasons: (i) the formal origin of the $\bs{u}\bs{\cdot}\bs{\nabla}\pcr$ term is the fictitious energy source term $\Gamma^{\hat{0}}_{\hat{\imath} \hat{\jmath}} T^{\hat{\imath} \hat{\jmath}}_{\rm cr}$ in equation~\eqref{eq:rel_ecr_dot1}. Thus, this term is solely introduced by the transformation into the non-inertial comoving frame and not due to mechanical work associated with a force. (ii) equation~\eqref{eq:rel_fcr_dot3} shows that the kinematic CR pressure acts on the CRs themselves and not on the thermal gas. Only if CRs carry no mean momentum ($\bs{\fcr} / c^2=0$), this pressure acts formally on the MHD background because in this case $\bs{\nabla} \pcr = \bs{g}_{\rm Lorentz} + \bs{g}_{\rm gri,+} + \bs{g}_{\rm gri,-}$ according to equation~\eqref{eq:rel_fcr_dot3}. Only in this case the $P\rmn{d}V$ work done by CRs equals $\bs{u} \bs{\cdot} \bs{\nabla} \pcr$. Furthermore, we can understand why the term $\bs{u}\bs{\cdot}\bs{\nabla}\pcr$ in equation~\eqref{eq:rel_ecr_dot3} does not describe mechanical work by examining the situation as seen by a observer comoving with the thermal gas: in the comoving frame we have $\bs{u} = \bs{0}$ by definition, which implies vanishing kinetic energy $\rho \bs{u}^2/2$. Consequently, there is (i) no kinectic energy that could be changed by the Lorentz force and (ii) for any force $\bs{g}$ there is no work  done in this frame, $\bs{u} \bs{\cdot} \bs{g} = 0$.

\section{Lab-frame equations and energy and momentum conservation}
\label{app:lab}

We use the results of the previous appendices to derive the evolution equations for CR energy and momentum density in the lab frame, expressed in their comoving quantities. This enables us to discuss energy and momentum conservation in the semi-relativistic limit and to point out problems arising in this formulation. To this end, we assume that CRs are not gyrotropic and further define the vectorial form of the CR energy flux as
\begin{align}
\bs{\fcr} = \int {\rm d}^3 p \, E(p) \, \bs{\varv} f(p,\mu),
\end{align}
which coincides with our definition in equation~\eqref{def:fcr} if we project this vector onto the direction of the mean magnetic field, i.e., if we require $\bs{\fcr} \parallel \bs{b}$. We thus allow for CR mean motions that are oblique to the mean magnetic field as seen from an observer in the comoving frame. We postpone the discussion of the necessity of this more general definition to a later time, after the derivation of the evolution equations.

First, we transform the thermodynamical quantities $\ecr$, $\fcr$ and $\pcr$ into the lab frame. This can be accomplished by applying a Lorentz transformation to the rank-2 tensor in equation~\eqref{eq:Tcr}. The results are straightforwardly calculated and can be found in \citet{BookMihalas} in their equations (91.10) to (91.12). For completeness, we list the fully relativistic result and a semi-relativistic approximation. For the energy density, we find
\begin{align}
	\left.\ecr\right\rvert_{\rm lab} &= \gamma^2 \ecr + 2 \gamma^2 \frac{\bs{u} \bs{\cdot} \bs{\fcr}}{c^2} + \gamma^2 \pcr \frac{\bs{u}^2}{c^2} = \ecr + 2 \frac{\bs{u} \bs{\cdot} \bs{\fcr}}{c^2} + \mathcal{O}(u^2 / c^2),
    \label{eq:ecr_lab}
\end{align}
which contains a non-trivial correction even in the semi-relativistic limit. Transforming the comoving CR momentum density (times $c^2$) into the lab frame yields
\begin{align}
	\left.\bs{\fcr}\right\rvert_{\rm lab} &= \gamma \bs{\fcr} + \gamma^2 \ecr \bs{u} + \gamma \pcr \bs{u} + \gamma (\gamma - 1) \pcr \bs{u} + \left(\gamma^2 \frac{\bs{u} \bs{u}}{c^2} + \gamma (\gamma - 1) \frac{\bs{u} \bs{u}}{\bs{u}^2} \right) \bs{\cdot} \bs{\fcr} \\
    &= \bs{\fcr} + \bs{u} (\ecr + \pcr) + \mathcal{O}(u^2 / c^2).
    \label{eq:fcr_lab}
\end{align}
Similarly, the CR pressure transforms as
\begin{align}
	\left.\mat{P}_{\rm cr}\right\rvert_{\rm lab} &= \pcr \mat{1} + \gamma \frac{\bs{\fcr} \bs{u} + \bs{u} \bs{\fcr}}{c^2} +\gamma^2 \ecr \frac{\bs{u}\bs{u}}{c^2} + 2 \frac{\gamma (\gamma - 1) \bs{u} \bs{\cdot} \bs{\fcr}}{c^2} \frac{\bs{u}\bs{u}}{\bs{u}^2} + (\gamma - 1)^2 \frac{\bs{u}\bs{u}}{\bs{u}^2} \pcr + 2(\gamma - 1)  \pcr \frac{\bs{u}\bs{u}}{\bs{u}^2}, \\
    &= \pcr \mat{1} + \frac{\bs{\fcr} \bs{u} + \bs{u} \bs{\fcr}}{c^2} + \mathcal{O}(u^2 / c^2),
    \label{eq:Pcr_lab}
\end{align}
which is now a (non-trivial) symmetric rank-two tensor.

Because the (generalised) CR mean momentum can have components that are perpendicular to the mean magnetic field, the equilibrium argument from Section~\ref{sec:coupling} does not hold any longer and we have to take the associated Lorentz force into account. With our definition of the CR mean momentum in terms of the CR energy flux density, we multiply the Lorentz force term in the Vlasov equation~\eqref{eq:vlasov} by $E(p) \bs{\varv}$ and integrate the result over momentum space to obtain
\begin{align}
c^2\bs{g}_{\rm Lorentz} &= \int {\rm d}^3p  E(p) \bs{\varv} Ze\,\frac{\bs{\varv} \bs{\times} \bs{B}}{c} \bs{\cdot} \frac{\upartial f}{\upartial \bs{p}} = -\bs{\Omega}' \bs{\times} \bs{\fcr},
\end{align}
where we have used the ultra-relativistic limit, $\bs{\Omega}' = \Omega' \bs{b}$ and $\Omega'$ is defined in Section~\ref{sec:streaming}. This expression was previously derived by \citet{1974Forman}, see their equation~(A10). Note that the gyromotion happens on kinetic time-scales and is thus fast compared to any hydrodynamical time-scale. While this expression is valid for any non-equilibrium motion perpendicular to the mean magnetic field, the CRs can be considered to be in equilibrium after averaging over a longer time that extends over multiple gyro-orbits. This argument was used in our discussion in Section~\ref{sec:coupling} that eventually resulted in equation~\eqref{eq:final_cr_lorentz_force}, justifying the approximation used.

Accounting for perpendicular CR mean momenta also complicates their interaction with Alfv\'en waves, because the small-scale Lorentz forces do not need to be aligned with the magnetic field as in equation~\eqref{eq:rel_fcr_scatter} for a gyrotropic CR distribution. The treatment by \citet{1989Schlickeiser} of this interaction, which is the basis for our theory in Section~\ref{sec:scattering}, formally only holds for a gyrotropic CR distribution and thus only for the case of $\bs{\fcr} \parallel \bs{b}$. None the less, following the preceding discussion, gyrotropy can be assumed on hydrodynamical time-scales so that we can extent the validity of equation~\eqref{eq:rel_fcr_scatter} to nearly gyrotropic CR distributions with non-vanishing perpendicular CR mean momenta. We proceed with the Lorentz forces due to small-scale magnetic field fluctuations,
\begin{align} 
c^2 \bs{g}_{\rmn{gri}, \pm} &= \bar{\nu}_{\rm wave, \pm} \bs{b} \bs{b} \bs{\cdot} \left[ \bs{\fcr} \mp \bs{\varv}_{\rm a} (\ecr + \pcr) \right],
\end{align}
which are equivalent to equation~\eqref{eq:rel_fcr_scatter} for the assumption of a gyrotropic CR distribution with $\bs{\fcr} = \fcr \bs{b}$.

With those equations in place, we can finally turn to the evolution equations for lab-frame quantities. Inserting the semi-relativistic limits of $\ecr$, $\fcr$, and $\pcr$ of equations~\eqref{eq:ecr_lab}, \eqref{eq:fcr_lab}, and \eqref{eq:Pcr_lab} into the conservation equation~\eqref{eq:lab_ene_mom} and evaluating the time-like and space-like components of that equation results in: 
\begin{align}
\frac{\upartial}{\upartial t} \left( \ecr + 2 \frac{\bs{u} \bs{\cdot} \bs{\fcr}}{c^2} \right) + \bs{\nabla} \bs{\cdot} \left[ \bs{\fcr} + \bs{u} (\ecr + \pcr) \right] &= -\left[\bs{u} \bs{\cdot} \bs{g}_{\rm Lorentz} + (\bs{u} + \bs{\varv}_{\rm a}) \bs{\cdot} \bs{g}_{\rmn{gri}, +} + (\bs{u} - \bs{\varv}_{\rm a}) \bs{\cdot} \bs{g}_{\rmn{gri}, -}\right],
\label{eq:ecr_cons_lab}\\
\frac{1}{c^2}\frac{\upartial}{\upartial t} \left[ \bs{\fcr} + \bs{u} (\ecr + \pcr) \right] + \bs{\nabla} \bs{\cdot} \left( \pcr \mat{1} +\frac{\bs{\fcr} \bs{u} + \bs{u} \bs{\fcr}}{c^2} \right) &= -\left(\bs{g}_{\rm Lorentz} + \bs{g}_{\rmn{gri}, +} + \bs{g}_{\rmn{gri}, -}\right).
\label{eq:fcr_cons_lab}
\end{align}
This result directly enables us to demonstrate energy and momentum conservation of our equations in the semi-relativistic limit. Adding equations~\eqref{eq:final_gas_energy}, \eqref{eq:eaw} and \eqref{eq:ecr_cons_lab} yields energy conservation and adding equations~\eqref{eq:final_gas_euler} and \eqref{eq:fcr_cons_lab} results in momentum conservation in the lab frame, respectively:
\begin{align}
\frac{\upartial}{\upartial t} \left( \varepsilon_{\rmn{tot}} + 2 \frac{\bs{u} \bs{\cdot} \bs{\fcr}}{c^2} \right) + \bs{\nabla} \bs{\cdot} \left[ \bs{u} (\varepsilon_{\rm tot} + P_{\rm tot}) + \bs{\fcr} + \bs{\varv}_{\rm a} (\varepsilon_{\rm a, +} - \varepsilon_{\rm a, -}) - \bs{B}\, (\bs{u} \bs{\cdot}\bs{B}) \right] &= 0, \label{eq:energy_cons_lab}\\
\frac{1}{c^2}\frac{\upartial}{\upartial t} \left[ \bs{\fcr} + \bs{u} (\rho c^2 + \ecr + \pcr) \right] + \bs{\nabla} \bs{\cdot} \left( P_{\rm tot} \mat{1} + \rho\bs{u} \bs{u}  - \bs{B} \bs{B} + \frac{\bs{\fcr} \bs{u} + \bs{u} \bs{\fcr}}{c^2}\right) &= \bs{0}, \label{eq:momentum_cons_lab}
\end{align}
where $\varepsilon_{\rm tot}$ and $P_{\rm tot}$ are given by equations~\eqref{eq:etot} and \eqref{eq:ptot}, respectively. 

To simplify the discussion on the applicability of the lab-frame equations, we neglect the relativistic corrections to the CR energy and pressure on the left-hand sides of equations~\eqref{eq:ecr_lab} and \eqref{eq:fcr_lab} and obtain:
\begin{align}
\frac{\upartial \ecr}{\upartial t} + \bs{\nabla} \bs{\cdot} \left[ \bs{\fcr} + \bs{u} (\ecr + \pcr) \right] &= -\left[\bs{u} \bs{\cdot} \bs{g}_{\rm Lorentz} + (\bs{u} + \bs{\varv}_{\rm a}) \bs{\cdot} \bs{g}_{\rmn{gri}, +} + (\bs{u} - \bs{\varv}_{\rm a}) \bs{\cdot} \bs{g}_{\rmn{gri}, -}\right], \\
\frac{1}{c^2}\frac{\upartial}{\upartial t} \left[ \bs{\fcr} + \bs{u} (\ecr + \pcr) \right] + \bs{\nabla} \bs{\cdot} \left( \pcr \mat{1} \right) &= -\left(\bs{g}_{\rm Lorentz} + \bs{g}_{\rmn{gri}, +} + \bs{g}_{\rmn{gri}, -}\right). \label{eq:lab_mome_eq}
\end{align}
As outlined above, the Lorentz force in this equation introduces a kinetic time-scale which is challenging to resolve in numerical simulations of macroscopic systems. To remedy the situation for numerical simulations we could use the near-equilibrium assumption and remove the kinetic time-scale by averaging over it. We adopted this approach in the comoving frame by using the gyroaveraged Fokker-Planck equation in Section~\ref{sec:focused_CRs} and consequently neglected any perpendicular inertia of CRs in Section~\ref{sec:coupling} to infer the time-averaged large-scale Lorentz force. This procedure proves to be difficult in the lab frame: here, the momentum equation carries all the information about the energy transport perpendicular to the mean magnetic field. 

This is in contrast to its comoving counterpart, where the momentum equation only describes the motion relative to the gas frame,  which solves a few conceptional problems of the lab frame. This additional complication precludes a clear separation of motions perpendicular and parallel to the mean magnetic field. A Chapman--Enskog expansion, which was used in Section~\ref{sec:perpendicular_forces}, yields the same result for the Lorentz force as in the comoving frame, namely equation~\eqref{eq:final_large_lorentz_force}. But this expansion neglects any perpendicular contributions in the time-derivative terms in the momentum equation~\eqref{eq:lab_mome_eq}. This equation states that CRs are not transported perpendicular to the mean magnetic field with the gas in the lab frame or $[\bs{\fcr} + \bs{u} (\ecr + \pcr)]_{\perp} = \bs{0} $. This contradicts the idea of gyroaveraged evolution where CRs are rapidly gyrating along field lines and are thus transported perpendicular to $\bs{B}$ with the gas by definition. Hence, we cannot use the Chapman--Enskog expansion in the lab frame the same way as we did in the comoving frame.

Attempting to circumvent this expansion by directly assuming $\bs{\fcr} = \fcr \bs{b}$ leads to difficulties, too. Using this assumption in equation~\eqref{eq:lab_mome_eq} leaves terms in the time derivative that are perpendicular to $\bs{B}$. This perpendicular component balances the term $\bs{\nabla}_{\perp} \pcr$ and the Lorentz-force term. It thus acts on kinetic time-scales at worst -- a property that we tried to avoid at all costs when deriving a CR hydrodynamics theory for macroscopic astrophysical scales. We thus conclude that it is difficult to treat the Lorentz force in the lab frame on hydrodynamical time-scales. Contrarily, this can be easily achieved in the comoving frame via a straightforward projection operation that is followed by gyroaveraging the CR distribution function.


\bsp	
\label{lastpage}
\end{document}